\documentclass[preprint,Biermann2017,12pt]{elsarticle}

\usepackage{lscape}
\usepackage{graphics}
\usepackage{epsfig}
\usepackage{graphicx}
\usepackage{color}
\usepackage{deluxetable} 
\usepackage{amsmath, amsthm} 

\journal{Advances in Space Research}

\begin{document}
\baselineskip=24pt

\newcommand{\D}{\displaystyle} 
\newcommand{\T}{\textstyle} 
\newcommand{\SC}{\scriptstyle} 
\newcommand{\SSC}{\scriptscriptstyle} 

\newcommand{\be}{\begin{eqnarray}}
\newcommand{\ee}{\end{eqnarray}}

\definecolor{yellow}{rgb}{0.95,0.75,0.1}
\definecolor{orange}{rgb}{0.95,0.4,0.1}
\definecolor{red}{rgb}{1,0,0}
\definecolor{green}{rgb}{0,1,0}
\definecolor{blue}{rgb}{0,0.5,1}

\definecolor{lblue}{rgb}{0,0.8,1}
\definecolor{dblue}{rgb}{0,0,1}
\definecolor{dgreen}{rgb}{0,0.65,0}
\definecolor{lila}{rgb}{0.8,0,0.8}
\definecolor{violet}{rgb}{1,0,0.9}
\definecolor{grey}{rgb}{0.3,0.3,0.3}

\definecolor{contoura}{rgb}{0,0,1}
\definecolor{contourb}{rgb}{0,1,1}
\definecolor{contourc}{rgb}{0,1,0}
\definecolor{contourd}{rgb}{0.95,0.75,0.1}
\definecolor{contoure}{rgb}{1,0,0}
\definecolor{contourf}{rgb}{1,0,1}

\newcommand\cred[1]{\textcolor{red}{#1}}
\newcommand\cblue[1]{\textcolor{blue}{#1}}
\newcommand\ccyan[1]{\textcolor{contourb}{#1}}
\newcommand\cgreen[1]{\textcolor{green}{#1}}
\newcommand\cyellow[1]{\textcolor{yellow}{#1}}
\newcommand\cviolet[1]{\textcolor{violet}{#1}}
\newcommand\cmagenta[1]{\textcolor{magenta}{#1}}
\newcommand\cgrey[1]{\textcolor{grey}{#1}}

\newcommand\cconta[1]{\textcolor{contoura}{#1}}
\newcommand\ccontb[1]{\textcolor{contourb}{#1}}
\newcommand\ccontc[1]{\textcolor{contourc}{#1}}
\newcommand\ccontd[1]{\textcolor{contourd}{#1}}
\newcommand\cconte[1]{\textcolor{contoure}{#1}}
\newcommand\ccontf[1]{\textcolor{contourf}{#1}}

\newcommand {\black} {\color{black}}
\newcommand {\violet}{\color{violet}}
\newcommand {\blue} {\color{blue}}
\newcommand {\dblue} {\color{dblue}}
\newcommand {\cyan} {\color{cyan}}
\newcommand {\green} {\color{green}}
\newcommand {\dgreen} {\color{dgreen}}
\newcommand {\magenta} {\color{magenta}}
\newcommand {\red} {\color{red}}
\newcommand {\orange} {\color{orange}}

\def\AJ{{\it Astron. J.} }
\def\ARAA{{\it Annual Rev. of Astron. \& Astrophys.} }
\def\ARNPS{{\it Annual Rev. of Nucl. \& Part. Sci. } }
\def\ApJ{{\it Astrophys. J.} }
\def\ApJL{{\it Astrophys. J. Letters} }
\def\ApJS{{\it Astrophys. J. Suppl.} }
\def\ApP{{\it Astropart. Phys.} }
\def\AA{{\it Astron. \& Astroph.} }
\def\AAR{{\it Astron. \& Astroph. Rev.} }
\def\AAL{{\it Astron. \& Astroph. Letters} }
\def\AASu{{\it Astron. \& Astroph. Suppl.} }
\def\AN{{\it Astron. Nachr.} }
\def\IJMP{{\it Int. J. of Mod. Phys.} }
\def\JCAP{{\it J. of Cosmol. and Astrop. Phys.} }
\def\JGR{{\it Journ. of Geophys. Res.}}
\def\JHEP{{\it Journ. of High En. Phys.} }
\def\JPhG{{\it Journ. of Physics} {\bf G} }
\def\MNRAS{{\it Month. Not. Roy. Astr. Soc.} }
\def\Nature{{\it Nature} }
\def\NewAR{{\it New Astron. Rev.} }
\def\PASP{{\it Publ. Astron. Soc. Pac.}Ê}
\def\PhFl{{\it Phys. of Fluids} }
\def\PLB{{\it Phys. Lett.}{\bf B} }
\def\PR{{\it Phys. Rev.} }
\def\PRD{{\it Phys. Rev.} {\bf D} }
\def\PRL{{\it Phys. Rev. Letters} }
\def\RMP{{\it Rev. Mod. Phys.} }
\def\RPP{{\it Rep. Pro.Phys.} }
\def\Science{{\it Science}Ê}
\def\ZfA{{\it Zeitschr. f{\"u}r Astrophys.} }
\def\ZfN{{\it Zeitschr. f{\"u}r Naturforsch.} }
\def\etal{{\it et al.}}

\hyphenation{mono-chro-matic  sour-ces  Wein-berg title
chang-es Strah-lung dis-tri-bu-tion com-po-si-tion elec-tro-mag-ne-tic
ex-tra-galactic ap-prox-i-ma-tion nu-cle-o-syn-the-sis re-spec-tive-ly
su-per-nova su-per-novae su-per-nova-shocks con-vec-tive down-wards
es-ti-ma-ted frag-ments grav-i-ta-tion-al-ly el-e-ments me-di-um
ob-ser-va-tions tur-bul-ence sec-ond-ary in-ter-action
in-ter-stellar spall-ation ar-gu-ment de-pen-dence sig-nif-i-cant-ly
in-flu-enc-ed par-ti-cle sim-plic-i-ty nu-cle-ar smash-es iso-topes
in-ject-ed in-di-vid-u-al nor-mal-iza-tion lon-ger con-stant
sta-tion-ary sta-tion-ar-i-ty spec-trum pro-por-tion-al cos-mic
re-turn ob-ser-va-tion-al es-ti-mate switch-over grav-i-ta-tion-al
super-galactic com-po-nent com-po-nents prob-a-bly cos-mo-log-ical-ly
Kron-berg Berk-huij-sen}
\def\simle{\lower 2pt \hbox {$\buildrel < \over {\scriptstyle \sim }$}}
\def\simge{\lower 2pt \hbox {$\buildrel > \over {\scriptstyle \sim }$}}
\def\intunits{{\rm s}^{-1}\,{\rm sr}^{-1} {\rm cm}^{-2}}
%
%
\begin{frontmatter}
\title{{Supernova explosions of massive stars and cosmic rays}}
\author{Peter L. Biermann$^{1,2,3,4}$}
\address{$^{1}$ MPI for Radioastr., Auf dem H{\"u}gel 69, D-53121 Bonn, Germany;\\
$^{2}$ Dept. of Phys., IETP, Karlsruhe Inst. for Tech. (K.I.T.), Wolfgang-Gaedestr. 1, D-76131 Karlsruhe, Germany;
$^{3}$ Dept. of Phys. \& Astron., U. Alabama, Box 870324, Tuscaloosa, AL 35487-0324, USA; {$\;\;\;\;$}
$^{4}$ Dept. of Phys. \& Astron., Univ. Bonn, Nu{\ss}allee 12, D-53115 Bonn, Germany;}
\ead{plbiermann@mpifr-bonn.mpg.de}
\author{Julia Becker Tjus$^{5}$, Wim de Boer$^{2}$, Lauren{\c t}iu I. Caramete$^{6}$, Alessandro Chieffi$^{7}$, Roland Diehl$^{8,9}$, Iris Gebauer$^{2}$, L{\'a}szl{\'o} \'A. Gergely$^{10}$, Eberhard Haug$^{11}$, Philipp P. Kronberg$^{12,13}$, Emma Kun$^{10}$, Athina Meli$^{14, 15}$, Biman B. Nath$^{16}$, Todor Stanev$^{17}$}
\address{$^{5}$ Fak. Phys. \& Astron., Ruhr-Universit{\"a}t Bochum, Universit{\"a}tsstra{\ss}e 150
D-44801 Bochum, Germany;
$^{6}$ Institute for Space Sciences, P.O.Box MG-23, Ro 077125 Bucure{\c s}ti-M{\v a}gurele, Rom{\^ a}nia;
$^{7}$ Ist. Naz. di Astrofis.-Ist. di Astrofis. e Planet. Spaz., Via Fosso del Cavaliere 100, I-00133 Roma, Italy;\\
$^{8}$ MPI extraterr. Phys., D-85748 Garching, Germany;
$^{9}$ Excellence Cluster Universe, Technische Universit{\"a}t M{\"u}nchen, D-85748, Garching, Germany;
$^{10}$ Inst. of Phys., Univ. Szeged, D{\'o}m t{\'e}r 9, H-6720 Szeged, Hungary;
$^{11}$ Inst. for Astron. \& Astrophys., Univ. T{\"u}bingen, Auf der Morgenstelle 10,
D-72076 T{\"u}bingen, Germany;
$^{12}$ Dept of Phys., Univ. Toronto, 60 St George Street, Toronto, ON, M5S 1A7, Canada; 
$^{13}$ Visiting Scholar, Theoretical Division, MS B285, Los Alamos National Laboratory, Los Alamos, NM 87545, USA;
$^{14}$ Dept. of Physics and Astronomy, Univ. of Gent, Proeftuinstraat 86, B-9000, Gent, Belgium;
$^{15}$  IFPA Group, Inst. d'Astrophys. et de G{\'e}ophys., Universit{\'e} de Li{\`e}ge, All{\'e}e du 6 Ao{\v{u}}t, B-4000, Belgium;
$^{16}$ Raman Res. Inst., Sadashiva Nagar, Bangalore 560080, India;
$^{17}$ Bartol Research Inst. \& Dept of Phys. and Astron., Univ. of Delaware, Newark, Delaware DE 19716, USA}

\begin{abstract}

Most cosmic ray particles observed derive from the explosions of massive stars.
Massive stars from slightly above about $10 \, {\rm M_{\odot}}$ explode as supernovae  via a mechanism which we do not know yet: two not mutually exclusive main ideas are an explosion driven by neutrinos, or the magneto-rotational mechanism, in which the magnetic field acts like a conveyor-belt to transport energy outwards for an explosion.  Massive stars above about $25 \, {\rm M_{\odot}}$, depending on their heavy element abundance,  commonly produce stellar black holes in their supernova explosions.  When two such black holes find themselves in a tight binary system they finally merge in a gigantic emission of gravitational waves, events that have now been detected.  The radio interferometric data demonstrate that all of these stars have powerful magnetic winds.  After an introduction (section 1) we introduce the basic concept (section 2):  Cosmic rays from exploding massive stars with winds always show two cosmic ray components at the same time: (i) the weaker polar cap component only produced by Diffusive Shock Acceleration, showing a relatively flat spectrum, and cut-off at the knee, and (ii)  the stronger $4 \pi$ component, which is produced by a combination of Stochastic Shock Drift Acceleration and Diffusive Shock Acceleration, with a down-turn to a steeper power-law spectrum at the knee, and a final cutoff at the ankle.  In section 3 we use the Alpha Magnetic Spectrometer (AMS) data to differentiate these two cosmic ray spectral components; these two cosmic ray components excite magnetic irregularity spectra in the plasma, and the ensuing secondary spectra can explain anti-protons, lower energy positrons, and other secondary particles.  Cosmic ray electrons of the polar cap component interact with the surrounding photon field to produce positrons by triplet pair production, and in this manner may explain the higher energy positron AMS data. In section 4 we test this paradigm with a theory of injection based on a combined effect of first and second ionization potential; this reproduces the ratio of Cosmic Ray source abundances to source material abundances.  We can interpret the abundance data using the relation of the total number of ions enhanced by $Q_{0}^2 \, A^{+2/3}$, where $Q_{0}$ is the initial degree of ionization, and $A$ is the mass number.  This interpretation implies the high temperature as observed in the winds of blue super-giant stars; it also requires that cosmic ray injection happens in the shock travelling through such a wind. Most injection happens at the largest radii before slowing down due to interaction with the environment. In section 5 we interpret the compact radio source 41.9+58 in the starburst galaxy M82 as a recent binary black hole merger, with an accompanying gamma ray burst. The tell-tale observational sign is the conical cleaning sweep of the relativistic jet during the merger, observed as an open cone with very low radio emission. This can also explain the Ultra High Energy Cosmic Ray (UHECR) data in the Northern sky.  Thus, by studying the cosmic ray particles, their abundances at knee energies, and their spectra, we can learn about what drives these stars to produce the observed cosmic rays.

\end{abstract}


\end{frontmatter}

\vskip2.0cm

\section{Introduction}

The origin of Cosmic Rays (CRs), directly observed energetic particles, is still not fully understood. But with a large number of experiments we now have a basis to ask better questions.  These particles, discovered in 1912 by Hess, extend in energy up to $10^{21}$ eV. They have characteristic spectral features, referred to as the {\it knee}, where the overall spectrum turns down around a few $10^{15}$ eV, as well as another transition near $3 \cdot 10^{18}$ eV, referred to as the {\it ankle}, where the overall spectrum turns up.  This turn-up is often believed to be the likely transition between a cosmic ray origin in our Galaxy and an origin in extragalactic sources. The spectrum shows a final turn-down around $10^{20}$ eV.  Today we also have  a variety of observational support for energetic particles elsewhere in the universe. In our Galaxy as well as in other galaxies and active galactic nuclei (AGN) we observe non-thermal radio emission, optical emission, X-ray emission, gamma-ray emission, and high energy neutrinos.  Are these populations actually the same and if not, in what way are these different CR populations related?  Already in 1934 Baade \& Zwicky argued that supernova explosions are the most plausible source of Galactic Cosmic Rays (GCRs).  In 1949 and 1954 Fermi showed how repeated interactions with a converging magnetic flow could lead to a power-law in energy or CR intensity.  Some relevant recent reviews and books on energetic particle physics are Beck et al. (1996), Aharonian (2004), Stanev (2010), Letessier-Selvon \& Stanev (2011), Diehl et al. (2011), Bykov et al. (2012), Diehl (2013, 2017),  Blasi (2013), Gaisser et al. (2016a), Kronberg (2016), Amato \& Blasi (2017). Earlier work is Berezinsky et al. (1990), with an extensive review by Ginzburg \& Ptuskin (1976).  Some older fundamental books are Heisenberg (1953), Ginzburg \& Syrovatskij (1963), Hayakawa (1969), with a review by L. Biermann (1953) and the original article collection edited by Rosen (1969).  A more advanced theory, called Diffusive Shock Acceleration (DSA), was then derived in a series of papers by Krymsky (1977), Axford et al. (1977), Blandford \& Ostriker (1978), and Bell (1978a, b). A thorough review was given by Drury (1983).   All these early studies are based on a supernova shock advancing through an ionized and magnetic medium.  Jokipii (1982, 1987) then added drift acceleration (here used as Stochastic Shock Drift Acceleration, StSDA); today we believe that both, DSA and StSDA,  are important (e.g., Lee et al. 1996, Ball \& Melrose 2001, Meli 2012, Le Roux et al. 2015, Zank et al. 2015, Li et al. 2017, Matsumoto et al. 2017).  We note that in the StSDA the shock transition itself is important only to define two different regimes of gradient and curvature drift. It also defines a time scale given by large scale turbulence to switch between upstream drifts and downstream drifts (see, e.g., Meli 2012, Li et al. 2017, Matsumoto et al. 2017).  As a third ingredient such shocks are unstable (Zank et al. 1990, Zakharian et al. 1999, Li et al. 2004):  CRs and magnetic fields mixing with normal plasma are discussed in Biermann (1994b).  The unstable shocks were again emphasized in Bell \& Lucek (2000, 2001) and Amato \& Blasi (2006), who show how magnetic fields can be strongly enhanced, increasing violent motions.  We also note that this turbulence leads to energy gains from higher curvature drift.  So all stars that explode as Supernovae (SNe) may produce CRs: Stars between about 10 and about $25 \; M_{\odot}$ Zero Age Main Sequence (ZAMS) masses explode into the Inter-Stellar Medium (ISM); Red and Blue Super Giant stars (RSG and BSG stars), with a higher ZAMS mass, explode into their wind.  Today additional sources are considered, such as pulsar wind nebulae (e.g. Bykov et al. 2017a), micro-quasars (Mirabel 2004, 2011), Gamma Ray Bursts (GRBs; Piran 2004), SNe type Ia (Scalzo et al. 2014), neutron star mergers (Abbott et al. 2017c), black hole mergers (Kun et al. 2017), etc.; these may be overlapping classes:  The contribution of GRBs was recognized a while ago, first qualitatively by analogy with jets from AGN (Biermann 1994a, b), then quantitatively (Milgrom \& Usov 1995, 1996; Vietri 1995, 1996, 1998; Waxman 1995, Miralda-Escud{\'e} \& Waxman 1996, Waxman \& Bahcall 1997; Zhang \& M{\'e}sz{\'a}ros 2004; Piran 2004).  An interpretation of IceCube data casts doubt on a connection between GRBs and high energy neutrinos (Abbasi et al. 2012, Aartsen et al. 2017b), as does a corresponding ANTARES analysis (Albert et al. 2017a).  Today it is recognized that GRBs and relativistic SNe are a special and small subset of very massive stars (e.g., Woosley et al. 2002, Heger et al. 2003, Soderberg et al. 2010a, b, Kamble et al. 2014).   Thoudam et al. (2016) show that to the understand the CR spectrum and chemical composition a contribution from very massive stars such as Wolf-Rayet (WR) stars is required, stars that explode into their winds.  This implies that we also include RSG stars (e.g., Woosley et al. 2002). The Fermi and H.E.S.S. observations of cosmic ray interaction with ambient gas producing $\gamma$-rays at GeV to TeV gamma energies (Abramowski et al. 2014) are consistent with a production by nuclei at relatively high energy such as given by RSG and BSG star explosions.  The SN rate in our Galaxy (Diehl et al. 2006, 2010), is approximately one per 75 years for all SNe of stars of more than 10 $M_{\odot}$ Zero Age Main Sequence (ZAMS) mass.  The explosions of BSG stars are a fraction of order 15\% of such SNe.  If we assume a causal connection these could all be stars above a ZAMS mass of about 33 $M_{\odot}$ (Heger et al. 2003, Chieffi \& Limongi 2013).  They are assumed to constitute efficient CR accelerators beyond the knee (Stanev et al. 1993, Todero Peixoto et al. 2015); then we expect about one such explosion per galaxy and per 600 years. We infer from observations many more powerful radio supernovae (RSNe) in the starburst galaxy M82 (Kronberg et al. 1985, Allen \& Kronberg 1998) at a distance of only about 3 Mpc; this can be attributed to a higher SN rate, as well as to the higher pressure in M82's ISM.  In a transition from a wind-SN to a Sedov-SN (i.e. energy conserving SN shocks) as the shock hits the surrounding medium, the luminosity scales with the magnetic field $B$ as $B^{1.7}$ (Biermann \& Strom 1993): Hence the lower ISM pressure and magnetic field in our galaxy imply in a steady state, supernova evolution in M82 occurs at a luminosity many powers of ten above that in our Galaxy.   At the rate of one such explosion about every 600 years and a very short duration of high radio emission, the fact that we have never identified such an explosion in about 70 years of radio observations is understandable.  Including stars down to about 25 $M_{\odot}$, so RSG star explosions as well, shortens this time scale and changes the rate to one every 400 years. 

It has been recognized early that propagating CR particles interact in the interstellar medium (ISM). Already in the 1940s it was clear that their propagation is chaotic, confined by magnetic fields, and adequately described by some sort of diffusion process (Chandrasekhar 1943, L. Biermann 1950, L. Biermann \& Schl{\"u}ter 1950, L. Biermann \& Schl{\"u}ter 1951).  Predictions were made in the 1980s for anti-protons and positrons, and other secondary particles, by, e.g., Protheroe et al. (1981), Protheroe (1981, 1982).  One well known process is the formation of neutral or charged pions, which decay into electrons, positrons, neutrinos, and/or photons (e.g. Stecker 1970, 1971).  Another process is, e.g., the formation of unstable nuclei that emit either a positron, an electron or a $\gamma$-photon upon decay. As an example Alexis et al. (2014) discussed the 511 keV annihilation emission, based on nuclear $\beta^{+}$ decays with emission of positrons.  Morlino (2009) discussed the injection of CR-electrons from ionization of CR-atoms.  The relevance of specific classes of massive stars for CR abundances has been recognized from the 1980s by Prantzos et al. (Prantzos 1984, 1991, 2012a, b; Prantzos \& Cass{\'e} 1986, Prantzos \& Diehl 2011, Prantzos et al. 1993, 2011).   Based on earlier work, Strong \& Moskalenko  developed a general CR propagation code (GALPROP:  Moskalenko \& Strong 1998, Moskalenko et al. 1998, 2002, 2003a, b, 2006, Strong et al. 1997, 2007, 2009).  Cowsik et al. explain the latest AMS data, using a nested leaky box (Cowsik et al. 2014, Cowsik \& Madziwa-Nussinov 2016, and Cowsik 2016).  Strong \& Moskalenko (as summarized in Strong et al. 2007) have explored in great detail the constraints given by interaction and propagation in the ISM in our galaxy, e.g. J{\'o}hannesson et al. (2016); they reaffirm that the new AMS data require different paths of interaction for different spallation products such as Lithium, Beryllium, Boron, anti-protons and positrons.  Based on the irregularity spectrum in the ISM, we use here a diffusion coefficient with energy $E$ dependence of $E^{1/3}$ to describe the CR transport and derive the observed secondary CR spectrum (Kolmogorov 1941a, b, Armstrong et al. 1995, Goldstein et al. 1995, Haverkorn et al. 2013, Iacobelli et al. 2013, 2014, Fraternale et al. 2016, summarized in Biermann et al. 2015).  However, as the SN-shock hits the wind-shell around a star, we use a locally excited spectrum of magnetic irregularities (e.g. Biermann 1998, Biermann et al. 2001, 2009). This constitutes a local ``nested leaky box".  The locally excited spectrum leads to an energy dependence of $E^{-5/9}$ of the secondary/primary ratio in the low-energy regime. This is very close to the often used $E^{-0.6}$. At higher energy this locally excited spectrum yields the energy dependence $E^{-1/3}$ (Kolmogorov 1941a, b).  Our inner nested box model is derived from observed properties of massive stars and their environment (V{\"o}lk \& Biermann 1988, Biermann 1993, Biermann \& Cassinelli 1993, Biermann \& Strom 1993, Stanev et al. 1993, Biermann 1998, Biermann et al. 2001, 2009).  As discussed in Stanev et al. (2014) and Dembinski et al. (2017) with the new IceTop data given in Gaisser et al. (2016b) new fits to the data were described.  

In this article we will address how the recent observational data can be included in a theoretical description of cosmic ray production: 





1a)  We first describe the radio data of SNe of very massive stars, and the young radio-supernovae (RSNe) in the starburst galaxy M82. Using the approaches of Parker (1958) and Cox (1972) we then derive the radial dependence of the magnetic field and the shock speed for explosions into a stellar wind (for an early discussion see V{\"o}lk \& Biermann 1988), differentiating between RSG stars with a slow, dense wind and BSG stars with a fast, tenuous wind.

1b)   We rederive the  two-component CR model (e.g. Biermann 1993), in which (in its simple limit) cosmic ray particles stemming from a SN-shock running through stellar winds always have two populations, the weak polar cap component and the much stronger $4 \, \pi$-component:  The source spectrum below the knee of the polar cap component is $E^{-2}$ and for the $4 \, \pi$-component it is $E^{-7/3}$.  Due to its flatter spectrum the polar cap component always dominates at higher energies.  Beyond the knee the polar cap component fades, and the $4 \, \pi$ component takes over, with a steeper power-law  and drop in intensity at the ankle energy.  

2)  The AMS data on positrons, anti-protons, Lithium, Beryllium, Boron, and more common elements such as Hydrogen, Helium, Carbon, Oxygen, etc. can be explained with propagation and interaction of freshly accelerated cosmic rays in a turbulent wave-field excited by the two components of the cosmic rays themselves, in the immediate neighborhood of the exploded star and its wind.

3)  The ratio of Galactic Cosmic Ray Source (GCRS) abundances to source material abundances  (e.g. Murphy et al. 2016) as a function of atomic mass number $A$ is  explained as the interplay of the first and second ionization potential with injection and acceleration.

4)  One speculative interpretation of the compact radio source 41.9+58 in the starburst galaxy M82 (Kronberg et al. 1985) is the merger of two stellar mass black holes with the powerful emission of gravitational waves (GWs). A subsequent GRB powers the emission of Ultra High Energy Cosmic Ray (UHECR) protons, as detected by the Telescope Array (TA).  This scenario has some similarity to the neutron star merger GW170817 in the galaxy NGC4993 (Abbott et al. 2017c, d, Albert et al. 2017b).

With our treatment we aim to establish relevant scales for the violent motions in the cosmic ray/magnetic fields/ionized plasma shock system. These scales are important for deducing the proportions of gradient and curvature drift that contribute to the stochastic shock drift acceleration (StSDA), in addition to the diffusive shock acceleration (DSA), e.g. Jokipii (1982), Drury (1983).  We recognize the spectrum of magnetic irregularities is excited by the CRs themselves.  Important physical ingredients to consider are (i) pitch angle scattering, (ii) feeding into shock acceleration, and (iii) the dependence of these two processes on charge $Q_{0}$ and mass $A$ of an ion.  We propose a cosmic ray description that can be tested with further observations.

\vskip2.0cm

\section{Massive star explosions observed at radio frequencies: M82 candidates and young radio supernovae}




For many years, data giving us insight on the magnetic field in exploding massive stars had been rare, but lately the situation has been greatly improved, and the limited data suggest, surprisingly, rather common properties.

Fortunately there are now interferometric radio data on massive stars, stars that explode into their predecessor winds (see V{\"o}lk \& Biermann 1988, for an early discussion of the consequences for cosmic ray (CR) acceleration, and earlier data).  These observations yield magnetic field, shock speed, and energetic electron spectrum as a function of radius and time; modeling these data also yields information on the prior wind mass loss.  The following tables give information on these observations:

\begin{table}[h!]
\begin{center}
\caption{Some radio supernova (RSN) data}
\begin{tabular}{|c|c|c|c|c|c|c|}
\hline
\hline
Name&type&progen.&$\log\left(\frac{R_{sh}}{\rm cm}\right)$&$\log\left(\frac{B_{sh}}{\rm Gau{ss}}\right)$&$\log(\Gamma_{sh} \, U_{sh}/c)$&ref.\\
\hline
1993J  &IIb  & RSG & 16.0 & +0.8 & -1.2 & (r1)\\
2003L&Ibc    & BSG & 15.7 & +0.6 & -0.7 & (r2)\\
2003bg&Ic/Ibc& BSG & 16.2 & +0.2 & -0.6 & (r3)\\
2007gr&Ic   & BSG & 15.7 & -0.4 & -0.7 & (r4)\\
2008D&Ibc    & BSG & 15.5 & +0.4 &$<$-0.2&(r7)\\
2008iz& II   & RSG & 16.0 & 0.0  & -1.1 & (r9,r11,r14)\\
2011dh&IIb   & BSG & 15.6 & +0.1 & -1.0 & (r5,r13,r16)\\
2011ei&IIb/Ib& BSG & 15.5 & -0.2 & -0.9 & (r6)\\
2012au&  Ib    & BSG & 16.2 & -0.4 & -0.7 & (r12,r15)\\
2013df&IIb   & RSG & 16.0 & -0.1 & -1.2 & (r8)\\
1998bw& rel. & BSG & 16.8 & -0.4 & +0.3 & (r12)\\
2012ap& rel.(i)  & BSG & 16.0& 0.0 &-0.3& (r10)\\
2012ap& rel.(ii) & BSG & 16.0& 0.0 & 0.0& (r10)\\
\hline
\hline
\end{tabular}
\label{tableRSNkneeI}
\end{center}
\end{table}

Here the references are, in sequence, r1: Fransson \& Bj{\"o}rnsson (1998); r2: Soderberg et al. (2005); r3: Soderberg et al. (2006);  r4: Soderberg et al. (2010a); r5: Krauss et al. (2012); r6: Milisavljevic et al. (2013a); r7:  Soderberg et al. (2008); r8:  Kamble et al. 2016; r9:  Kimani et al. (2016); r10: Chakraborti et al. (2015); r11: Muxlow et al. (2010); r12: Kamble et al. (2014); r13: Bietenholz et al. (2012); r14:  Brunthaler et al. 2010; r15: Milisavljevic et al. (2013b); r16:  Soderberg et al. (2012).  These papers also use data from other wave-lengths such as X-rays or optical, which never have the spatial resolution that the interferometric radio data can easily supply.  If the inferred wind velocity of the exploding star was estimated as about $10^3$ km/s, we classified it as Blue Super Giant (BSG) star, and if the estimated wind velocity was estimated to be of order 10 km/s, we classified it here as Red Super Giant (RSG) star.  In many cases the star was directly classified as a Wolf-Rayet (WR) star, which is a BSG star  (WR stars are a subset of the BSG stars, see Maeder et al. 2005).

\begin{table}[h!]
\begin{center}
\caption{Some radio supernova (RSN) data and results}
\begin{tabular}{|c|c|c|c|c|c|}
\hline
\hline
Name&type&progen.&$\log\left(\frac{\dot{M}_{\star}}{ M_{\odot} yr^{-1}}\right)$&$\log\left(\frac{R_{sh} \, B_{sh}}{\rm Gau{ss} \, cm}\right)$&$\log\left(\frac{R_{sh} \, B_{sh} \, \{U_{sh}/c\}^2}{\rm Gau{ss} \, cm}\right)$\\
\hline
1993J&IIb    & RSG & -5.0 & 16.8 & 14.4\\
2003L&Ibc    & BSG & -5.1 & 16.3 & 14.9\\
2003bg&Ic/Ibc& BSG & -3.5 & 16.4 & 15.2\\
2007gr&Ic   & BSG & -6.2 & 15.3 & 13.9\\
2008D&Ibc    & BSG & -5.1 & 15.9 & 15.5\\
2008iz& II 	 & RSG & -4.4 & 16.0 & 13.8\\
2011dh&IIb   & BSG & -4.5 & 15.7 & 13.7\\
2011ei&IIb/Ib& BSG & -4.9 & 15.3 & 13.5\\
2012au& Ib     & BSG & -5.4 & 15.8 & 14.4\\
2013df&IIb   & RSG & -4.1 & 15.9 & 13.5\\
1998bw& rel. & BSG & -6.6 & 16.4 & ... \\
2012ap& rel.(i)  & BSG & -5.2& 16.0 & 15.4 \\
2012ap& rel.(ii) & BSG & -5.2& 16.0 & 16. \\
\hline
Mean and & mean & error & $- 5.1 \pm 0.2$ & $15.9 \pm 0.2$&$14.3 \pm 0.2$\\
 & & & & 2008D vel. &  limit used;  \\
 & & & &  no rel.&  case used   \\
\hline
\hline
\end{tabular}
\label{tableRSNkneeII}
\end{center}
\end{table}

In table \ref{tableRSNkneeI} we show from left to right: in the first column the name of the supernova (SN) explosion; then type of star which exploded: a RSG or a BSG star; next the radius $R_{sh}$ at which the determinations were made, usually close to $10^{16}$ cm, in the form $\log\left(\frac{R_{sh}}{\rm cm}\right)$, which is the typical radius for radio data due to optical depth effects; next the magnetic field determined in the shocked region $B_{sh}$ in the form $\log\left(\frac{B_{sh}}{\rm Gau{ss}}\right)$; then the velocity of the shock including a possible Lorentz factor of the shock $\Gamma_{sh} U_{sh}/c$ in the form $\log(\Gamma_{sh} \, U_{sh}/c)$; and finally the references used. In table \ref{tableRSNkneeII} we first show again the name as an identifier; then again the type of star exploded; then the stellar mass loss $\dot{M}_{\star}$ determined from a model used by the observer teams (expanded on below) in the form $\log\left(\frac{\dot{M}_{\star}}{ M_{\odot} yr^{-1}}\right)$; then the derived quantity $R_{sh} \; B$ in the form $\log\left(\frac{R_{sh} \, B_{sh}}{\rm Gau{ss} \, cm}\right)$; and finally the derived quantity $R_{sh} \; B \; (U_{sh}/c)^2$ in the form $\log\left(\frac{R_{sh} \, B_{sh} \, \{U_{sh}/c\}^2}{\rm Gau{ss} \, cm}\right)$.  At the bottom of the table we list the average (with mean error) for three of these quantities, $\log\left(\frac{\dot{M}_{\star}}{ M_{\odot} yr^{-1}}\right)$; $\log\left(\frac{R_{sh} \, B_{sh}}{\rm Gau{ss} \, cm}\right)$; and finally $\log\left(\frac{R_{sh} \, B_{sh} \, \{U_{sh}/c\}^2}{\rm Gau{ss} \, cm}\right)$.

The average (with mean error) for  the mass loss is $\log(\dot{M}_{\star}/\{ M_{\odot} yr^{-1}\})$ is $- 5.1 \pm 0.2$.  The average and mean error for $\{\log(R_{sh} \, B_{sh} \{U_{sh}/c\}^2)\}$ is $14.3 \pm 0.2$, while the average and mean error for $\{\log (R_{sh} \, B_{sh})\}$ is $15.9 \pm 0.2$.  These latter two quantities correspond to  characteristic energies, $ (1/8) \, e \, Z \, R_{sh} \, B \, (U_{sh}/c)^2$ and $(1/8) \, e \, Z \, R_{sh} \, B$, where $e$ is the elementary charge, and $Z$ is the numerical charge of a CR nucleus considered. We interpret these two quantities below as knee and ankle.  The radio data, including those from the compact radio sources in M82 (almost all of which are interpreted as exploded BSG stars; see below), are consistent with an interpretation that the shocks running through winds have a radial run of the magnetic field using $r^{-1}$, corresponding to a density dependence of $r^{-2}$ as long as the shock velocity is constant; the late-time dependence of these data also shows that, prior to the explosion, the wind sometimes changed.  These arguments here emphasize the SN-shock racing through a stellar wind with such properties; below we will discuss the ensuing complications when the piston driving the shock runs out of steam, and also how the compact radio sources in M82 help us understand the long term evolution.
 
Soderberg et al. (2010b) focussed on the early phase, when the non-thermal radio emission becomes optically thin at peak luminosity $L_{\nu, p}$, and showed both the distribution of the shock speed as well as the associated peak radio luminosity for a sample of radio SNe.  Their distributions confirm that $U_{sh}/c$ is typically about 0.1, the magnetic field is typically 0.4 Gau{ss} at the associated radius $r$ of $3 \cdot 10^{16}$ cm.  This is consistent with 1.2 Gau{ss} at $10^{16}$ cm at $B \, \sim \, r^{-1}$, at their choice of nominal parameters: These parameters are (i) the fraction of post-shock energy density in electrons relative to that in magnetic fields $\epsilon_{e}/\epsilon_{B}$; (ii) the filling fraction $f/0.5$; and (iii) the observed peak radio luminosity $L_{\nu, p}/\{10^{28} \, {\rm erg \, Hz^{-1} \, s^{-1}}\}$ at a chosen reference frequency of $\nu/\{5 \, {\rm GHz}\}$. The derived product (using the expressions in Soderberg et al. 2010b) of radius and magnetic field is only weakly dependent on the input numbers, with 

\begin{equation}
r \, B \; \simeq \; 1.3 \cdot 10^{16} \, {\left(\frac{\epsilon_{e}}{\epsilon_{B}}\right)}^{-4/19} \, {\left(\frac{L_{\nu, p}}{10^{28} \, {\rm erg \, s^{-1} \, Hz^{-1}}}\right)}^{7/19} \, {\left(\frac{f}{0.5}\right)}^{-4/19} \, {\rm Gauss \, cm} \, ,
\end{equation}
\noindent independent of the associated radio frequency.  Their model is rather general, based on earlier work by Chevalier (1982, 1996, 1998),  Chevalier \& Blondin (1995), Chevalier \& Li (2000), and Chevalier \& Fransson (2006).  Our interpretation uses a predecessor stellar wind, as is often seen in the radio data. All very massive stars that explode have a wind before the explosion, but of course, it is open to question how this impacts what we observe (see, e.g., V{\"o}lk \& Biermann 1988, Biermann 1993, Biermann \& Cassinelli 1993, Biermann \& Strom 1993, Biermann 1994b, Meli \& Biermann 2006, Biermann 2014, Todero Peixoto et al. 2015).   

We can rederive the expression using explicitly a wind as in the modelling used below in the next section (using an approximated CR spectrum of $E^{-3}$ close to what is observed for electrons in the early phase):

\begin{equation}
r \, B \; \simeq \; 3. \, 10^{16} \, {\left({\left(\frac{L_{\nu}}{10^{28} \, {\rm erg \, s^{-1} \, Hz^{-1}}}\right)} \, {\left(\frac{1 + k_{CR}}{101}\right)} \,  {\left(\frac{r}{10^{16} \, {\rm cm}}\right)} \, {\left(\frac{\nu}{ 5 \, {\rm GHz}}\right)}\right)}^{1/4} \, {\rm Gau\ss \, cm} \, ,
\end{equation}
\noindent where $k_{CR}$ is the ratio of the energy density in energetic nucleons relative to electrons, using the observed ratio of 100.  We assume equipartition between energetic nucleons and magnetic fields. We adopt the optically thin case, and yet at the nominal parameter values this yields a very similar result as in Soderberg et al. (2010a), as cited above.  Applying this expression to the compact sources in M82 yields about the same value for the product $r \, B$, supporting the view that the expansion from a radial scale of $10^{16}$ cm to pc scale is unimpeded, with indeed $B \, \sim \, r^{-1}$, suggesting these were all BSG stars with tenuous fast winds. We discuss necessary piston masses further below.  This finding will be important further below, when we discuss CR injection.

We translate these observed numbers from the table into accelerated particle energies (see Biermann 1993, and below): This gives (i)  at nominal parameters an energy of $10^{17.5 \pm 0.2} \, Z$ eV which we interpret as the {\it ankle} of CRs, as $(1/8) \, e \, Z \, r \, B$: here $Z$ is again the charge of the CR nucleus; (ii) the second energy is $10^{15.9 \pm 0.2} \, Z$ eV, as $(1/8) \, e \, Z \, r \, B \, (U_{sh}/c)^2$, which we interpret as the {\it knee}.  This relationship in acceleration is due to the dependence of the Jokipii (1987) perpendicular scattering coefficient, $\kappa \sim E \, r \, U_{sh} / Z$ in a magnetic wind; balancing the acceleration time $\kappa/U_{sh}^2$ with the flow time $r/U_{sh}$ then allows the spatial limit (Hillas 1984) to be reached (see also below, and Biermann 1993).  We note here, that we do not assume that the magnetic field is parallel to the shock normal (see, e.g., Jokipii (1982, 1987), Biermann (1993), Meli \& Biermann (2006)); this assumption is implicitly made in Chakraborti et al. (2011), a follow-up paper to Soderberg et al. (2010b). In fact, the Parker (1958) limit solution ($B_{\phi} \, \sim \, r^{-1}$) taken together with the radio observations shown above (Table \ref{tableRSNkneeI} and Table \ref{tableRSNkneeII}), and the M82 compact source observations (Kronberg et al. 1985; explained below) demonstrate that the magnetic field runs as $r^{-1}$ all the way out for BSG star winds, the most favorable case.  This implies that a SN-shock is perpendicular, so fulfilling the assumption in Jokipii's (1987) argument.

We can summarize these radio data of observed massive stars exploding into their wind:  (i)  Reference radial distance $r_0$ for such radio observations $r$ (due to optical depth effects); typical is $r_0 = 10^{16}$ cm.  (ii)  Upstream shock velocity $U_{sh,1}$; typical is $U_{sh,1} \simeq 0.1 \; c$. (iii)  Magnetic field $B(r_0)$; typical is $B(r_0) \simeq $ 1 Gauss.  (iv)  Radial dependence; typical is $B(r) \sim r^{-1}$.  All these numbers and relationships are derived from radio observations.  However, we need to caution that this behavior is determined over only a very limited range of radii, and quite uncertain. Only the comparison with the RSNe in the starburst galaxy M82 allows us to conclude that this radial behavior extends quite far out. (v)  Electron spectral index; typically  -3, steeper since the electrons required to explain the radio emission have such a low energy, that they only experience drift energy gains. They do not ``see" the shock (see below); however, here optical depth effects can confuse a clear determination of the original spectral index.  (vi)  The same numbers for the $B(r_0)$, $U_{sh,1}$ values are true by order of magnitude both for BSG star winds with low density  (or WR stars, see Maeder et al. 2005), as well as for RSG star winds with high density:    However, the data set is quite limited here, with 9 BSG stars and 3 RSG stars, and needs to be expanded, as Soderberg et al. (2010b) have tried to do.

We can conclude the following from these data:  This just seems to exclude the Bell-Lucek mechanism (2000, 2001) to explain the magnetic field as arising from the supernova shock. The mechanism depends on density; the densities in the winds of RSG stars and BSG stars are very different, while the shock velocity in the supernova explosion is about the same.  However, if the magnetic field were to depend on the ionized upstream density, then the Bell-Lucek mechanism (but see Amato \& Blasi 2006) just might be viable for both kinds of stars, provided that the degree of ionization is kept approximately constant at all radii around a value of $10^{-2}$ by cosmic ray action for RSG stars. In such a case the relevant ionized density in the wind would be similar to that in BSG stars.  However, as noted below, the magnetic field in the stellar wind being traversed by the SN-shock cannot easily have a magnetic field as strong as required, so that the magnetic field seen post-shock could simply come from the density enhancement in a perpendicular shock. We speculate on other possibilities below.

On the other hand, scaled to the surface using the Parker limit (i.e. $B_{\phi} \, \sim \, r^{-1}$), these magnetic fields are observed to be much stronger than on the main sequence, where they are detected at about few hundred Gau{ss}  on the surface of the stars (Martins et al. 2010; Wade et al. 2011, 2016; Hubrig et al. 2014; Kholtygin et al. 2015).  So a serious enhancement of the magnetic fields in the wind going from the main sequence to the super-giant star pre-SN phase may be a solution to this conundrum: This may partially derive from a pre-SN activity commonly observed for such stars (Svirski \& Nakar 2014a, b; Gal-Yam et al. 2014, Ofek et al 2014, Strotjohann et al. 2015, Tartaglia et al. 2016).  Another option is indeed some enhancement of the magnetic field during the supernova shock advance through wind, possibly due to the Bell-Lucek (2000, 2001) mechanism (but see also, e.g., Amato \& Blasi 2006, Fraschetti 2013, Mizuno et al. 2014, ).  An alternative could be that the magnetic fields are pulled along by the piston from the highly magnetized layers deep inside pre-SN star, now exposed and visible in star's wind abundances, and mixed into the post-shock region. Isotope abundances in cosmic rays observed might shed light on this speculation.   This latter mechanism might allow us to understand, why the magnetic field is always the same order of magnitude.

\subsection{Supernova shocks in stellar winds}

Consider a shock driven by a supernova explosion running through the wind of the predecessor star (Parker 1958, Weber \& Davis 1967, V{\"o}lk \& Biermann 1988, Biermann \& Cassinelli 1993, Biermann 1997, Seemann \& Biermann 1997). We assume that this wind has a density structure of $r^{-2}$ (a steady wind), a magnetic field of $B \, \sim \, r^{-1}$ (Parker 1958:  lines of force of the magnetic field coincide with the stream lines), a constant wind velocity of $V_{W}$ (again a steady wind), an associated Alfv{\'e}n velocity $V_A \, < \, V_W$ (Weber \& Davis 1967, otherwise there would be excessive angular momentum transport), as well as a piston driving it of mass $M_{piston}$.  The shocked region in the wind comprises the radial fraction of $1/4$ from the Rankine-Hugoniot shock conditions.  We examine these assumptions below.  The accumulated mass from sweeping up wind material can be written as

\begin{equation}
\Delta M_{W} \; = \; 4 \, \pi \, r^2 \, \frac{r}{4}  \, \rho_{W, 0} \, 4 \, \left(\frac{r_0}{r}\right)^2 \; = \; 4 \, \pi \, r \, {r_0}^2 \, \rho_{W, 0} \, ,
\end{equation}
\noindent where $r$ is given by $\dot{r} \; = \; U_{sh}$, now allowed as a variable.  The factor of $1/4$ derives from the thickness of the shell of $r/4$, and the factor of $4$ derives from the density jump in the shock (strong shock with adiabatic gas constant 5/3).  As reference we use again $r_0 \, = \, 10^{16} \, {\rm cm}$.  The accumulated mass slowly rises linearly with radius $r$.

The energy equation can be written more generally as

\begin{equation}
\left(M_{piston} + \Delta M_{W}\right) \, \frac{1}{2} \, {\dot{r}}^2 \; = \; E_{SN} \, ,
\end{equation}
\noindent where $\dot{r} \; = \; U_{sh}$.  The energy can be written as an initial condition, since at first the accumulated wind mass is negligible, so that

\begin{equation}
M_{piston} \frac{1}{2} \, U_{sh, init}^2 \; = \; E_{SN} \, ,
\end{equation}
\noindent where $U_{sh, init}$ is the initial shock velocity; however, we will assume that this velocity is constant, until the accumulated mass exceeds the piston mass.  This equation can be integrated to give

\begin{equation}
4 \, \pi \, r \, {r_0}^2 \, \rho_{W, 0} \; = \; \left({\left(6 \, \pi \, {r_0}^2 \, \rho_{W, 0} \left(2 \, E_{SN}\right)^{1/2} \, t + \, M_{piston}^{3/2}\right)}^{2/3} \, - \, M_{piston}\right) \, .
\end{equation}
\noindent It can be immediately seen that this expression has the correct limits for $M_{piston} \, -$$> \, 0$.  We obtain

\begin{equation}
r \, \sim \, t^{2/3} \, ,
\end{equation}
\noindent and for $M_{piston}$ large we obtain

\begin{equation}
r \, = \, {\left(\frac{2 \, E_{SN}}{M_{piston}}\right)}^{1/2} \, t \, = \, U_{sh} \, t \, ,
\end{equation}
\noindent assuming here, that the shock velocity $U_{sh}$ is constant, with the switch-over at 

\begin{equation}
4 \, \pi \, r \, {r_0}^2 \, \rho_{W, 0} \; \simeq \; M_{piston} \, .
\end{equation}
\noindent Now to put numbers into this, let us take the example of a BSG star, with wind velocity $V_W \, = \, 2000 \, V_{W, 8.3} \, {\rm km/s}$, and a stellar mass loss of $10^{-5} \, {\dot{M}}_{-5} \, {\rm M_{\odot} \, yr^{-1}}$.  The reference density $\rho_{W, 0}$ is given by

\begin{equation}
10^{-5} \, {\dot{M}}_{-5} \, {\rm M_{\odot} \, yr^{-1}} \; = \; 4 \, \pi \, r_0^2 \, \rho_{W, 0} \, V_W \, ,
\end{equation}
\noindent giving with our example

\begin{equation}
\rho_{W, 0} \; = \; 10^{-20.6} \, {\dot{M}}_{-5} \, V_{W, 8.3}^{-1} \, {\rm g \, cm^{-3}} \, ,
\end{equation}
\noindent and an accumulated mass of 

\begin{equation}
\Delta M_{W}(r) \, = \; 4 \, \pi \, r \, {r_0}^2 \, \rho_{W, 0} \; = \; 10^{-4.8} \, {\rm M_{\odot}} \, \frac{r}{r_0} \, .
\end{equation}

These numbers can be checked using observations of binary stars, in which one partner is a super-giant star, and the other partner is a compact star:  Such objects have been seen in gamma rays and X-rays, allowing their column density to be determined (e.g. Walter \&  Zurita Heras 2007, Butler et al. 2009, Tomsick et al. 2009, Manousakis \& Walter 2010).  The large numbers for the column density deduced are consistent with the values implied here ($10^{21}$ to $10^{24}$ cm$^{-2}$), depending on whether we observe a compact object circling a BSG or an RSG star.  Furthermore, the observations demonstrate that these winds are clumpy, consistent with expectations (e.g. Owocki et al. 1988).

Assuming the wind itself to be super-Alfv{\'e}nic requires the magnetic field in the unperturbed wind to be (written as a constraint on the magnetic field at radial distance $r_0$, and assuming that the magnetic field runs as $r^{-1}$):

\begin{equation}
B_{0} \; < \, 10^{-1.45} \, {{\dot{M}}_{-5}}^{1/2} \, {V_{W, 8.3}}^{1/2} \, {\rm Gau{ss}} \, .
\end{equation}
\noindent We note again that a sub-Alfv{\'e}nic wind would transport angular momentum excessively (Parker 1958, Weber \& Davis 1967, as well as Seemann \& Biermann 1997).

This requires an extra factor of enhancement of the magnetic field in the shock of at least $10^{0.85}$, i.e. beyond the simple enhancement in a strong perpendicular shock; if the Alfv{\'e}nic Mach number is assumed  to be of order 3 (see Seemann \& Biermann 1997 for an argument on this number based on instabilities in driving the wind), then the enhancement required is correspondingly higher, at $10^{1.35}$.  So some enhancement by a mechanism such as the Bell-Lucek (2000, 2001) concept (see also, e.g., Fraschetti 2013, Mizuno et al. 2014) is required (see especially the criticism of Amato \& Blasi 2006).  One obvious option already noted above here is that the wind got stronger in activity episodes just prior to the SN-explosion, a possibility suggested by observing the structure of the wind (e.g., Svirski \& Nakar 2014a, b; Gal-Yam et al. 2014, Ofek et al 2014, Strotjohann et al. 2015, Tartaglia et al. 2016).  Of course we cannot know if the piston material and magnetic field keep mixing into the post-shock region, disturbing simple shock arguments on magnetic fields in shocks.  One could speculate that the piston material could contain a rather strong magnetic field that keeps slowly getting mixed in with the post-shock material.

This means for a possible final wind radius of 3 pc, that in the case of BSG stars the piston needs to exceed

\begin{equation}
M_{piston} \, > \, 10^{-1.8} \, {\rm M_{\odot}} \, \frac{r}{{\rm 3 \, pc}} \, ,
\end{equation}
\noindent which is well within the uncertainties.  The expression ``final wind radius" is that radius when the shock stalls due to encountering the wind-shell, built up during the lifetime of the stellar wind; occasionally we just use ``final" when the application is clear from the context.  Below we show from the gamma ray line that the piston mass is in fact about $0.1 \, M_{\odot}$, and so leads to a high energy.

This translates into a kinetic energy (using free expansion all the way out, as implied by the M82 sources) of 

\begin{equation}
E_{piston} \, = \, E_{SN} \, = \, 10^{51.0} \, {\rm erg} \, ,
\end{equation}
\noindent implying that the SN energy itself needs to be larger, consistent with many other arguments, as noted below.  

Using the $^{26}$Al line at 1809 keV we can estimate the piston mass:  The observed lines have a half width of about 300 km/s (Diehl 2017; full width 593 km/s).  In our picture we interpret this number as momentum conservation of the ejecta, so that 

\begin{equation}
\left(M_{piston} \, + \, \Delta M_{W}(r_{final})\right) \, \frac{c}{10} \, \simeq \, 300 \, {\rm km/s} \, \times \, 10 \, M_{\odot} \, ,
\end{equation}
\noindent where $10 \, M_{\odot}$ is an uncertain estimate of the wind mass already ejected earlier (Woosley et al. 2002, Fig. 16) for the most abundant stars that eject heavy nuclei.  Using 3 pc as the final radius gives for $\Delta M_{W}(r_{final})$ a mass of $10^{-1.8} \, M_{\odot}$.  Hence this condition gives then $10^{-1.} - 10^{-1.8} \, \simeq \, 10^{-1.1} \, M_{\odot}$, so much larger than the wind loss mass $\Delta M_{W}(r_{final})$, that the velocity can be kept up until $r_{final}$ can be reached, then at ``full steam".  
 
Assuming then the same piston mass for the conditions for a RSG star wind, with a density about 100 times higher, implies that equality is reached far below 3 pc:
 
\begin{equation}
\Delta M_{W}(r) \, = \; 10^{-2.8} \, {\rm M_{\odot}} \, \frac{r}{r_0} \, = 10^{-1.1} \, M_{\odot} 
\end{equation}
\noindent implies a radius of $10^{1.7} \, r_0 \, = \, 10^{17.7} \, {\rm cm}$, and beyond, the velocity goes down with $r^{-1/2}$, so up to 1 pc, for instance, would go down by $10^{-0.4}$; to 3 pc it would go down about $10^{-0.7}$.  These estimates are fairly uncertain, but the key consequence is that for RSG stars the final shock velocity hitting the wind shell is expected to be far below the initial velocity of $0.1 \, c$.  We will need this much lower shock velocity in RSG star winds later in our explanation of the anti-protons observed by AMS.

At this point it helps to note that in explosions into the interstellar medium (ISM) we have first a free-expansion phase, when the piston mass dominates over the accumulated mass from the environment, then a Sedov-phase, i.e. the stage when the accumulated mass dominates over the piston mass, but the energy is still constant (e.g. Cox 1972), before cooling sets in.  Analogously we can distinguish a free expansion phase, a wind-Sedov phase, and a final phase for explosions into winds.  Since these explosions occur into a wind of density run $\rho \; \sim \; r^{-2}$, all the dependencies on radius $r$ for the wind-Sedov differ from the normal ISM-Sedov case.

There might be a useful analogy between the transition from a piston-dominated stage to a wind-Sedov shock phase argued here, to the transition in Solar ejections driven by powerful Solar flares (Pinter \& Dryer 1990) from piston-driven to energy-conserving shocks.

We do identify here the compact sources in M82 with the slightly later stages of the RSNe discussed above (Kronberg et al. 1985, Golla et al. 1996, Allen \& Kronberg 1998, Kronberg et al. 2000).  At the full and sustained speed of the shock at about $0.1 \, c$, the time scale to go from $10^{16}$ cm to a pc is only 30 years, and so this implies that most, if not all, of the compact RSNe in M82 are this old or somewhat older.  This is consistent with the estimated SN-rate in M82 of 1 every 5 years, requiring for 42 compact sources about 200 years, about one order of magnitude above the simple estimate of 30 years. In Allen \& Kronberg (1998) a lower rate is suggested of about 1/60 years. However, correcting for the SN expansion velocity now known (see the table above), and possible selection effects (we may see only BSG star explosions), this rate is consistent with the higher rate, and in fact the rate might be even much higher, if we indeed detect only BSG star explosions:  The inferred rate from the compact radio sources is 1/10 years, correcting for the velocity; if these compact sources are only BSG star explosions, i.e. only stars above about $33 \, M_{\odot}$ then the total rate for SN explosions of stars above $10 \, M_{\odot}$ is about 7 times higher.  This does not take into account any SN Ia explosions.  This implies a SN rate of order 1/yr, surely an upper limit.  If we allow RSG stars to contribute to the detected compact radio sources, then the rate comes down to about 1 per 1.5 years.

If the magnetic field strength were due to the shock itself, then 

\begin{equation}
\frac{B^2}{8 \, \pi} \, = \; \epsilon_B \, \rho_W (r) \, {\dot{r}}^2 \, .
\end{equation}

In the free expansion case, the radial run of the magnetic field $B$ is the same as in the Parker limit (Parker 1958).  However, the difference is strong between BSG stars, and RSG stars, since their wind density is very different (e.g. Hirschi et al. 2005, Crowther 2007, Maeder \& Meynet 2012); so for the same shock velocity the magnetic field would be very much stronger, if derived from instabilities in the shock.  What is seen in the few well-observed cases is that the magnetic field is about the same.

In the wind-Sedov limit (i.e. low piston mass $M_{piston}$) this is readily rewritten as

\begin{equation}
B \, \sim \; r^{-3/2} \, ,
\end{equation}
\noindent so quite a bit steeper than in the free expansion case, when it runs as $r^{-1}$.    

One other speculative possibility is that the magnetic field derives from the interior of the stars, since the wind-base does expose (for BSG stars, at least) already the deeper layers, as visible in the chemical abundances.

One can also estimate how far the wind-Sedov case may reach in case of a Red Super Giant (RSG) star, when the wind velocity is about 100 times lower than in the Blue Super Giant (BSG) stars, and so at the same rate of stellar mass loss the density correspondingly 100 times higher:

\begin{equation}
\Delta M_{W, RSG} \; = \; 10^{-2.8} \, {\rm M_{\odot}} \, \frac{r}{{\rm 10^{16}} \, cm} \, .
\end{equation}

The associated kinetic energy at full, observed shock velocity for RSG stars is

\begin{equation}
\Delta E_{W, RSG} \; = \; 10^{49.2} \, {\rm erg} \, \frac{r}{{\rm 10^{16} \, cm}} \, ,
\end{equation}
\noindent suggesting that a RSG supernova shock may remain rather strong and super-sonic to pc scale, if the wind went that far and the energy were available up to $10^{52}$ erg; Soderberg et al. (2010a) show that the energies involved may reach $10^{52}$ ergs for Gamma Ray Bursts (GRBs), and invoking the magneto-rotational mechanism of Bisnovatyi-Kogan (1970; many later papers, e.g. Moiseenko \& Bisnovatyi-Kogan 2015) would suggest that the same energy is involved in all SN explosions of stars in the upper mass range, consistent with other observations. As the winds of BSG stars are about $10^{2}$ lower density at the same mass loss rate, the energy requirement is correspondingly reduced.  As noted above, even in the high pressure environment of the starburst galaxy M82, it is probable that the BSG explosions do go that far; in a lower pressure environment they would  obviously be capable of going further.  At that point the SN-shock would reach and hit the wind-shell, built up over the active lifetime of that strong wind preceding the SN-explosion (Heger et al. 2003), all inside an OB-star-super-bubble.  A different point of view has been taken by Cardillo et al. (2015).

This immediately leads to another constraint: Does the piston solve the problem of the abundances to actually supply all the cosmic ray particles of enriched abundances?  The core-collapse SNe above a Zero Age Main Sequence (ZAMS) mass of about 25 $M_{\odot}$ are about 1/5 of all supernovae above a ZAMS mass of 10 $M_{\odot}$, for which the rate in the galaxy is 1 every 75 years (Diehl et al. 2006, 2010).  The total energy in CRs required to explain the observations is about $10^{41.5}$ erg/s, using a high estimate; of this about 1/3 is in nuclei heavier than protons, so $10^{41}$ erg/s.  This implies that every SN of a star originally above $25 \, M_{\odot}$ has to provide $10^{51.1}$ ergs in CRs, so probably $10^{51.5}$ erg in kinetic energy, a larger number than those derived above. This in turn suggests that the total SN energy for all such stars may well be of order $10^{52}$ erg.  The material in the piston is about $0.1 \, M_{\odot}$, as demonstrated above using the RSNe as well as the compact radio sources in M82 as older RSNe, in conjunction with the gamma-ray line data. This can provide all the CR injection material. One key is that the piston adds additional material to the wind as source material for injection of CRs; another key is that we have a mix of abundances from RSG and BSG stars: RSG star winds have approximately ISM abundances, while BSG star winds have a range of abundances, some of which are highly enriched. Despite the slowing down of the shock in the wind-shell the SN-shock still accelerates further material.  Since the shock is at $0.1 \, c$ initially, and, as the M82 sources show - probably all the way out to pc scale - the estimate above shows that this can be satisfied for BSG explosions.  It does imply that these SNe of massive stars are more energetic than those at lower ZAMS mass, by about an order of magnitude.

There is the constraint from the width of the observed $\gamma$-line of $^{26}$Al (Diehl et al. 2017) of about 300 km/s (half width) beyond galactic rotation, already used above.  This also sets a limit on how fast all this happens. At the density in the wind the cooling time is shorter than the lifetime of the star as $\tau_{cool} \, = \, \{10^{10.7} \}/n \; {\rm s}$ using the density $n$ in $\rm cm^{-3}$ at the peak of the cooling curve at $T \, = \, 10^{5.4} \, {\rm K}$. The lifetime of the star is about $10^{6.5} \, {\rm  yrs} \, = \, 10^{14} \, {\rm s}$ and so this allows a pre-shock density of $n = 10^{-3.9} \, {\rm cm^{-3}} $, lower than the density given above at 3 pc, of $10^{-2.8} \, {\rm cm^{-3}}$.  This gives a cooling time of about $10^{5}\, {\rm yrs}$.  Therefore the density in the wind shell can go quite high, allowing drastic cooling when the SN-shock hits, and so running the remnant quickly into the cooling limit.  This then also limits the lifetime of the radio emission, since the shock rapidly slows down due to this extreme cooling and so fails to accelerate electrons.  This probably limits the radio emitting lifetime to a few hundred years; this in turn lets us understand the scarcity of such sources in our galaxy, and the abundance in M82.  This suggests that the Pevatron source detected by H.E.S.S. is in fact a very recent RSN (Abramowski et al. 2016), as the data show a spectral index of about - 2.4, and a kink or cutoff in the original spectrum of order PeV.

Another constraint comes from the non-observation of $^{59}$Ni in CRs as noted by Wiedenbeck et al. (1999), and Binns et al. (2006, 2008): The time has to exist between the production and the acceleration of $^{59}$Ni nuclei for them to decay by electron capture to $^{59}$Co.  However, the underlying stellar evolution models (e.g. Chieffi \& Limongi 2013, Neronov \& Meynet 2016) are still too uncertain to make this argument really conclusive.

Consider the wind-shell in high pressure environments such as the inner regions of the starburst galaxy M82, and in galaxies like ours.  In our Galaxy it has been convincingly argued that the SN-shock expands into an environment already shaped by earlier SN-explosions, and so forms an OB-star-super-bubble (e.g. Mac Low \& McCray 1988 for an early discussion).  In such OB-star-super-bubbles the SN shock runs out of steam, becomes subsonic and just heats up the interior of the super-bubble, until the next OB star explodes.  This is the scenario discussed at length by Binns et al. (e.g. Murphy et al. 2016), after Higdon \& Lingenfelter (Higdon \& Lingenfelter 2003, 2005, 2006, Higdon et al. 2004, Lingenfelter \& Higdon 2007).  However, all these arguments follow observations, and explosions of the most massive stars in our galaxy are so rare, that none has happened during the time of modern observations. We observe the explosions of the most massive stars best in other galaxies, such as M82, where SN explosions are also much more frequent.  On the other hand, CRs accumulate their particles right up to the escape time from the galaxy, thus much longer even for very energetic particles than human observations exist.  We propose that these most massive stars live such a short time, that the wind-shell is only disrupted by the explosion itself. After the explosion, given some more time, it merges into the environment of other earlier explosions, an OB-star-super-bubble.

The constancy of the observed radio emission is expected as soon as the wind-shock reverts to a Sedov expansion in the local ISM shaped by the earlier wind stages of the star.  This then gives (i) a near constant magnetic field, (ii) an energy density of the CR particles produced as $r^{-3}$, and (iii) a volume covered as $r^{+3}$ (Biermann \& Strom 1993). Thus the total synchrotron luminosity is constant with radial distance $r$ or, equivalently, time.  As soon as cooling becomes relevant, the radio emission ought to decrease rapidly (Kronberg et al. 2000).

Observations support this conclusion.  Observations of WR stars (Marston et al. 2015), i.e. very massive stars before explosions, suggest that most in our galaxy are in fact surrounded by dusty material: This is possibly a combination of the wind shell and the material out of which the stars formed.  $\gamma$-ray observations indeed suggest (de Boer et al. 2017) that the clumps out of which massive stars form are well defined through their mass-magnetic flux ratio long before we can discern any star.  This material easily survives the short lifetime of very massive stars, and hides, it seems, most of them.  Observations of massive star SNe (Tinyanont et al. 2016) are consistent with the hypothesis that these shell which surround massive star SNe with embedded dust survive for years.

\subsection{Polar cap and {$4 \, \pi$} regimes, knee and ankle}

\begin{figure}[h!]
\centering
\includegraphics[bb=0cm 0cm 29.7cm 21.0cm,viewport=2.0cm 0.0cm 28.5cm 21.0cm,clip,scale=0.6]{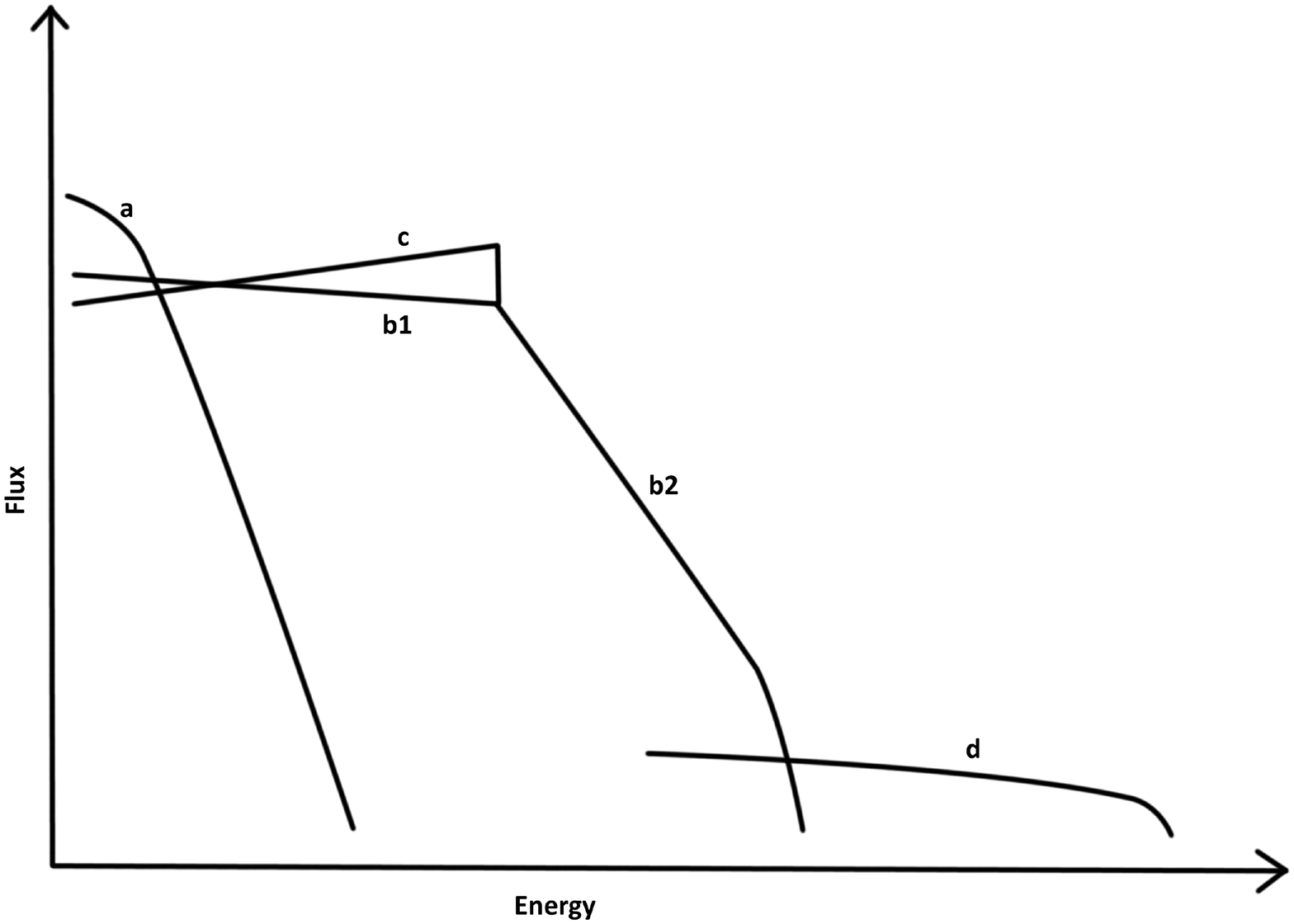}
\caption{Here we show schematically the various components of the CR model for primary cosmic rays, with the ISM-SN-CR component (a), the wind-SN-CR $4 \, \pi$ component below the knee (b1), and above the knee (b2), the wind-SN-CR polar cap component (c), and the extragalactic CR component (d), in a graph of $E^{2.1} \times {\rm CR \; flux}$ versus particle energy $E$ using the source spectra proposed in a log-log plot. So for wind-SN-CRs there are two CR components, here labelled (b) and (c), and this paper focusses on those two. Note, that between component (b1) and (c) there is always an upturn in the total spectrum.  We argue below that this upturn has been detected by both CREAM and AMS.}
\label{Alias_Fig1_CR_IV_Plot_LIC}
\end{figure}

The schematic concept includes four primary CR components: (a) An ISM-SN-CR component with a power-law at source of about $E^{-2.4}$ and a cutoff well below the knee energy of a few PeV; this component is briefly discussed first in the spallation section for comparison and then in the summary section of this review.  (b1) The $4 \, \pi$ component from wind-SN-CRs with a source spectrum of $E^{-7/3}$ up to the knee, and then (b2) about $E^{-2.8}$ beyond until a cut-off near the ankle. (c) The polar cap component also from wind-SN-CRs of at source $E^{-2}$ up to the knee, and then a sharp cutoff  (these spectra are all those at the source).  We note that plasma physics allows much sharper cut-offs than the exponential form.  (d) The extra-galactic UHECRs (Hillas 1984, Biermann \& Strittmatter 1987, Rachen \& Biermann 1993, Rachen et al. 1993, and more recently Biermann et al. 2011, and Biermann et al. 2016; recent reviews are by Kotera \& Olinto 2011, and by Letessier-Selvon \& Stanev 2011).  These UHECRs are only addressed here in the context of the starburst galaxy M82; UHECRs from radio galaxies, radio-quasars and blazars are discussed in Biermann et al. (2016).  The paper here addresses almost exclusively components (b) and (c), i.e. two components of CRs from the same origin, the very massive stars with winds. A figure describing these four components is Fig. 1 in Stanev et al. (1993), and here that figure is reproduced schematically in Fig.\ref{Alias_Fig1_CR_IV_Plot_LIC}.  

We differentiate the ISM-SN-CRs from the wind-SN-CRs, the polar cap component from the $4 \, \pi$ component, and the RSG winds from BSG winds.  Both the cut-off energy of the polar-cap component and the turn-down-energy of the $4 \pi$ component are proportional to $(U_{sh}/c)^2$ to within a factor of order unity in this theory, so are essentially the same, which we identify with the {\it knee} energy.

We emphasize here and below that all these spectral shapes and spectral indices have theoretical error bars (see Biermann 1993, Biermann \& Cassinelli 1993, Biermann \& Strom 1993, and Stanev et al. 1993): of course they need to be tested against observations as, e.g., done in Wiebel-Sooth et al. (1998), and again in Biermann et al. (2010).

The SN-explosion produces shocks racing through the winds of the massive stars; from the radio observations we know the magnetic fields in the shock regions, and so we can now work out the characteristic particle energies corresponding to the magnetic fields.  We note here that the Rankine-Hugoniot conditions at a strong shock give a density jump of a factor of 4 for strong shocks, a corresponding velocity jump of also a factor of 4 in the shock frame, and as a result a thickness scale of the post-shock region of $r/4$ (all for an adiabatic gas constant of 5/3).  As we demonstrated above, in a perpendicular shock (i.e. magnetic field direction perpendicular the shock normal) using the Jokipii (1987) scattering coefficient gives a maximal energy just limited by the available spatial scale, the Hillas (1984) limit.

The Hillas (1984) limit is just the spatial limit, valid in the case that the magnetic field is perpendicular to the shock normal, or parallel to the shock surface.  This is the standard case in the asymptotic magnetic field configuration in a magnetic wind, as the higher multi-poles of any multi-pole structure of the magnetic field just add as a factor (see Parker 1958, after eq. 26, p. 673), and the magnetic field-lines coincide with the flow lines.  But only this magnetic field component is enhanced in a shock, and so in the post-shock region, the transverse component dominates strongly.

\begin{equation}
E_{ankle} \;  = \; \frac{1}{8} \, Z \, e \, B(r) \, r  \, ,
\end{equation}
\noindent  where $Z \, e$ is the charge of particle, $r$ radial distance, $B(r)$ magnetic field as function of $r$, and we use $B \, = \, B_0 (r_0/r)$.  Here we use the characteristic radial extent of the shocked shell in a wind of $r/4$ as a spatial limit; this uses a strong shock, for which $U_{sh,1}/U_{sh,2} \, = \, 4$ for an adiabatic gas constant $5/3$:  We define $U_{sh,1}$ as the upstream velocity in frame of shock, and $U_{sh,2}$ as the corresponding downstream velocity. We also require the Larmor diameter (twice radius) of a gyrating particle to fit into this space. For simplicity we often refer to $U_{sh, 1} $ as $U_{sh}$.  However, it needs to be shown that this energy can be reached at all against all the various loss processes.

The scattering coefficient in a configuration most perpendicular to the shock normal (for random direction of the magnetic field prevalent) has a limit (Jokipii 1987, eq. 10) of

\begin{equation}
\kappa \, = \, \frac{E \, r}{Z \, e \, B_0 \, r_0} \, U_{sh} \, ,
\end{equation}
\noindent and we adopt this limit here.  A large part of this acceleration is due to Stochastic Shock Drift Acceleration (StSDA).  The acceleration time is then limited by turbulence time across the region, so $(r/4)/(U_{sh}/4)$, using the post-shock scales for both distance and velocity. Setting the acceleration time scale using this scattering coefficient equal to this limiting time scale allows the limit to be written as

\begin{equation}
\frac{E \, r}{Z \, e \, B_0 \, r_0} \frac{8}{U_{sh}}\; = \; \frac{r}{U_{sh}} \, ,
\end{equation}
\noindent reproducing the energy limit derived above.  Using the observed numbers this gives

\begin{equation}
E_{ankle} \; = \; Z \, 10^{17.5 \pm 0.2} \; {\rm eV} \, ,
\end{equation}
\noindent which we identify with the {\it ankle}.  

In a small fraction of space and time, which might be called magnetic islands, the magnetic field is parallel to the shock normal, and the acceleration is temporarily purely diffusive shock acceleration (DSA). The scattering coefficient in this case is given by

\begin{equation}
\kappa \; = \; \frac{1}{3} \, \frac{E}{Z \, e \, B} \, c \, \frac{B^2/\{8 \, \pi\}}{I(k) \, k} \, ,
\end{equation}
\noindent where  $I(k)$ is the energy density of resonant fluctuations in the magnetic field, so that $k^{-1} \, \sim r_g \, = \, (E)/(Z \, e \, B)$.  In the Bohm case we take this factor $(B^2/\{8 \, \pi\}(/(I(k) \, k)$ to be a constant, requiring $I(k) \, \sim \, k^{-1}$ in what is called saturated turbulence, quite different from Kolmogorov turbulence; we define this factor to be $b \, > \, 1$. We adopt for the limit itself, $b \, = \, 3$, which renders the expression maximally simple; we note that the integral of the irregularity spectrum can be maximally equal to the overall energy density of magnetic fields.

The acceleration time is then given in the case of a strong shock by (Drury 1983) 

\begin{equation}
\tau_{acc} \; = \; \frac{8 \, \kappa}{U_{sh, 1}^2} \, .
\end{equation}
\noindent This requires that the ratio of scattering coefficients $\kappa \, \sim \, 1/B$ scale as the velocities on the two sides of the shock - this can be justified by noting that in the perpendicular case the magnetic field $B$ scales inversely with the velocity on the two sides of the shock.

Here the limiting time is shorter than the turbulent time, since particles can just escape along the magnetic field lines in $r/c$ (in the perpendicular case they cannot), and so here the limit is

\begin{equation}
\frac{E \, r}{Z \, e \, B_0 \, r_0} \, \frac{8 \, c}{U_{sh}^2} \; = \; \frac{r}{c} \, ,
\end{equation}
\noindent where $r_0$ is a reference radial scale, here using $10^{16} \, {\rm cm}$, and $B_0$ the magnetic field strength at that radius, chosen because we have radio data giving these numbers (see above).  This gives a maximal energy $E_{knee}$ in this case of

\begin{equation}
E_{knee} \;  = \; \frac{1}{8} \, Z \, e \, B_0 \, r_0 \, \frac{U_{sh}^2}{c^2} \, .
\end{equation}
\noindent Using the observations listed above we obtain

\begin{equation}
E_{knee} \; = \; Z \, 10^{15.9 \pm 0.2} \; {\rm eV} \, ,
\end{equation}
\noindent which we identify with the {\it knee} energy.  The two expressions for the energies $E_{ankle}$ and $E_{knee}$ differ in their formal expression by $(U_{sh}/c)^2$, but we do not use the average from the data separately for $B_{0} \, r_{0}$ and $(U_{sh}/c)^2$, but use the average of the combined expressions $B_0 \, r_0 \, \frac{U_{sh}^2}{c^2}$, and $B_0 \, r_0$.  To emphasize the difference, the {\it knee} energy is thus interpreted as the limit for the parallel case (magnetic field parallel to shock-normal), and also as the turn-down energy for the perpendicular case (magnetic field parallel to shock surface, or perpendicular to shock normal), that energy where the spectrum turns from one power-law to a steeper power-law.   The {\it ankle} energy is the maximum energy reached overall.

This is consistent with other arguments:  The most important aspect is that $E_{knee}$ is just the limit for DSA, as StSDA is faster (Jokipii 1982, 1987; Meli \& Biermann 2006; Matsumoto et al. 2017).  This is the limit for the polar cap cosmic ray contribution, the small part of momentum space $4 \, \pi$, where the magnetic field is locally and temporarily radial, using only DSA.  The combination of StSDA and DSA has a reduction in efficiency of acceleration at that same energy, $E_{knee}$, so its spectrum steepens to a steeper power-law, and cuts off only at $E_{ankle}$, as is argued below.

The fast Jokipii limit of acceleration (Jokipii 1987) implies $\kappa \, = \, r_g \, U_{sh}$, and thus determines how fast particles gain energy.  The large-scale Jokipii limit ($ \kappa \, = \, (1/4) \, r \, U_{sh}$ upstream: Biermann 1993) determines the spectrum.

\subsubsection{The kink in the spectrum at the knee}

At this stage we need to dig a bit deeper into what scales are relevant in such a turbulence in a shocked layer (Biermann 1993).  The basic conjecture is that a scattering coefficient downstream can be constructed from the relevant length scale $r \, U_{sh, 2}/U_{sh}$ and the velocity difference across the shock $U_{sh} - U_{sh, 2}$; in a strong shock the ratio $U_{sh}/U_{sh, 2}$ is 4 and so we obtain

\begin{equation}
\kappa_{2} \; = \; \frac{1}{3} \, r \, \frac{U_{sh, 2}}{U_{sh}} \, (U_{sh} - U_{sh, 2}) \; = \; \frac{1}{16} \, r \, U_{sh} \, .
\end{equation}
\noindent For upstream we conjecture that the scattering is $U_{sh}/U_{sh, 2}$ times stronger, and so we calculate

\begin{equation}
\kappa_{1} \; = \; \frac{1}{3} \, r \,  (U_{sh} - U_{sh, 2}) \; = \; \frac{1}{4} r \, U_{sh} \, .
\end{equation}
\noindent There is a maximal lateral diffusion coefficient (Biermann 1993) constructed from the velocity difference across the shock squared, times the residence time of

\begin{equation}
\kappa_{\perp \perp , max} \; = \; \frac{3}{16} {\left( \frac{U_{sh,1}}{c}\right)}^2\, r \, c \, ,
\end{equation}
\noindent for the normal energy-dependent lateral diffusion coefficient, given by

\begin{equation}
\kappa_{\perp \perp} \; = \; \frac{\epsilon}{3} \frac{E}{Z \, e \, B(r)} c \, ,
\end{equation}
\noindent referring to curvature drift, so with a factor of $\epsilon \, > \, 1$ to enhance curvature from the inverse of the pure radial scale $1/r$ to the thickness of the shocked layer $4/r$ or possibly even more. So the corresponding partial energy gain is strongly reduced, giving the corresponding break energy as

\begin{equation}
E_{break} \; = \; \frac{9}{16 \, \epsilon} \, Z \, e \, B(r) \, {\left( \frac{U_{sh,1}}{c}\right)}^2 \, .
\end{equation}
\noindent With $\epsilon \, = \, 9/2$ this yields the same expression as above, so $E_{break}$ can be equal to $E_{knee}$, and in fact has to be quite close to it. So in the concept introduced here, these two energies are very close, and differ by at most some factor of order unity.

This implies that the $4 \, \pi$ component, driven by a combination of StSDA and DSA,  has a break to a steeper power law at the same energy, as the polar cap component (only driven by DSA) cuts off.  The data suggest that at the energies where the energy content of the spectrum maximizes, near about $2 \, A \, m_p \, c^2$, a ratio between these two components is of order 3 - 10 at injection.  Note that this refers only to the CR-components produced by massive stars with winds, when they explode. As noted earlier (Stanev et al. 1993) the normal proton component of CRs is probably dominated by lower mass stars that explode into the interstellar medium (but see also below).

\subsection{Knee and ankle energies, implications}

However, as commented on already elsewhere (Biermann 1995), this leads directly to several questions; the first one is whether all such stars could be the same in terms of their magnetic field properties.  It is possible, but hard to believe that a single star dominates all the cosmic ray particles contributing to the knee.  Such a hypothesis would imply that the number of contributing supernovae slowly decreases going from GeV energies towards the knee, ending at one supernova. It is hard to argue that such a concept would not give a significant bump in the spectrum of cosmic rays, by going from a few contributors to just one - but it is clearly not impossible.  If, as seems likely, many supernovae contribute to the knee observed in cosmic rays, then all these explosions must be relatively similar, since we can still discern in the data the spread of the various chemical elements without substantial extra smearing in energy (e.g. Todero Peixoto et al. 2015).  We estimate now how many supernovae may contribute.  On Earth we observe cosmic ray particles coming from a distance of 1 - 2 kpc, given by the estimated escape scale height; since the time for them to meander to us is about $10^{7} \, {\rm yrs} \, (E/GeV)^{-1/3}$, at the knee this meandering time scale is about $10^{5}$ yrs.  We have a supernova rate of about one every 75 years in the Galaxy overall (Diehl et al. 2006, 2010: only counting core collapse SNe, that is, type Ib/c and type II supernovae), with an active radial range in the Galaxy of about 10 kpc; within 1 - 2 kpc from Earth, the time-scale between supernovae is about $5^2$ to $10^2$ longer, about 1,250 yrs to 2,500 yrs.  The stars which contribute at the knee are only the really massive stars, above about 25 $M_{\odot}$, giving a fraction of about 1/5.  In our neighborhood of the Galaxy we may have only one half of all supernovae, and so the rate of supernovae within 1 - 2 kpc, only from stars above 25 $M_{\odot}$ is about 1 in 6,250 yrs to 1 in 12,500 yrs.  If we were to limit the mass range even further, to stars above $33 \, M_{\odot}$, so the BSG stars, would change these numbers by another factor of 1.5, so result in a rate of 1 in about 10,000 years, to 1 in 20,000 years. As we discuss below, RSG stars do not maintain the magnetic field at maximal level ($B(r) \, r \, \sim \, const$) to the outer shell of the wind-bubble. Hence for a discussion of what types of stars contribute at the knee it is justified to only use BSG stars. So for BSG stars,  this implies about 10 to 5 stars within $10^5$ yrs. Extrapolating this line of reasoning towards the ankle gives of order one star's explosion contributing.

It is perhaps of interest that this mass range is the same that possibly produces black holes (Heger et al. 2003). That suggests a common mechanism for the magnetic field and the explosion, connecting the explosion to the magnetic field; and the one mechanism that might connect the two is the magneto-rotational mechanism of Bisnovatyi-Kogan (1970), further expounded in many papers (e.g. Moiseenko \& Bisnovatyi-Kogan 2015).  Obviously, a neutrino mechanism assisted substantially by magnetic fields may be another option to correlate magnetic fields and the explosion. This might work in a manner described by Seemann \& Biermann (1997): Magnetic fields increase the signal speed in a plasma and the coupling of light interacting particles such as neutrinos and the plasma runs via excitation of waves, with better coupling if the wave-speed is high.  We speculate here that neutrino coupling may similarly be enhanced in a very dense plasma by magnetic fields.  

There is another hint about the mechanism of the explosion.  We have seen in SN 87A that there is extreme mixing up from deep inside the star (many papers, also Biermann et al. 1990).  Since we have argued that the piston mass is high enough to sustain the shock velocity at sustained speed and also provide sufficient material to account for the cosmic rays observed, this implies that the piston mixes in with the material to contribute to the cosmic ray particle population.  Material that might survive from the depth of the location of the original piston consists of isotopes of atoms formed at the extreme temperatures and densities expected there. The most interesting hypothesis would be that isotopes are pulled into acceleration, and so blocked from decay by Lorentz boosting.  Such unstable cosmic ray isotopes might help us to understand what happens at these depths, close to the budding black hole.  In a later section we give a table of the isotopes which might be of interest to consider, albeit very difficult to observe.

\subsection{Cosmic ray particle spectra below and above the knee}

Here we re-derive the extra energy gain from shock drifts (Biermann 1993):
\noindent The drift velocity is given by

\begin{equation}
V_{\perp, d} \; = \; c \, f_d \, \frac{E}{Z \, e \, B \, r} \, .
\end{equation}
\noindent Here we assume as in Biermann (1993) that this is a combination of curvature and gradient drift.  We refer everything to the case $f_d \, = \, 1$ for simplicity.  Thus the energy gain due to drifts can be calculated by the drift velocity, using the electric field induced, and the residence time.  Working this out upstream gives

\begin{equation}
E \, \frac{U_{sh, 1}}{c} \, ,
\end{equation}
\noindent with the corresponding expression for downstream being

\begin{equation}
E \, \frac{U_{sh, 2}}{c} \, ,
\end{equation}
\noindent and so for a strong shock, for which $U_{sh, 2}/U_{sh, 1} \, = \, 1/4$ we obtain that the total energy gain from drifts 

\begin{equation}
E \, \frac{5}{4} \, \frac{U_{sh, 1}}{c} \, ,
\end{equation}
\noindent and adding in the energy gain from standard first order Fermi gives an extra term ${U_{sh, 1}}/{c}$: Combining these two expressions gives a total numerical factor of $x \, = \, 9/4$.  

Since the shock injects particles from a density law of $r^{-2}$, we have a parameter for this power-law of $b = - 2$ radial power index for injection in wind density.  Similarly we have a parameter for dimensionality to adequately describe adiabatic losses, $d = 3$. $\kappa_{rr,1} = (1/4) \, r \, U_{sh,1}$ in the Jokipii limit in a wind, using the shocked layer thickness.

Here we need to emphasize that the spectrum is determined by the maximal time scale of a particle going back to the shock, while the acceleration rate is given by the shortest time scale.  The fastest scattering (Jokipii 1982, 1987) is given by $\kappa \, = \, r_g \, U_{sh}$, while the slowest is given by $(1/4) \, r \, U_{sh}$  (both upstream; Biermann 1993).  These two rates differ in perpendicular shocks, and are the same for parallel shocks.  Note that the acceleration time back and  forth across a shock is proportional to the scattering coefficient $\kappa$ (this is the time scale to establish a spectrum and maximal particle energy), while the diffusion time out of a region is given by the inverse of the scattering coefficient (this is the time scale relevant for producing anti-protons). These two scattering coefficients differ by a factor of order $E/E_{max}$, where $E_{max}$ is the maximal particle energy that can be contained.

The spectral index is then given by 

\begin{equation}
\frac{3 \, U_{sh,1}}{U_{sh,1} - U_{sh,2}} \, \left(\frac{U_{sh,2}}{U_{sh,1}} \left[\frac{1}{x} -1\right] + \frac{2}{x} \,(b + d) \, \frac{\kappa_{rr,1}}{r \, U_{sh,1}}\right) \; .
\end{equation}
\noindent This is given here as the difference in spectral index to -2; a positive value implies a steeper spectrum (this is eq. (39) in paper CR-I 1993, based on work by Krymskii \& Petukhov 1980, Prishchep \& Ptuskin 1981, Drury 1983).

The parameter values entering here are (i) $x \, = \, 9/4$, describing the addition of DSA and StDSA; (ii) the radial density power-law $b \, = \, -2$; (iii) the dimensionality $d \, = \, 3$; and (iv) the strong shock condition ${U_{sh,2}}/{U_{sh,1}} \, = \, 1/4$.  In a wind this equation gives a number of $1/3$, by which the spectrum is steeper than $E^{-2}$, so that the $4 \, \pi$ source spectrum is $E^{-7/3}$.  The polar cap source spectrum is $E^{-2}$, since there $x \, = \, 1$ and ${\kappa_{rr,1}}/{r \, U_{sh,1}} \, << \, 1$.  

To recap and proceed further, below the knee we used maximum curvature $4/r$ and argued that half of this would be average, so that the total energy gain is characterized by

$$x = 1 \, + \, \frac{5}{4} \, \left(\frac{2}{3} \, + \, \frac{1}{3}\right) \, = \, 9/4 \, .$$
\noindent Beyond the knee we use no turbulence-induced curvature and in fact allow that the natural curvature also occasionally goes to zero due to very large scale motion; and so we use half the normal curvature rather than twice the curvature.  This implies that the term $2/3$ goes to $1/6$, and thus we take

$$x = 1 \, + \, \frac{5}{4} \, \left(\frac{1}{6} \, + \, \frac{1}{3}\right) \, = \, 13/8 \, .$$
\noindent This gives $x \, = \, 1.625$, and for the spectral index

$$\frac{3}{x} + 1 = 2.8462 \, ,$$
\noindent which is (Kolmogorov added) 3.1795, resulting in $E^{-3.1795}$ as the predicted observable spectrum beyond the knee.  

\subsubsection{The acceleration of electrons by drifts alone}

The observations reveal that the spectrum of the electrons in the RSNe of very massive stars is about $E^{-3}$, perhaps slightly steeper even.  The question then arises, why the spectrum is just steeper by about unity compared to the spectra of nuclei.  At first sight this suggest the Kardashev (1962) loss limit, or even secondaries. Inserting numbers demonstrates that the Lorentz factor of the electrons associated with the radio emission is so low that the energies are below the rest mass of protons, even though these electrons are relativistic.  For any CR spectrum of nuclei, with a spectrum steeper than $E^{-2}$, the energy density of the population maximizes around a small multiple of the rest mass. Therefore one might well expect that the associated Larmor radius provides the main scale for the thickness of the shock transition layer.  It follows that the electrons do not even ``see" the shock transition, and only experience the shock drifts as acceleration.  There is some similarity to the ``shock surfing mechanism" discussed by, e.g., Lee et al. (1996), Li et al. (2017) and Matsumoto et al. (2017).  The process as described here obeys all the limiting conditions for electron drifts discussed in Ball \& Melrose (2001).  In the language of paper CR-I (Biermann 1993), also used above,  this implies that the parameter $x$, denoting the total energy gained per cycle relative to the energy gain from undergoing pure DSA (see, e.g., Drury 1983) is of order $5/4$.  However, the ``missing" shock transition can be thought of as another additional drift energy gain, using only the gradient, which can be crudely estimated as follows:  We average the drift energy gain both downstream and upstream, so take 1/2 of the sum (in paper CR-I this is eq.23).  Only 1/3 of this was from gradient drifts, so obtain 1/6 of the combined energy gain. This implies

\begin{equation}
x \; = \; \left( 1 + \frac{U_{sh,2}}{U_{sh}}\right) \, \left[1 + \frac{1}{6}\right] \; = \; (5/4)*(7/6) \; = \; (35/24) \; = \; 1.4583..  \, .
\end{equation} 
\noindent The spectral index is given by the expression given above, here simplified

\begin{equation}
\frac{3}{x} +1 \, = \, 3.0571.. \; .
\end{equation}
\noindent A clear prediction is then obviously, that the radio spectrum should become flatter as soon as the energy of the relevant electrons becomes large enough so that the electrons ``see" the shock.  This is confirmed by the compact sources in M82.

\subsection{Summary of the wind-SN-component cosmic ray spectra}

In this section we have described how to explain {\it knee} and {\it ankle} energies, and the spectrum both below and above the {\it knee} energy, and below the {\it ankle} energy, focussing on both RSG and BSG star winds.  In the following two sections we will ignore for didactic simplicity all the uncertainties of these spectral indices, and just use them directly, taking $E^{-2}$ for the polar cap component, and $E^{-7/3}$ for the $4 \, \pi$ component below the knee.  However, at every step we ought to be mindful of the underlying uncertainties: There are uncertainties due to the fact that we use strong shocks, with additional terms proportional to the inverse of the shock's Mach number squared (e.g. eq. 2.46 in Drury 1983).  Other uncertainties pertain to the ratio of the wind velocity relative to the shock velocity, of order 1/10.  The main uncertainty, of course, relates to the underlying non-linear model.  All these uncertainties can be tested with data.  We will also repeat various conventions, so that each section can be read independently of the others.

The data require both components, a component with a source spectrum close to $E^{-2}$ as well as another component with a source spectrum close to $E^{-7/3}$ (Boer et al. 2014, 2017).  In Biermann et al. (2010), and earlier papers (e.g., Wiebel-Sooth et al. 1998, 1999, Biermann et al. 2001) as well, the data were compared with the model proposal.  Spectral hardening has been detected by CREAM (Ahn et al. 2010a, Yoon et al. 2017) and again at much higher energy by IceCube/IceTop (Gaisser et al. 2016b, Aartsen et al. 2017a), consistent with what has been proposed here (Biermann et al. 2010, Todero-Peixoto et al. 2015) for both energies.  AMS now detects such a hardening as well (Kounine et al. 2017, Aguilar et al. 2017, 2018), shown for protons, Helium, Carbon, Nitrogen, Oxygen, and also the secondary Lithium, Beryllium, Boron.  We define the transition energy as that energy where the fluxes of a steeper power-law at lower energy and flatter power-law  a higher energy are equal:   Then this transition energy is at that energy, when each component provides half the flux, or, in other words, when the total flux rises to a factor of 2 above the lower power-law flux.  This is then at much higher energy than is suggested by a logarithmic-linear plot, and so the AMS data appear to be consistent with the CREAM data.

We have not addressed here the additional component of cosmic rays produced in SNe, that explode into the ISM, nor the interaction, that CRs suffer propagating throughout the ISM (Biermann \& Strom 1993, Biermann et al. 2001, Nath et al. 2012).  Based on data the latter is thought to be small (Nath et al. 2012, Cowsik et al. 2016), and the former may only contribute lower energy protons, and some Helium nuclei (Biermann et al. 2010).  Thoudam et al. (2016) confirm that this division is consistent with all the latest data.

Finally, are there any predictions arising from this section?  We list two:  (i)  Our argument says that GW events such as GW170104 (Abbott et al. 2017a)  with discrepant individual Black Hole (BH) spins ought to occur relatively often, at a very crudely estimated frequency of order 1/100 of BSG SN explosions, or one in several hundred RSG+BSG explosions.  In a case of a GW event with discrepant original BH spins (as in GW170104, Abbott et al. 2017a), but at much smaller distance, the compact radio emission might be detectable showing both the original jet or jets and the sweeping of the environment; corrected for projection the angles ought to match the GW analysis. (ii)  For BSG explosions the compact radio emission (again at sufficiently small distance) ought to be detectable all the way to the stalling of the shock, allowing a check on whether the magnetic field indeed follows $B (r) \times r \, \simeq \, const$, and whether the value indicates the energies derived here.  For RSG explosions, the transition from free expansion to wind-Sedov expansion ought to be detectable.  We return to these ideas in the section on the compact radio source 41.9+58 in the starburst galaxy M82 further below.

\vskip2.0cm

\section{AMS data:  The spectrum of anti-protons, positrons, and other energetic particles}



\subsection{Introduction}

In this section we focus on secondary particle $N_s$ production resulting from interaction of primary accelerated particles $N_p$. The scenario we envisage is that a supernova shock racing through a wind is loaded with energetic particles when it hits the wind-shell.  These particles interact the entire time while going through the stellar wind, but the shock stalls when hitting the wind shell; for this reason it will be in this phase when most interactions can be expected (see, above, for the discussion of the compact radio sources in the starburst galaxy M82, all thought to be in such a stage). Both the secondaries and primaries escape from this region and populate the cosmic ray disk of the host galaxy. As noted above this constitutes a ``nested leaky box approach".

A general energetic particle transport equation is given, for example, in Ginzburg \& Ptuskin (1976), their eqs. 3.18 and 3.36:

\begin{equation}
\frac{\partial \, N_i}{\partial \, t} \; - {\rm div} \left(\kappa \, \nabla \, N_i \right) \, + \, n_{ISM} \, c \, \sigma_i \, N_i \; = \; Q_i \, ,
\end{equation}
\noindent where (i) $N_i$ is the population number of an energetic nucleus, as a function of particle momentum, space and time; (ii)  $\kappa$ is the scattering coefficient, usually also a function of momentum; (iii) $n_{ISM}$ is the density of the environment; (iv) $\sigma_i$ is the cross-section to spallate a nucleus and turn it into other nuclei, and (v)  $Q_i$ is the source function, including the spallation of other, heavier nuclei.  Another form of this equation is given in Strong et al. (2007); there it corresponds to their eq. 1, which includes scattering in momentum, adiabatic compression and expansion, as well as direct losses.

The balance equation describes the interaction of primary nuclei producing secondary nuclei:

\begin{equation}
\frac{\partial \, N_s}{\partial \, t} \; = \; \frac{N_p}{\tau_{int}} - \frac{N_s}{\tau_{esc}} \, ,
\end{equation}
\noindent with the stationary solution expressed as

\begin{equation}
N_s \; = \; N_p \, \frac{\tau_{esc}}{\tau_{int}} \, ,
\end{equation}
\noindent where the two time-scales $\tau_{int}$ and $\tau_{esc}$ describe interaction and escape.  We assume here for simplicity that the interaction and escape time scales are both significantly shorter than the lifetime of this phase.

For simplicity and didactic reasons we approximate the interaction as energy-independent; we have explored this approximation in a large number of calculations in de Boer et al. (2017) fitting the $\gamma$-ray data of the Galaxy, using proper numerical interaction codes.

The main energy dependence of the secondary/primary ratio depends on the escape time, which is given by

\begin{equation}
\tau_{esc} \; = \; \frac{H^2}{2 \, \kappa} \, ,
\end{equation}
\noindent where $H$ is the radial scale of the interaction region.  This includes the wind-shell and its extension into the pre-supernova stellar environment in a molecular cloud; it may also involve an already evolved wind-bubble produced by the pre-supernova HII-regions, stellar winds, and earlier supernova explosions.

The scattering coefficient $\kappa$ in turn is given (see also above) by 

\begin{equation}
\kappa \; = \; \frac{1}{3} \, \frac{E}{Z \, e \, B} \, c \, \frac{B^2/\{8 \, \pi\}}{I(k) \, k} \, ,
\end{equation}
\noindent where  $I(k)$ is again the energy density of resonant fluctuations in the magnetic field, so that $k^{-1} \, \sim r_g \, = \, (E)/(Z \, e \, B)$.  In the following we focus on determining this wave-field as resulting from particle-wave interaction with the existing energetic particle population. A given energetic particle population with a certain spectrum excites a wave-field; this wave-field then gives a scattering coefficient, which in turn leads to the energy dependence of the escape time. In final consequence this allows us to determine the energy dependence of the secondary to primary ratio, the key answer we are seeking.

One question which needs to be raised is whether the secondary particles in turn interact so much that they do not even escape.  We can estimate this by checking the spectra of those nuclei that interact the most, Iron nuclei.  As the time to interact is given by the escape time, lower energy nuclei will interact more, and so we expect that such an effect would render the primary particle spectrum slightly flatter (see, e.g., Wiebel-Sooth et al. 1998).  Indeed the Iron spectra are slightly flatter, suggesting that a significant fraction interact, but still much less than one hundred percent.  A fortiori this then must also be true for the secondary nuclei.

The recent flurry of AMS data for anti-protons, positrons, and atomic nuclei  have prompted a long series of theoretical attempts to explain them (Accardo et al. 2014, Aguilar et al. 2013, 2014a, b, c, 2015a, b, 2016a, b, 2017, 2018).  The positron fraction seems to defy many conventional models, the anti-proton over proton ratio is flat over a large energy range, and the other secondaries, such as Lithium, behave differently again (but see Cowsik 2016b for a nested leaky box model which has some similarities to ours).  Here we outline an attempt to understand these spectra.  As above, the focus is on the properties of the stars that explode into their own winds, again emphasizing explosions of Red Super Giant (RSG) stars that have a slow, dense wind, and explosions of Blue Super Giant (BSG) or Wolf-Rayet (WR) stars that explode into a tenuous, fast wind.

\subsection{Anti-protons and other secondaries}






To understand the production of secondary particles by primary particle interaction, we need to know the time of interaction available to energetic particles in the dense regions.  This is (i) behind the shock, all the way out towards the boundary of wind zone, or, possibly more efficiently (ii) in the zone created by the wind shock itself and its immediate environment, which may include remnants of the molecular cloud (see de Boer et al. 2017) out of which the star formed. 

The mass in the wind shell is the accumulated mass loss over the star's lifetime up to the Supernova(SN)-explosion, and so is a substantial fraction of the stars Zero Age Main Sequence (ZAMS) mass, of order $10 \, M_{\odot}$; we conclude that more interaction happens (i) in the wind shell, and possibly even more (ii) in the surrounding region, the leftover material from earlier explosions of other stars and so earlier massive stars in the neighborhood, as well as (iii) in the molecular cloud out of which these massive stars formed (see de Boer et al. 2017).

This shows that we have several sources of turbulence:  First of all the spectrum of magnetic irregularities produced in the shock, as discussed above, first of all the Bohm case $I(k) \, k \, \sim \, const(k)$, and second the Jokipii case $I(k) \, k \, \sim \, k^{-1}$. This turbulence is rapidly fading as the shock weakens running through all this material of the wind shell and the more distant environment.  But we also have the newly created turbulence produced by the cosmic ray (CR) particles themselves.   That allows it to dominate over any shock-related turbulence. We let observations guide us here.

Kulsrud \& Cesarsky (1971), Bell (1978a) and Drury (1983) show that excitation and damping of Alfv{\'e}n waves can also be vastly different in dense media, as damping (at $10^4$ K; for $10^2$ K and $10^3$ K the factor changes from 8 to 1.5 and 3, respectively) is given by 

\begin{equation}
\Gamma_{damp} \; = \; 8 \cdot 10^{-9} \, n_H \, s^{-1} \, ,
\end{equation}
\noindent where $n_H$ is the local neutral hydrogen density ($n_e$ is the corresponding electron density, both in $\rm cm^{-3}$); this is a damping of Alfv{\'e}n waves by neutral-ion collisions (Kulsrud \& Cesarsky 1971).  A second damping mechanism is the cascade into sound waves, which happens whenever the speed of sound is less than the Alfv{\'e}n velocity (Bell 1978a), which can easily be the case in a shock racing through a stellar wind of a BSG star (see above):

\begin{equation}
T \, < \; 10^{8.5} \, {\left(\frac{B}{30 \mu \, Gau{ss}}\right)}^2 \, {\left(\frac{n_e}{\rm cm^{-3}}\right)}^{-1} \, {\rm K} \, ,
\end{equation}
\noindent is easily satisfied in the pre-shock region due to cooling, and possibly also satisfied in the post-shock region due to the strong enhancement of the magnetic field already observed, but barely due to equipartition.  This leads to a damping only if the the sound speed is very much smaller than the Alfv{\'e}n velocity, thus is not really applicable here.  This is important since we need Alfv{\'e}n waves to scatter CR particles resonantly as a key element of their diffusive transport (see, e.g., eq 3.18 in Ginzburg \& Ptuskin 1976, used above).

The density is (see above, and using the stellar mass loss and wind velocity adopted as an example there) about $10^{-3} \, cm^{-3}$ at 3 pc for BSG star winds, and after going through a wind shock, $4 \cdot 10^{-3} \, cm^{-3}$, without any enhancement, due to rapid cooling. The corresponding numbers for RSG star winds are about $10^2$ higher.   This enhances the excitation rate, and lowers the damping rate.  The shock velocity is typically $0.1 \, c$, and we have used that for scaling.  If the energy density of the energetic particles is equal to the energy density of the magnetic field, at 3 pc, then we could have an enhancement over the ambient galactic cosmic rays of a factor of about $10^5$, and so the excitation could be that much faster, or, at the same rate, extend to energies $10^3$ times larger than GeV, which is actually required.

The greatest amount of cosmic ray particle injection (referring to the maximal shock speeds) happens in the last phase of the shock before it hits the wind-shell.  We consider a shock region fully loaded with cosmic ray particles, of the two spectra $E^{-2}$ (the polar cap component; less common) and $E^{-7/3}$ (the $4 \, \pi$  component), and then calculate the excitation of a magnetic irregularity spectrum by these energetic particles.

The spectrum of magnetic irregularities excited by a given cosmic ray particle spectrum has been treated by Bell (1978a, b), and reviewed by Drury (1983).  The cosmic ray particle spectrum excites a magnetic irregularity spectrum $I(k)$ in resonant wave-number $k$ (Bell 1978a,b, Drury 1983, Biermann et al. 1998, 2001), with $k \sim (Z \, e \, B)/(p c)$, with momentum $p$

\begin{equation}
\frac{\partial}{\partial t} \left(\frac{I(k)}{4 \pi k^2}\right)  -  \frac{1}{k^2} \, \frac{\partial}{\partial k} \left[  \frac{k^4}{3 \, \tau_k} \,  \frac{\partial}{\partial k} \, \left( \frac{I(k)}{4 \pi k^2} \right)\right]  =  A \delta(k- k_0) \, ,
\end{equation}
\noindent with the cascade time scale given by 

\begin{equation}
\frac{1}{\tau_k} \; = \; k {\left(\gamma_{eff} \frac{I(k) k}{\rho}\right)}^{1/2} \, ,
\end{equation}
\noindent with an effective adiabatic index $\gamma_{eff}$, with $\rho$ here the affected density, the ionized density.  $\gamma_{eff}$ is for relativistic fluids $4/3$, such as dominant turbulent magnetic fields, or photon dominated gas as in very massive stars; for normal gas $\gamma_{eff}$ is 5/3, and for a mixed ionized gas it can be in between. The affected density can incorporate also some neutral fraction, depending on relaxation time scales in resonant motion between neutral and ionized particles in resonance with waves.  This gives, for instance, the Kolmogorov law as one natural solution in a steady state. This uses some long wave-length, or low wave number with a delta-function in wave-number as an injection; the irregularities cascade down in wave-length, or up in wave-number.

However, instead of using a single wave-length as a source of excitation, we can also consider excitation at all wave-lengths or wave-numbers together, say, by a spectrum of CR particles (Bell 1978a, b):


\begin{equation}
\frac{\partial}{\partial t} \left(\frac{I(k)}{4 \pi k^2}\right) -  \frac{1}{k^2} \, \frac{\partial}{\partial k} \left[  \frac{k^4}{3 \, \tau_k} \,  \frac{\partial}{\partial k} \, \left( \frac{I(k)}{4 \pi k^2} \right)\right] \; = \; \left(\sigma_{excit} - \Gamma_{damp}\right) \,  \left(\frac{I(k)}{4 \pi k^2} \right) \, ,
\end{equation}
\noindent with (Bell 1978a, b; Drury 1993)

\begin{equation}
\sigma_{excit} \; = \; \frac{4 \pi}{3} \, \frac{V_A}{I(k) k} \, p^4 \, v \, \frac{\partial f}{\partial r} \, ,
\end{equation}
\noindent where $f$ is the distribution function in momentum phase space $p$, with power index steeper by 2 as compared with $E^{-7/3}$. Here $v$ is the particle velocity, $V_A$ the Alfv{\'e}n velocity. The relativistic case corresponds to $v = c$.  This formula essentially compares the energy density in resonant waves per log bin $I(k) \, k$ with the energy density in the particles in resonance $ \, f(p) \, p \, c \, 4 \, \pi \, p^3$, in the relativistic limit and again per log bin.  The factor is the Alfv{\'e}n velocity, divided by the radial scale, and so depends on density.

Using as the cosmic ray spectral index $\alpha$, and the magnetic irregularity turbulence spectral index $\beta$ we have then

\begin{equation}
 - \frac{1}{2} - \frac{3}{2} \beta \; = \; \alpha - 5 \, .
\end{equation}

\noindent For $4 \, \pi$ component with an $E^{-7/3}$  (i.e. $\alpha \, = \, 7/3$) spectrum $I(k) \, k \, \sim \, k^{-4/9}$ (i.e. $\beta \, = \, 13/9$), for the polar cap component with an $E^{-2}$ (i.e. $\alpha \, = \, 2$) spectrum $I(k) \, k \, \sim \, k^{-2/3}$ (i.e. $\beta \, = \, 5/3$), same as Kolmogorov, with excitation at all wavelengths.  

At this stage we note, that all these relationships are for relativistic particles only, and only apply when the diffusion approximation still holds.  As discussed in Biermann et al. (2001), at particle velocities well below the speed of light $c$, the relationships change. At very high energy the escape time is given by a convective velocity, resulting in energy independent interaction. Then the secondaries quasi automatically follow the primaries closely, and only the intrinsic variation of the production cross-section and multiplicity remain.  This is equivalent to supersonic turbulence with $I(k) \, \sim \, k^{-2}$, and the time scale of interaction independent of energy.

Table \ref{tableCRexcit} shows the systematics of an application of this expression.  This table contains in the first line the two cases of the exciting CR spectrum; in the second line the corresponding spectrum of the irregularities as function of wave-number $k$; in the third the same, but using log bins of the wave-number (since in the scattering coefficient we have the term $I(k) k$ corresponding to energy per log bin); in the fourth line we are writing this irregularity spectrum as a function of particle energy; in the fifth the secondary CR production time scale energy dependence;  in the sixth the secondary spectrum; and in the seventh the secondary/primary energy dependence for action on a $E^{-2}$ spectrum.  We observe in CR particles the sum of different source classes and interactions, so we need to compare with perhaps other contributions.  This is done in the next few lines for specific cases, and then the set of comparisons is done again for the case of acting on a $E^{-7/3}$ spectrum.

Applying these results to the secondary/primary ratio in cosmic ray particle interaction we obtain for the $4 \, \pi$ excitation $E^{-5/9}$ (Biermann et al. 1998, 2001),  and for the polar cap excitation $E^{-1/3}$.  The excitation rate depends on the spectrum of the exciting energetic particles, and the spectrum of the excited turbulence, written as a function of the resonant particle energy.  To take the example of the $k^{-5/3}$ wave spectrum excited by an $E^{-2}$ particle spectrum: The energy density of the waves has then a corresponding energy dependence of $E^{+2/3}$, while in this example the energy density of the energetic particles is flat with energy. The ratio then runs as $E^{-2/3}$. Clearly they meet at a pivot energy, which is where weakest excitation takes place. This pivot energy depends on the density of the ionized matter through the Alfv{\'e}n velocity dependence.  For the $4 \, \pi$ component the energy density of the waves runs as $E^{+4/9}$ with the energetic particle energy density running as $E^{-1/3}$, and the ratio running as $E^{-7/9}$.  At the pivot energy, by definition, the ratio of the two energy densities is of order unity. Hence the number in front, basically the inverse of the Alfv{\'e}n-time over the inverse of the flow time using it as the limit, determines where that pivot energy is.  We are assuming here that overall instabilities of the flow determine that scale, so that the number in front is basically the inverse of the Alfv{\'e}nic Mach-number times some numerical factor of order unity, which we have to determine empirically.  Here we focus on is on the difference between the waves excited by the polar cap component, and the waves excited by the $4 \, \pi$ component. On the other hand we determine how the competition plays out between BSG star winds with their low density, and the RSG star winds with their high density.  We summarize these connections between exciting cosmic ray spectrum all the way to the energy dependence of the secondaries and their comparison in Table \ref{tableCRexcit}:

\begin{table}[h!]
\begin{center}
\caption{CR spectra, excitation spectra and secondary spectra, all at source, for relativistic particles and in the diffusion limit.  At very high energy the interaction is no longer diffusive, but convective, so that the secondary production time scale is essentially independent on energy. $E^{-2}$ is the polar cap component, and $E^{-7/3}$ is the $4 \, \pi$ component. Here B/C is the Boron/Carbon and Li/C is Lithium/Carbon ratio.}
\begin{tabular}{|c|c|c|c|}
\hline
\hline
& & & \\
{\bf Exciting CR spectrum:} & $E^{-\alpha}$ &$E^{-2}$ & $E^{-7/3}$\\
\hline
Excited wave-spectrum,&  & & \\
 en. per wavenumber $k$  &  & $k^{-5/3}$ & $k^{-13/9}$\\
 \hline
Excited spectrum, & & & \\
energy per log bin of $k$ & & $k^{-2/3}$ & $k^{-4/9}$\\
\hline
Excited spectrum, & & & \\
energy per log bin of $E$ & & $E^{+2/3}$& $E^{+4/9}$ \\
\hline
Secondary CR spectrum& & & \\
 production time scale & & $\sim \, E^{-1/3}$& $\sim \, E^{-5/9}$ \\
 \hline
Secondary acting on & & & \\
{\boldmath \bf $E^{-2}$ CR primary} & & $E^{-7/3}$ & $E^{-23/9}$\\
\hline
Secondary/primary ratio & & & \\
energy dependence & & $E^{-1/3}$ &  $E^{-5/9}$ \\
\hline
Comparing with & & $E^{-7/3}$ primary & \\
\hline
Observed in RSG explos.& & antiprotons & \\
So sec/prim comparison & & $E^{+0}$ & \\
\hline
Comparing with & & $E^{-2}$ primary & \\
\hline
Prediction for BSG explos. & & B/C and Li/C & \\
at high energy & & also $E^{+0}$ & \\
\hline
Prediction for BSG explos.& & B/C and Li/C & \\
at yet higher energy & & $E^{-1/3}$ & \\
\hline
Secondary acting on & & &\\
{\boldmath \bf $E^{-7/3}$ CR primary} & & $E^{-8/3}$&  $E^{-26/9}$\\
\hline
Secondary/primary ratio& & & \\
 energy dependence & & $E^{-1/3}$ &  $E^{-5/9}$ \\
 \hline
 Comparing with & & $E^{-7/3}$ primary & \\
\hline
Observed in BSG explos. & & B/C high energy & B/C low energy \\
\hline
From BSG explos. & & & \\
possibly observed in & & & low energy protons/Helium\\
\hline
\hline
\end{tabular}
\label{tableCRexcit}
\end{center}
\end{table}

Next we calculate the normalization between polar cap component and $4 \, \pi$ component:  For BSG star wind primaries such as Carbon and Oxygen nuclei in energetic particles we can discern in the data a transition between the $4 \pi$ cosmic ray component with spectrum $E^{-7/3}$ at source and the polar cap component with spectrum $E^{-2}$ at source at  $\sim$ 5 $Z$ TeV: the spectrum flattens and can be explained by this transition (Ahn et al. 2010a, Biermann et al. 2010, Aguilar et al. 2017, 2018).  This is a transition between two primary components, that gives the normalizations between the polar cap component and the $4 \, \pi$ component.

For BSG secondaries we observe a transition between excitation to give $I(k) \, k \sim \, k^{-4/9}$ (the $E^{-7/3}$ $4 \, \pi$ component) and the $I(k) \, k \sim \, k^{-2/3}$ (the $E^{-2}$ polar cap component) working on the $4 \, \pi$ spectrum of $E^{-7/3}$ at source at  $\sim$ 20 $Z$ GeV observed; the spectral energy dependence of the ratio goes from - 5/9 to - 1/3 (Ptuskin et al. 1999; Aguilar et al. (2016b) and Oliva et al. (2017) in B/C).  There may be a hint of this secondary/primary ratio tending towards a constant near TeV, so possibly signifying a transition to a polar cap primary.  However this is not certain at present.  For the BSG star winds these observed transition energies set constraints on the scenario.  Setting the flux ratio between polar cap component and $4 \, \pi$ component from the 5 TeV observed gives a factor of $\sim \, 10^{-1.3}$, which we refer to as $10^{-1.3} \, f_{-1.3}$, rounding to the first decimal place again.  Setting the normalization of the ratio at GeV energies to $A \, > \, 1$ we have a decrease of the ratio cosmic ray energy density to wave energy density as $E^{-2/3}$ for the excitation of the $I(k) k \, \sim \, k^{-2/3} \, \sim \, E^{+2/3}$ spectrum to unity at the pivot energy given by $10^{-1.3} \, f_{-1.3}\, A \, (E/{\rm GeV})^{-2/3} \, = \, 10^{-1.3} \, f_{-1.3}$, so $E_{pivot, pc} \, = \, {\left(10^{-1.3} \, f_{-1.3} \, A \right)}^{3/2} \, {\rm GeV}$.  For all these arguments to work in BSG stars up to TeV scale requires $A \, \simge \, 10^{+3.3} \, {\rm GeV}$.  The spectrum of irregularities in terms of energy density goes from that energy down along $E$ with $E^{+2/3}$.  An irregularity spectrum, which has most of its energy at one wave-number, corresponding to a well defined energy in resonance with a particle spectrum, which has equal energy per log bin, could be problematic. There could be an extra factor $f_{pc} \, > \, 1$ for the amplitude of the irregularity spectrum.  We use here $10^{-1.3} \, f_{-1.3} \, A \, (E/{\rm GeV})^{-2/3} \, = \, 10^{-1.3} \, f_{-1.3} \, f_{pc}$.  We normalize everything at the $4 \, \pi$ component and at GeV energies to unity. The cosmic ray particles exciting these irregularities must dominate in energy, thus we write the index $dom$.  So with normalization this spectrum of irregularities $I(k) \, k \; \sim \; I(E) \, E$ is proportional to

\begin{equation}
I(k) \, k \; \sim \; I(E) \, E \; \sim \; 10^{-1.3} \, f_{-1.3} \, A_{dom}^{-1} \, f_{pc, dom} \, {\left(\frac{E}{{\rm GeV}}\right)}^{2/3} \, ,
\end{equation}
\noindent relevant at higher energies.  Considering then the other excited spectrum, we have it reach the pivot energy given by the condition $A_{dom} \, (E/{\rm GeV})^{-7/9} \, = \, 1$, thus expressing its pivot energy by $E_{pivot, 4 \pi} \, = \, A^{9/7} \, {\rm GeV}$.  This then gives a spectrum of

\begin{equation}
I(k) \, k \; \sim \; I(E) \, E \; \sim \; A_{dom}^{-1} \, {\left(\frac{E}{{\rm GeV}}\right)}^{4/9} \, ,
\end{equation}

\noindent relevant at lower energies.  These two irregularity spectra are equal at 

\begin{equation}
E_{cross} \, = \, {\left(\frac{10^{+1.3}}{f_{pc, dom}}\right)}^{9/2} \, {\rm GeV}.
\end{equation}


The efficiency of producing secondaries may have different time scales for the polar cap component and the $4 \, \pi$ component (Biermann et al. 2009).  However, the transition between the secondaries of the $E^{-2}$ component and secondaries of the $4 \, \pi$ component may be different from the transition in the irregularity spectrum;  this factor is $f_{eff} \, \simeq c/\{3 \, U_{sh}\} \, > \, 1$.  Importantly this factor differs in RSG star winds from BSG star winds, since in RSG star winds the shock slows down due to the much higher density.   The transition from secondary to primary ratio of cosmic rays switching from a $E^{-5/9}$ dependence to an $E^{-1/3}$ dependence, is of order 20 GeV/nucleon, judging from secondaries (Ptuskin et al. 1999; Aguilar et al. (2016b) and Oliva et al. (2017) in B/C), and this constrains the various terms, especially $f_{eff, BSG}$. Including this extra factor $f_{eff}$ we obtain for the four reasonable combinations of spectrum and primary CR spectrum acted upon the relationships

\begin{equation}
{\rm secondary \; CR \; spectrum} \; \sim \; 10^{-1.3} \, f_{-1.3} \, A_{dom}^{-1} \, f_{pc, dom} \, f_{eff, RSG} \, X_{RSG, i} \, {\left(\frac{E}{{\rm GeV}}\right)}^{- 7/3} \, ,
\end{equation}
\noindent is the secondary CR spectrum for the $E^{2/3}$ irregularity spectrum acting on the polar cap $E^{-2}$ CR component; there is a related expression for element/isotope $i$ for the example of a BSG star.  Here $X_{RSG, i}$ is the relative abundance of element/isotope $i$ in an RSG star, relative to the dominant ion. 

\begin{equation}
{\rm secondary \; CR \; spectrum} \; \sim \; 10^{-1.3} \, f_{-1.3} \, A_{dom}^{-1} \, f_{pc, dom} \, X_{RSG, i} \, {\left(\frac{E}{{\rm GeV}}\right)}^{- 8/3} \, ,
\end{equation}
\noindent is the secondary CR spectrum for the $E^{2/3}$ irregularity spectrum acting on the $E^{-7/3}$ CR $4 \; \pi$ component; again a similar expression holds for any element
/isotope $i$ from a BSG star.

\begin{equation}
{\rm secondary \; CR \; spectrum} \; \sim \; A_{dom}^{-1} \, X_{RSG, i} \, {\left(\frac{E}{{\rm GeV}}\right)}^{- 2 - 5/9} \, ,
\end{equation}
\noindent is the secondary CR spectrum for the $E^{4/9}$ irregularity spectrum acting on the $E^{-2}$ CR polar cap component; again a similar expression holds for any element/isotope $i$ from a BSG star.

\begin{equation}
{\rm secondary \; CR \; spectrum} \; \sim \; A_{dom}^{-1} \, X_{RSG, i} \, {\left(\frac{E}{{\rm GeV}}\right)}^{- 7/3 - 5/9} \, ,
\end{equation}
\noindent is the secondary CR spectrum for the $E^{4/9}$ irregularity spectrum acting on the $E^{-7/3}$ CR $4 \, \pi$  component; again a similar expression holds for any element/isotope $i$ from a BSG star.

To summarize: The irregularities produced by the CR polar cap component always appear at higher energy, since their spectrum is flatter.  Secondaries from proton primary interaction ought to derive from RSG stars, while secondaries from Carbon and Oxygen primary interaction derive from BSG stars.  BSG stars are those for which the SN shock runs at full speed, about $0.1 \, c$, all the way to the wind boundary, as demonstrated by the compact radio sources in the starburst galaxy M82 (Kronberg et al. 1985); on the other hand, in RSG star winds the SN shock slows down significantly before getting to the wind boundary, due to the higher density.

The data show - given these choices - that for the low energy Boron CR component the fourth variant is relevant; for the slightly higher energy Boron CR component the second variant is relevant as for the Lithium CR and Beryllium component (Aguilar et al. 2018); while for anti-protons the first variant is relevant.  These choices are made due to the densities and energies involved, as explained in the following.

In RSG star winds the first component dominates over the second component, and in BSG star winds the second over the first.  The only difference between these two variants is the factor $f_{eff} \, = \, c/\{3 \, U_{sh}\}$ which would have to be much larger in RSG star winds than in BSG star winds.  Using the estimate of this factor in Biermann et al. (2009) of about $c/\{3 \, U_{sh}\} \, \simeq$ 2 to 3 in BSG star winds, we noted above that the shock speed is greatly reduced for RSG star winds, which are much denser.  This entails that $f_{eff}$ can be expected to be much larger for RSG star winds than for BSG star winds, strongly enhancing the secondary anti-proton production, making this term then likely dominant.  The other two terms, $A_{dom}$ and $f_{pc, dom}$, appear in both variants in the same way, and so again there ought to be no difference.

There are two transition energies, (i) first the energy of the transition for the primaries, e.g. energetic protons, from the $4 \, \pi$ component to the polar cap component for RSG stars, and (ii) second the transition energy between secondaries from the polar cap and $4 \, \pi$ primary components.  The Fermi data show a hint of a flattening for the protons (Ackermann et al. 2014) consistent with a transition at several TeV between the $4 \, \pi$ and the polar cap component, so rather similar to the BSG star winds; however, the AMS data do not confirm the numbers (Aguilar et al. 2015a), but suggested rather a steeper spectrum, consistent with a possibly pure ISM component.  This apparent conflict has not been resolved; the data are consistent with a pure $4 \, \pi$ component primary spectrum of $E^{-7/3}$ at source all the way to TeV.

The transition energy from secondary particles of the $E^{-7/3}$ $4 \, \pi$ primaries to secondary particles of the $E^{-2}$ polar cap primaries, using wave excitation of $E^{-2}$, so $k^{-5/3}$, for BSG stars and RSG stars is not known yet. ISS-CREAM may yield an answer (Seo et al. 2015).

Due to this uncertainty, the predictions in Table \ref{tableCRexcit} have no transition energies.  However, for example, the anti-protons in the model proposed run as secondaries from the polar cap component and are compared with the $4 \, \pi$ component. Therefore, it is clear from the energy, at which the polar cap dominant component itself becomes dominant for protons, that the anti-proton/proton fraction decreases with $E^{-1/3}$, all the way to the secondary antiprotons produced by the maximal energies of the RSG polar cap component.  This maximal energy is far below the knee energy, since by the same token, that the anti-proton production is strongly enhanced due to higher density and lower shock velocity, the final shock velocity is far below its initial value, and so the RSG knee energy is far below the BSG knee energy.

\subsection{The secondary cosmic ray particles}

Here we summarize the results for specific nuclei, and describe, what the model proposed predicts at higher energy.

\begin{figure}[h!]
\centering
\includegraphics[bb=0cm 0cm 38.92cm 24.13cm,viewport=0.8cm 0.0cm 38.92cm 24.13cm,clip,scale=0.43]{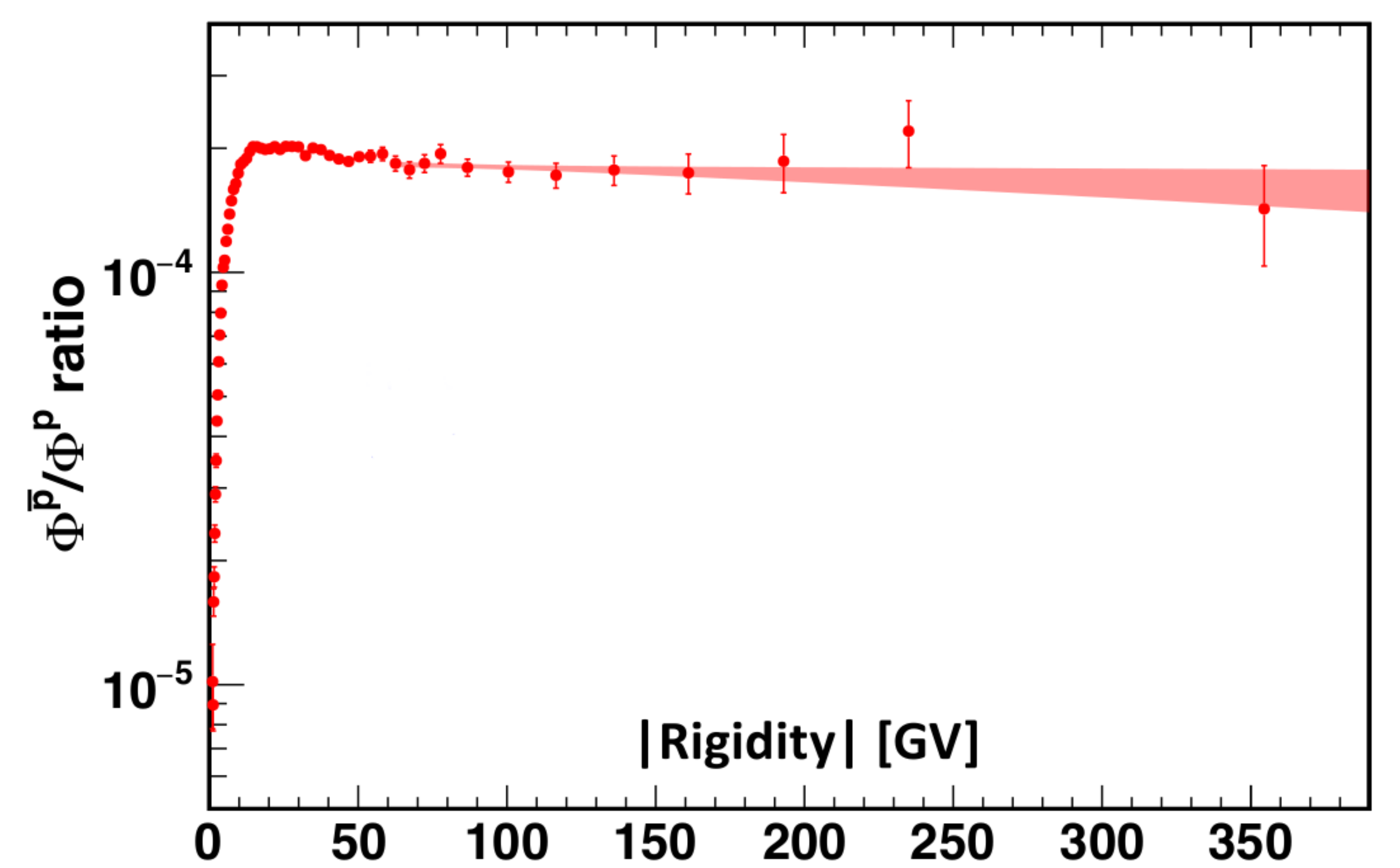}
\caption{The AMS antiproton fraction. Can be fitted with proton interaction, { protons from massive RSG star explosions.  Source: Aguilar et al. (AMS-Coll.) 2016a, modified by I. Gebauer; used with permission.  }}
\label{AMS2015antiprotonfraction}
\end{figure}

We propose to explain the anti-protons by using an enhanced production in the slowed down shock of a RSG star wind, so acting on the $E^{-2}$ polar cap component to produce a secondary component of $E^{-7/3}$ energetic particles. Since this is exactly the spectrum of the primary $4 \, \pi$ component, the ratio is flat, see 
Fig.\ref{AMS2015antiprotonfraction}.  The slow-down of the shock is the critical ingredient to enhance the production in the polar cap regions over the production in most of $4 \, \pi$.

There is a possibility of a component of secondary particles with the $E^{4/9}$ spectrum of irregularities acting on the $E^{-2}$ polar cap component;  this could increase the anti-protons at low energy.

The anti-proton spectrum resulting from interaction is slightly flatter than the primary CR-proton spectrum (see, e.g., the examples in de Boer et al. 2017);  we ignore this effect here.  A second effect also rendering the secondary spectrum flatter is optical depth: We note that the normal grammage numbers derived are based on the assumption, that the chemical composition of the interacting gas is ``cosmic" (e.g., Seo \& Ptuskin 1994, Ptuskin et al. 1999, Moskalenko et al. 2003, Strong et al. 2007). In contrast the chemical composition is enhanced with heavy elements in BSG star winds and therefore also in the wind-shells; this yields a higher real interaction column than indicated by the assumption of cosmic abundances (Sina et al. 2001).  

As a consequence the following result may be indicative.  There is a spectrum component of about $E^{-2.88}$ for the lower energy AMS data for protons and Helium (Aguilar et al. 2015a, b).  This spectral behavior is obtained when fitting the AMS data below several hundred GeV with a double power-law, and allowing for a hardening such as seen in the CREAM data (Ahn et al. 2010a, Yoon et al. 2017).  The running spectral indices (Aguilar et al. 2015b) show that they differ slightly and do not run perfectly in parallel; fitting a double power-law with a different transition energy (different for protons and Helium nuclei) allows us to understand such a pattern, as any sum of two power-laws with very similar indices influences the locally determined power-law index over many powers of ten in energy.  This seems to be consistent with the $E^{-2.85 \pm 0.05}$ spectrum shown in Dembinski et al. (2017) for both protons and Helium, although that paper focusses on energies above 100 GeV.  However, this contradicts the PAMELA results, which indicate a flatter spectrum (Karelin et al. 2016).
If experimentally confirmed, this could be a secondary rather than a primary component.  It is readily explainable by spallation of heavier nuclei such as Carbon or Oxygen, by the $E^{4/9}$ irregularity spectrum on the $E^{-2}$ polar cap CR component, and then steepened by another 1/3 during transport to Earth, thus resulting in a predicted spectrum of $E^{-2 - 5/9 - 1/3} \, = \; E^{-2.889}$.  Here the source would be BSG star winds, with some enhancement of production, since the shock is still going at full blast ($f_{eff} \, = \, c/\{3 \, U_{sh}\} \, \simeq \, 3$, see Biermann et al. 2009). One problem with such a speculation could be the flux required.   At this time, the uncertainties in the data do not support such a conclusion, and the explanation that these protons and Helium CR particles derive from ISM-SN-CRs is certainly consistent with all current data and their full uncertainties.  There is a related argument by Kappl et al. (2015), Winkler 2017, and Reinert \& Winkler (2018), who propose that the anti-protons are secondary from heavy nuclei spallation; in the framework proposed here this says that their proposal corresponds to secondaries of the $4 \, \pi$-component, using the turbulence spectrum excited by the polar cap component.

In addition, of course, there is the ISM-SN component of cosmic ray particles, mostly ignored in this article, which also contributes at lower energy experiencing a low grammage throughout the ISM (Nath et al. 2012, Cowsik 2014, Cowsik 2016a, b).  That is more relevant for protons and Helium, but seems unimportant for all heavier nuclei at the current stage of exploration.  Should the speculation above for protons and Helium be confirmed by a more specific analysis and more data, then the observed cosmic ray spectrum may show little evidence for the ISM-CR-component, a rather disconcerting notion.

At higher energy the primary polar cap component will take over and steepen the secondary/primary ratio to $E^{-1/3}$.  This will change as soon as the diffusive approximation breaks down and we go over to a convective limit approach, when interaction becomes energy independent to a reasonably good approximation.

Also, in the same vein, the dominant irregularity spectrum above a few tens of GeV is the $E^{2/3}$ spectrum (analogous to Kolmogorov) in BSG star winds, which gives a steepening by 1/3, so that the secondary/primary ratio here is $E^{-1/3}$, such as for the Lithium/Carbon or Boron/Carbon ratio.  The newest AMS data confirm this dependence with a slope of $-0.333 \, \pm \, 0.014 \, ({\rm fit}) \, \pm \, 0.005 \, ({\rm syst.})$ (Oliva et al. 2017).

Alternative astrophysical models to explain anti-protons and positrons have been given by, e.g., Cowsik et al. (2014), Cowsik \& Madziwa-Nussinov (2016), and Cowsik (2016). Constraints on nested propagation making extensive use of the code GALPROP have been offered by J{\'o}hannesson et al. (2016).  These models, like the one proposed here, use a nested leaky box (see, e.g. Cowsik \& Wilson 1973; Meneguzzi 1973; Peters \& Westergaard 1977). In these models the energy dependence of the inner propagation has some similarity to what we propose here, but the external propagation is different.  Aloisio et al. (2015) use self-generated waves to discuss propagation through the Galaxy.

At higher energy two things could happen: (i)  The polar cap primary component is known to become dominant around $5 \, Z \, {\rm TeV}$, interpreted here as due to CRs from explosions of BSG stars (Biermann et al. 2010).  (ii) But also the $E^{2/3}$ irregularity spectrum caused by the $E^{-2}$ CR spectrum will take over acting on the $E^{-2}$ primary CR component with some enhancement of secondary production in BSG star winds.  These two changes are unlikely to appear at the same energy, and we may observe either a flattening of the ratio, or a steepening of the ratio depending on which transition is first going up in energy.  But ultimately  - as long as we are in the diffusion limit - both happen and then the secondary/primary ratio is again $E^{-1/3}$.  The transition region could thus be different for different nuclei, depending on details of the production (the term $f_{eff} \, = \, c/\{3 \, U_{sh}\}$).  There is now evidence for an identical behavior of Lithium, Beryllium and Boron secondary nuclei (CERN AMS lectures 2015, Kounine et al. 2017, Aguilar et al. 2018), and they all show a flattening of their spectra beyond about 200 GeV/n relative to the primary CRs, fully consistent with the expectations here.
ISS-CREAM (Seo 2015) can be expected to throw light on these questions.

\vskip2.0cm

\subsection{Triplet pair production and positrons}




Here we propose that the polar-cap component of energetic electrons with a spectrum $E^{-2}$ in the source region interacts with the local radiation field to produce secondary electrons and positrons thus explaining most of the positrons detected by AMS (Accardo et al. 2014, Aguilar et al. 2013, 2014a, b); Fermi data give the same spectrum to within the errors (Ackermann et al. 2010).  This is a process known as ``triplet pair production", in which an energetic electron encounters a photon, and produces an extra electron-positron pair (Haug 1975, 1981, 1985, 2004).  So three leptons result from this process, hence the name. In such a calculation of the shape of the spectrum of secondary positrons there are only two free parameters (given the electron spectrum and assuming isotropic photons): (i)  the maximal energy of the electrons in the source region, and (ii)  the typical energy of the photons of its field. We assume that in a first approximation such an average energy of the photons adequately describes a rather complicated spectrum (see, e.g., Moskalenko et al. 2006, Strong et al. 2007).  The full expressions for the process have been used to obtain many numerical examples of such spectra. Here we show just one example, for an electron spectrum of $E^{-2}$ extending to 30 TeV, and an average photon energy of 3 eV;  this matches what is expected in the neighborhood of a massive star exploding.  In these numerical programs, care has been taken to include all angles relevant in the collision; no short-cut has been used, as is usually done when treating this process.  The spectral shape is a strong function of the energy above the threshold energy of this process.  The key parameter is the density of these photons; the test is whether we have enough photons in the region.  

The data show directly a spectrum of $E^{-3.0}$  from Fermi data (Abdo et al. 2009), referring to relatively high energy electrons, which obviously have ``seen" the shock, unlike the much lower energy electrons induced from the radio emission of a radio-supernova (RSN; see above); the spectrum seen in the Fermi data is clearly the reduction of the spectrum by losses (Kardashev 1962); at the source, the spectrum is deduced to be $E^{-2}$.  Any steeper spectrum would drastically reduce the level of the resulting secondary leptons, since the resulting flux depends on the fraction of electrons above threshold to produce pairs; thus it is a power of the lever-arm of the energy ratio between the threshold energy and the energy where the total energy of the population sits; this lever arm corresponds to very many powers of ten.  One important caveat here is that this observed spectrum refers to a combination of electrons and positrons, but we know from other data, that positrons make up about ten percent or less of the sum of electrons and positrons; therefore they should not influence the electron plus positron spectrum strongly (see the discussion and references in Biermann et al. 2009, Accardo et al. 2014).  In Aguilar et al. (2014b) the AMS data of the combined electron and positron spectrum is shown, and suggests a combined spectrum about -0.08 in slope steeper than $E^{-3}$ (at source corresponding to $E^{-2}$) considering the entire range from 30 GeV to 1 TeV. In the model here this difference is attributed to the consumption and production of electrons and positrons; in fact, below we use an estimate of a 30 percent reduction, quite consistent with the AMS numbers.  We conclude that the observational evidence is consistent with a flat component of $E^{-2}$ for electrons in the sources.

The features of the shape of the spectrum, as deduced from many new calculations, Haug (1975, 1981, 1985, 2004), are the following. 

\noindent The positron spectrum rises sharply at 

\begin{equation}
\gamma_{L,cut} \, \simeq \, 10^{4.45} \, {\left(\frac{h \, \nu}{5 \, {\rm eV}}\right)}^{-1} \, ,
\end{equation}
\noindent where $h \, \nu$ is the typical energy of the environmental photon field.  At higher energies the positron spectrum drops sharply at 

\begin{equation}
\gamma_{H,cut} \, \simeq \, 10^{6.25} \, \frac{E_{e,edge}}{\rm TeV} \, , 
\end{equation}
\noindent where $ E_{e,edge}$ is the maximal energy of the adopted cosmic ray electron spectrum of $E^{-2}$.

\noindent The ratio of these energies given by 

\begin{equation}
10^{1.80} \, (h \nu)/(5 \, {\rm eV}) \, (E_{e,edge}/{\rm TeV}) \, .
\end{equation} 
\noindent It describes the range $x_{th}$ above threshold to make positrons, 

\begin{equation}
(h \nu)/(m_e c^2) \times (E_{e,edge})/(m_e c^2) \; = x_{th} \; > \, 2 \, . 
\end{equation}
\noindent The spectrum multiplied by $\gamma_{+}^2$, the dimensionless energy of the positrons,  has a maximum given by 

\begin{equation}
\gamma_{2,max} \, \simeq \, 10^{5.3} \, \{(h \nu)/(5 \, {\rm eV})\}^{-2/3} \, \{(E_{e,edge})/({\rm TeV})\}^{+1/3} \, .  
\end{equation}
\noindent The ratio of $\gamma_{H,cut}$ and $\gamma_{2,max}$ is 

\begin{equation}
\frac{\gamma_{H,cut}}{\gamma_{2,max}} \, \simeq \, 10^{0.95} \, {\left(\frac{h \nu}{5 {\rm eV}}\right)}^{+2/3} \, {\left(\frac{E_{e,edge}}{\rm TeV}\right)}^{+2/3} \, ,
\end{equation} 
\noindent so it is also just a function of the range $x_{th}$ above threshold. 

For a reasonable fit, the data suggest $h \nu$ of order 1 eV or a bit larger (broad bump), and $E_{e,edge} \, >> \, 1 \, {\rm TeV}$; we use here 3 eV and 30 TeV in our example.
The AMS positron data suggest $\gamma_{L,cut} \, \simeq \, 10^{4.8}$ and $\gamma_{2,max} \, \simge \, 10^{5.5}$, so we used these approximate numbers to select the parameters for our plot.  The condition of using an appreciable fraction of energetic electrons locally to allow a slight distortion of the spectrum gives the lower limit required for the product of the photon density $n_{ph}$ and  time spent $\tau_{CR}$ - in an OB-star-super-bubble environment.

We note that the total cross-section for triplet pair production, Haug (1975, 1981, 1985, 2004) is

\begin{equation}
\sigma_{3,tot} \, = \, \alpha r_0^2 \, \left(\frac{28}{9} \, \ln \{2 \, x_{th}\} - \frac{218}{27} + \frac{1}{x_{th}}{\rm terms}\right) \, ,
\end{equation}
\noindent with  $\alpha r_0^2 \, = \, 10^{-27.3}  \, {\rm cm^2} $  and  $x_{th}$ defined above, with $x_{th} = 2$ as threshold. Obviously, the factor of the $\frac{1}{x_{th}}$-term must be such as to render the total cross-section positive above threshold.
\noindent Using $h \, \nu \, = \, 3.0 \, {\rm eV}$ and $E_{e,edge} \, = \, 30 \, {\rm TeV}$ gives as total cross section (using just the first term)  $\sigma_{3,tot} \, = \, 10^{-26.0} \, {\rm cm^2}$.

\noindent So to distort the CR-e  spectrum, the condition is

\begin{equation}
\sigma_{3,tot} \, \tau_{CR} \, c \, n_{ph} \; \simge \;  1/3 \; .
\end{equation}
\noindent This requires the photon density to obey the condition
\begin{equation}
\frac{\tau_{CR}}{10^6 \, {\rm yrs}} \, \frac{n_{ph}}{10^{1.5} \, {\rm cm^{-3}}} \; \simge \; 1 \; ,
\end{equation}
\noindent with a photon density just somewhat higher than in the Galactic Center (GC) region (Moskalenko et al. 2006). The other parameter entering here is the time-scale available, and this corresponds to the convective or diffusive turn-over time-scale at the outer edge and outside of a massive star wind, where most of the interaction takes place.  So this can be estimated to be of the order of many parsec divided by a velocity of order of km/s, perhaps 10 km/s: this corresponds to just $10^6$ yrs for 10 pc and 10 km/s. The diffusive time scale is longer, and may therefore not be relevant:  We use 30 TeV as the relevant particle energy, and the extrapolated magnetic field at 10 pc of an estimated $10^{-3.5}$ Gau{ss}, enhanced by the wind-shock but also diluted outside the transition from the wind boundary. We use these numbers to get a Larmor radius of $10^{14.5}$ cm, to give a scattering coefficient of order $10^{25.0}\, {\rm cm^2/s}$ using fully developed Bohm-like turbulence ($I(k) \, \sim \, k^{-1}$). Here $I(k)$ is the energy density per wavenumber $k$, in resonance with the Larmor motion of a particle of energy $E \, = \, p \, c$. Using relativistic approximations yields a minimal time of $10^{15.0}$ s or $10^{7.5}$ yrs.  However, we have shown above that the turbulence driven by the polar cap component of energetic particles in the winds of RSG stars itself has a Kolmogorov-type character corresponding to $I(k) \, \sim \, k^{-5/3}$, thus the scattering coefficient is $10^{1.5}$ smaller for 30 TeV versus 1 GeV as lever arm. The real effective diffusive time scale is in fact $10^{13.5}$ s, or $10^{6.}$ yrs.  We conclude that $10^{6}$ yrs is a crude, but plausible, estimate for the interaction time scale.  These two conditions, of photon density and time-scale available for interaction, are both plausible in any region of very many OB stars.  Therefore there may be sufficient interaction.

\begin{figure}[h!]
\centering
\includegraphics[bb=0cm 0cm 42.1cm 31.6cm,viewport=0.0cm 0.0cm 42.1cm 31.6cm,clip,scale=0.37]{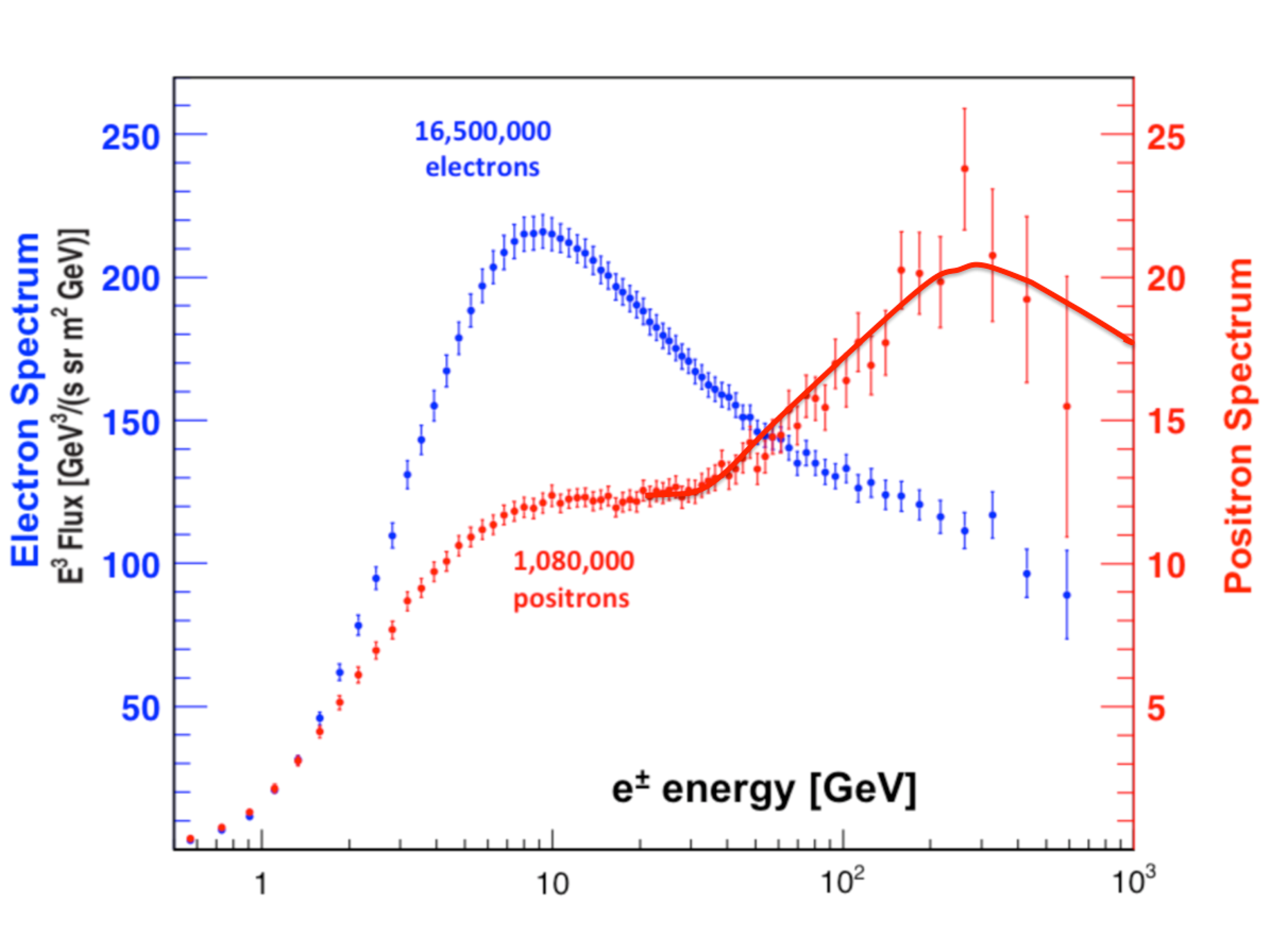}
\caption{Fitting the positrons from the triplet pair production using $(h \nu) \, = \,  3.0 \; {\rm eV}$, $E_{e,edge} \, = \, 30 \; {\rm TeV}$, and electron spectrum with slope $\delta \, = \, 2$ to the AMS data. For a comparison we note that at such energies the spectrum of both positrons and electrons is steepened by unity from losses in propagation (Kardashev 1962).  Sources E. Haug 2017, the AMS-talk by N. Zimmerman at the TeVPA conference 2016 at CERN, and Aguilar et al. (2014a); used with permission.}
\label{EberhardHaug2017tripl5positronsAMSfit}
\end{figure}

We used the properties of photon-electron interaction derived above to estimate  the key parameters from the published AMS data (Accardo et al. 2014, Aguilar et al. 2013, 2014a, b); they are the main photon energy and the maximal electron energy and produce the curve shown in Fig.\ref{EberhardHaug2017tripl5positronsAMSfit} as a fit to the AMS data.  Since in this model the initial electron spectrum is $E^{-2}$ at source, and both positrons and electrons are steepened in their spectrum by unity in propagation from losses (Kardashev 1962), we can read this diagram as a strong contribution to the observed positron spectrum.  Hence for the fit we note that a source spectrum weighted by the factor of $E^2$ is equivalent to raising the observed electron and positron spectra by $E^3$.  Once again we note an alternative model, also a nested leaky box model, by Cowsik (Cowsik et al. 2014, 2016).  A comparison with dark matter models is in de Boer (2017).

The electron and positron spectrum resulting from our model ought to reflect both the ``consumption" of primary electrons, as well as the production of secondary positrons and electrons.  The simultaneous increase of the flux of electrons and positrons relative to the lower energy trend, respectively, in Fig.\ref{EberhardHaug2017tripl5positronsAMSfit} above 50 GeV is just what is expected from the triplet production.  To be affirmative, this test may require even higher precision than currently achieved, but even at current precision such a test could be used to rule out our model.  The main uncertainty in such a calculation is the radiation field in the source region; since there may be a number of contributing source regions, this may not be a unique test. A second uncertainty is, of course, whether the injected electron spectrum is really well described by a pure $E^{-2}$ law.

\vskip2.0cm

\section{Cosmic ray injection, 2nd ionization potential effect ?}



\subsubsection{Scenario}

\begin{figure}[h!]
\centering
\includegraphics[bb=0cm 0cm 31.75cm 28.23cm,viewport=0.5cm 0.0cm 31.75cm 28.23cm,clip,scale=0.50]{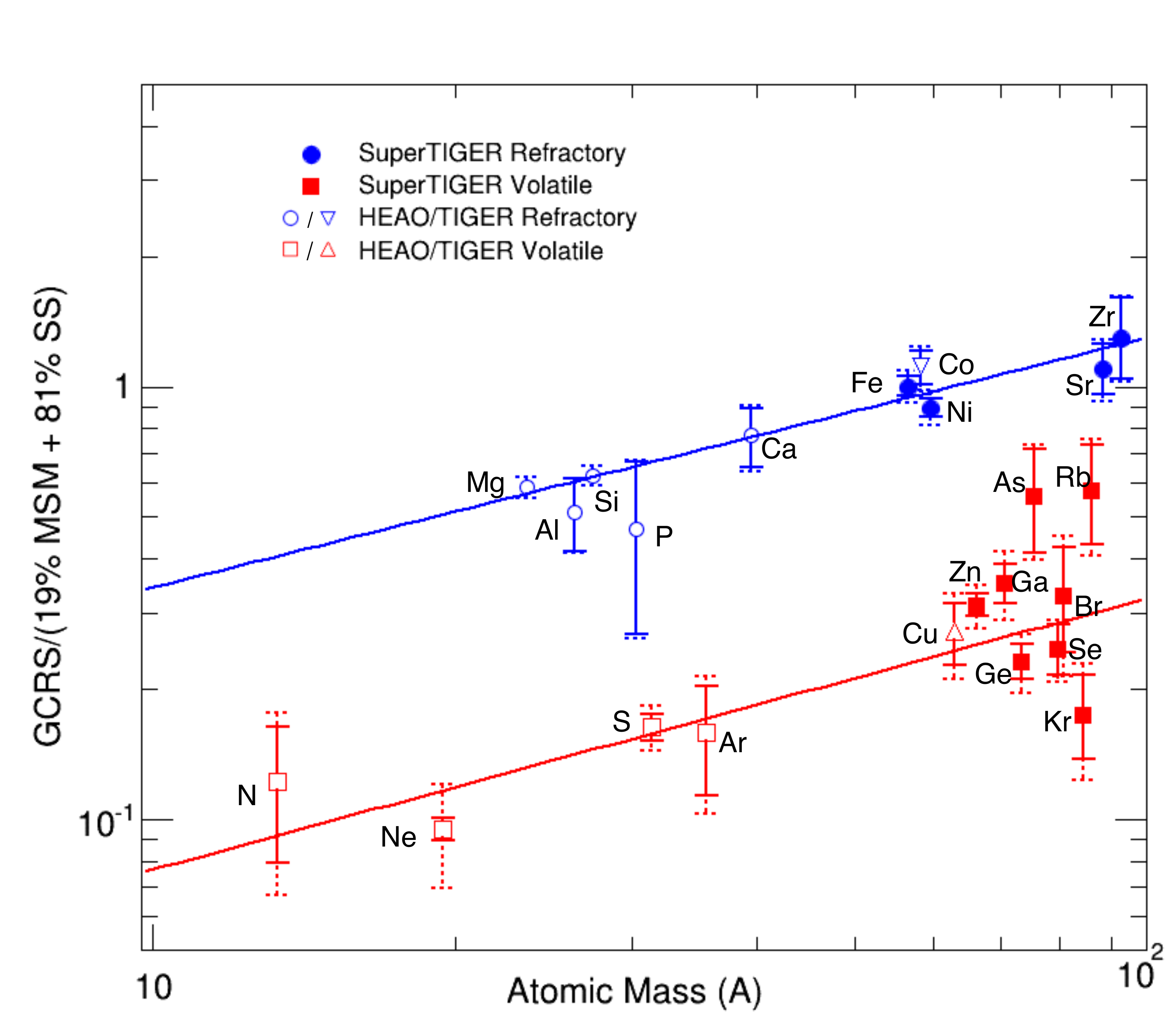}
\caption{Abundance ratio in deduced Cosmic Ray (CR) sources over source material model, versus atomic mass number, using Woosley \& Heger (2007) and Lodders (2003); analogous to Fig.9 in Murphy et al. 2016: source: W.R. (Bob) Binns 2017 and used with permission.}
\label{Binnsabundances2016}
\end{figure}

In the following we will outline an argument demonstrating that the pattern seen in a special plot, the Binns-diagram  (Fig.\ref{Binnsabundances2016}), can be explained.  The Binns-diagram shows that the ratio of Galactic Cosmic Ray Source (GCRS) abundances over source model abundances can be described by $Q_{0}^2 A^{2/3}$, where $Q_{0}$ is the initial degree of ionization of the element concerned, and $A$ is its mass number (see, e.g., Wiedenbeck et al. 1999; George et al. 2001; Binns et al. 2001, 2005, 2006, 2007, 2008, 2011, 2013; Wiedenbeck et al. 2003; Rauch et al. 2009) and Murphy et al. (2016).  In the model proposed to explain this dependence $Q_{0}^2 A^{2/3}$ we will once again consider a shock racing through a magnetic wind.  We use a magnetic field configuration that is perpendicular to the shock-normal.  In a random orientation this is the strongest average component;  in a Parker-limit wind, the perpendicular magnetic field component $B_{\phi}$ is by far the strongest component (Parker 1958, Weber \& Davis 1967).  We note (see above) that the compact radio sources in the starburst galaxy M82 demonstrated that in Blue Super Giant (BSG) star winds the Parker limit is maintained until the Supernova(SN)-shock hits the wind-shell, just when injection is maximized.

\subsubsection{Refractories and volatiles in 1st and 2nd I.P.}

The arguments by Binns et al.  rest on the concept that elements can be split into volatiles and refractories that determine their dust grain properties, and that the element selection in cosmic ray injection starting with Carbon depends on the properties of these dust particles (e.g., Ellison et al. 1997, Wiedenbeck et al. 1999, Higdon \& Lingenfelter 2003, 2005, 2006, Higdon et al. 2004, Lingenfelter \& Higdon 2007, Rauch et al. 2009, Binns et al. 2011, 2013, Murphy et al. 2016).  These properties also correlate with the second ionization potential, as shown in Table \ref{tableIPs}.  The significance of separating the unstable isotopes into those with various lifetimes derives from the fact, that with Lorentz boosting, some of these isotopes can survive as cosmic ray particles; a condition is that the time between creation and acceleration is sufficiently short. If, for instance, an isotope is made directly at high energy from high energy spallation collisions, then it may survive, given a sufficiently high energy.  However, if the time between its creation, for instance by nuclear reactions deep in the star, and its mixing into the acceleration zone and acceleration itself is too long compared with the decay time, then it may not survive.  The chemical composition of the piston, mixing in with the shocked material at a late point of the shock's evolution, may become detectable.  Given an observed shock speed of $0.1 \; c$, we assume that the speed survives to about 3 pc (based on the discussion above of the compact radio sources here traced to the late stages of the explosions of BSG stars in the starburst galaxy M82).  In such a case isotopes with a lifetime larger than about $10^{2}$ yrs may become detectable if they are produced in the piston material just before the explosion.  

\begin{table}[h!]
\begin{center}
\caption{First and Second Ionization Potentials (I.P.) in eV for isotopes. Isotopes most abundant in normal matter are marked with $^{+}$.  Under unstable isotopes those with a lifetime larger than 10 days are listed, with those with a lifetime larger than 1 year (but less than $10^2$ yrs) marked with $^{*}$, and those with a lifetime larger than $10^2$ yrs marked with $^{**}$. Those that decay only by electron capture, and are so effectively stable above a few 100 MeV/n, are marked with $^{e}$, with the other signs omitted then (NNDC 2017). }
\begin{tabular}{|c|c|c|c|c|c|c|}
\hline\hline
Z & A (stable) & unstable & name &  1st I.P. & 2nd I.P. & refr./vol.\\
\hline
6 & 12$^{+}$,13 & 14$^{**}$ & C & 11.3 & 24.4 & \\
7 & 14$^{+}$+15 & - & N  &   14.5 & 29.6  &vol.\\
8 & 16$^{+}$-18 & - & O & 13.6 & 35.1 & \\
9 & 19$^{+}$ & & F & 17.4 & 35.0 & vol.\\
10 & 20$^{+}$-22 & - & Ne &  21.6 & 41.0  &vol.\\ 
11 & 23$^{+}$ & 22$^{*}$& Na & 5.1 & 47.3 &  vol.\\
12 & 24$^{+}$-26 & - & Mg &   7.6 & 15.0  &refr.\\
13 & 27$^{+}$ & - & Al &  6.0 & 18.8 &refr.\\
14 & 28$^{+}$-30 & 32$^{**}$ & Si & 8.2	& 	16.3  &refr.\\ 
15 & 31$^{+}$ & 32+33 & P  &  10.5 &  19.8  &refr.\\ 
16 & 32$^{+}$-34,36 & 35 & S  &   10.4	& 	23.3 &vol.\\ 
18 & 36,38,40$^{+}$ & 37$^{e}$,39$^{**}$,42$^{*}$& Ar &  15.8 & 27.6	 &vol.\\ 
20 & 40$^{+}$,42-44,46 & 41$^{e}$,45& Ca & 6.1 	& 11.9  &refr.\\
26 & 54,56$^{+}$-58 & 55$^{e}$,59,60$^{**}$& Fe &  7.9		 & 	16.2  &refr.\\ 
27 & 59$^{+}$ & 56,57$^{e}$,58$^{e}$,60$^{*}$& Co &  7.9	& 17.1 &refr.\\ 
28 & 58$^{+}$,60-62,64 & 59$^{e}$,63$^{**}$& Ni &   7.6 & 18.2  &refr.\\
29 & 63$^{+}$+65 & - & Cu &  7.7	& 20.3  &vol.\\
30 & 64$^{+}$,66-68,70 & 65& Zn & 9.4 & 18.0	&vol.\\
31 & 69$^{+}$+71 & 67$^{e}$ & Ga &  6.0	& 20.5 &vol.\\
32 & 70,72-74$^{+}$,76 & 68$^{e}$,71$^{e}$ & Ge &  7.9 	& 15.9 &vol.\\
33 & 75$^{+}$ & 73$^{e}$+74& As &  9.8	& 18.6 &vol.\\
34 & 74,76-78,80$^{+}$,82 & 72$^{e}$,75$^{e}$,79$^{**}$& Se &  9.8  & 21.2 &vol.\\
35 & 79$^{+}$+81 & - & Br &  11.8	& 21.8 & vol.\\
36 & 78,80,82-84$^{+}$,86 & 81$^{**}$,85$^{*}$& Kr &  14.0 & 24.4 &vol.\\
37 & 85$^{+}$+87 & 83$^{e}$,84,86& Rb &  4.2 	& 27.3 &vol.\\
38 & 84,86-88$^{+}$ & 82$^{e}$,85$^{e}$,89,90$^{*}$& Sr &  5.7	& 11.0 &refr.\\
39 & 89$^{+}$    & 88,91 & Y  &  6.2  & 12.2 &refr.\\
40 & 90$^{+}$-92,94,96 & 88$^{e}$,93$^{**}$,95& Zr &  6.6  & 13.1 &refr.\\

\hline
\end{tabular}
\label{tableIPs}
\end{center}
\end{table}

The source for the nuclear data in Table \ref{tableIPs} is the 8th edition of the Karlsruhe Nuclear Tables (2012).  The source for the ionization potential numbers is the online version of the CRC Handbook of Chemistry and Physics (2003).  The main source for the corresponding dust properties is Lodders \& Fegley (1998).

The data in Table \ref{tableIPs} demonstrate the differentiation of refractories and volatiles into low and high 2nd Ionization Potentials (I.P.): a clear separatrix can be drawn between them.  We include here the elements higher than Nickel (28).  The elements Copper (29) to Zircon (40) also show the same separation, with the line in terms of the second I.P. separating the two regimes tending lower.  This may be due to a higher recombination rate of the second level of ionization for these massive atoms beyond Iron as compared to atoms below Iron, possibly due perhaps to the many more sub-levels available for the same recombination.

\subsection{First and second order Fermi acceleration}

The argument here rests on the injection of cosmic ray particles from the post-shock population, and pitch angle scattering to attain isotropy (first stage), moving up in momentum initially by first order Fermi scattering (second stage) in the near-shock region. Further on they increase their momentum slightly, possibly also by second order Fermi acceleration at the shock using the entire post-shock region, or/and then by first order Fermi acceleration (second stage), also using the entire post-shock region.  This entails that the initial first order Fermi acceleration uses the Bohm limit of magnetic irregularities ($I(k) k \, \sim \, 1$), while the third stage uses the Jokipii limit of magnetic irregularities ($I(k) k \, \sim \, k^{-1}$), equivalent to a picture of repeated shocks running through the region affected, steady only on average over longer time-scales.  We note from observations (also elsewhere in this paper) that these shocks are typically and on average at about $0.1 \, c$ (summarized in Biermann et al. 2016, and also above), where $c$ is again the speed of light.

The Jokipii limit case is equivalent to the scattering, independent of particle momentum (in the relativistic case), see Jokipii (1982, 1987) and the arguments inspired by Jokipii in Biermann (1993).  The corresponding magnetic irregularity spectrum of $k^{-2}$  spectrum can be thought of as the Fourier transform of a saw-tooth pattern of repeated shocks running through the region of interest, as would happen in an unstable shock region (cf. also Federrath 2013).  We will refer to these two cases again as Bohm limit and Jokipii limit.

\subsection{Toy model}

Here we describe a toy model to help us understand injection and acceleration, and how they depend on the nucleon number of the nucleus $A$, and its initial ionized charge number $Q_{0}$.  We will study the case of a strong shock racing through a stellar wind as our example, although the details do not critically depend on this; we will show below where a shock expansion into a homogeneous medium would modify the result.  As shown by Parker (1958, after eqs. 25 and 26 on p. 673) dissecting the magnetic field into multi-pole components allows the dominant term to be the $B_{\phi} \, \sim \, r^{-1}$ component, except for a polar cap, for which Maxwell's laws require $B_{r} \, \sim \, r^{-2}$.  The Rankine-Hugoniot conditions govern the scale, using a density jump of 4 in a strong shock (using the high Mach-number approximation in a gas of adiabatic gas index $5/3$); this implies a radial scale of $r/4$ in a wind, and $r/12$ is a homogeneous medium, where $r$ is the current radius coordinate of the shock (Biermann 1993, Biermann \& Cassinelli 1993, Biermann \& Strom 1993). This implies that in a wind, the characteristic ``long" time scale is $r/\{4 U_{sh, 2}\} \, = \, r/U_{sh, 1}$.  With a chaotic magnetic field, the strongest component is perpendicular to the shock normal, i.e. parallel to the shock surface.  We will focus on this component.

The first step in this model is the pitch angle scattering, starting from the torus-like configuration in momentum phase space just behind a shock (i.e. particles running along a Larmor circle, so also circling in momentum space).  The shock is highly unstable, producing a population of cosmic ray particles that can be thought of as a light fluid, being pushed by a heavy fluid (i.e. the thermal gas undergoing the shock transition).  Biermann (1994b) discussed the associated instabilities, using observations (e.g., Braun et al. 1987 with radio observations of the supernova remnant Cas A). This can be considered as similar to an irregularly moving wave, advancing then slowing down, and advancing again, but irregularly with a grand cycle time of $r/\{4 U_{sh, 2}\} \, = \, r/U_{sh, 1}$; the waves keep coming, and their average location is approximately constant.  In radio/X-ray images of Supernova Remnants (SNRs) this picture implies sharp edges to the emission, as observed, but a variable edge location corresponding to the different phases of the shock reforming, and moving ahead, again as observed. 

The second step is momentum scattering in the shock transition, so ``first order Fermi acceleration" in the immediate neighborhood of the shock.  Here we use the assumption of maximal scattering, so a Bohm-like scattering regime.  

The third step is acceleration over the entire region ($r/4$ in a wind) by all the disturbances.   At the lower energies this may still be in Bohm mode, and there we can allow for a contribution from second order Fermi acceleration, but at the higher energies this surely will be in Jokipii mode, which implies that the particles gain energy independent of charge $Q$ and mass $A$.  This then transitions into a first order Fermi acceleration process with Jokipii-like scattering (Jokipii 1982, 1987, Biermann 1993), the main process at this stage, involving also the entire region ($r/4$ in a wind).  In such a picture the polar cap component corresponds to the first regime, first order Fermi in a highly non-steady mode.  

The line of reasoning we are taking is the standard approach of following the flow of particles in appropriate phase space. We use the following nomenclature: The population of particles continuously freshly established as a torus in momentum phase space and highly anisotropic in pitch angle is called $N_{torus}$.  These particles are being provided from a freshly shocked particle population $N_{shock}$ coming in from upstream. Pitch angle scattering with a rate of $1/\tau_{\mu \, \mu}$ feeds into the first order Fermi process.  This particle population $N_{1FB}$ then feeds into the first order Fermi population accelerated using an magnetic irregularity spectrum of Bohm spectrum.  Its rate of momentum change is $1/\tau_{1FB}$.  As the next step the particle population$N_{1FJ}$ is accelerated by the first order Fermi using the Jokipii spectrum.  It gets fed from the population $N_{1FB}$.  We then obtain the two differential equations

\begin{equation}
\frac{\partial N_{1FB}}{\partial t} \; = \; \frac{N_{torus}}{\tau_{\mu \, \mu}} - \frac{N_{1FB}}{\tau_{1FB}} - \frac{N_{1FB}}{\tau_{conv}} \, ,
\end{equation}
\noindent as well as

\begin{equation}
\frac{\partial N_{1FJ}}{\partial t} \; = \; + \frac{N_{1FB}}{\tau_{1FB}} - \frac{N_{1FJ}}{\tau_{flow}} \, .
\end{equation}
\noindent The solution is straightforward:  Writing

\begin{equation}
\frac{1}{\tau_{\star}} \; = \; \frac{1}{\tau_{1FB}} \, + \, \frac{1}{\tau_{conv}} \, ,
\end{equation} 
\noindent and

\begin{equation}
\frac{1}{\tau_{\star \, \star}} \; = \; \frac{1}{\tau_{flow}} \, - \, \frac{1}{\tau_{\star}} \, ,
\end{equation}
\noindent we find

\begin{equation}
N_{1FJ} \; = \; N_{torus} \; \frac{\tau_{\star} \, \tau_{flow}}{\tau_{\mu \, \mu} \; \tau_{1FB}} \, \left[{\left( 1 \, - \, e^{-\frac{t}{\tau_{flow}}}\right)} - \frac{\tau_{\star \, \star}}{\tau_{flow}} \, {\left(e^{-\frac{t}{\tau_{\star}}} \, - \, e^{-\frac{t}{\tau_{flow}}}\right)}\right] \, ,
\end{equation}
\noindent with the limit  ${t}/{\tau_{flow}}$ and ${t}/{\tau_{\star}}$ both large compared to unity 

\begin{equation}
N_{1FJ} \; = \; N_{torus} \, \frac{\tau_{\star} \, \tau_{flow}}{\tau_{\mu \, \mu} \, \tau_{1FB}} \, ,
\end{equation}
\noindent and the condition, that

\begin{equation}
\frac{1}{\tau_{\star}} \; = \; \frac{1}{\tau_{1FB}} \, + \, \frac{1}{\tau_{conv}} \, \simeq \, \frac{1}{\tau_{conv}} \, ,
\end{equation}
\noindent so that the parameter dependence of $\tau_{1FB}$ is superseded; or in other words

\begin{equation}
\tau_{conv} \, < \, \tau_{1FB} \, .
\end{equation}
\noindent This entails then finally that

\begin{equation}
N_{1FJ} \; = \; N_{torus} \, \frac{\tau_{conv} \, \tau_{flow}}{\tau_{\mu \, \mu} \, \tau_{1FB}} \; .
\end{equation}
\noindent The time scales $\tau_{conv}$ and $\tau_{flow}$ are clearly independent of parameters charge $Q$ and mass $A$, since they refer to overall flow, like the irregularities excited by the dominant ions, or the scales determined by the Rankine-Hugoniot conditions of the shock.  This means that we have to work out the dependencies of the two rates $1/\tau_{\mu \, \mu}$ and $1/\tau_{1FB}$ on the key parameters $Q$ and $A$.  This then determines the scaling of $N_{1FJ}$ with mass $A$ and the initial value of charge $Q_{0}$.  We will have to ascertain that ionization is slow enough to allow this.

We note that these conditions are the same as always invoked in deriving non-thermal power-law spectra:  In incomplete Comptonization the escape time has to be of the same order of magnitude as the scattering to produce power-law spectra (e.g. Illarionov \& Sunyaev 1972a, b; Shapiro \& Lightman 1976; Katz 1976).  Analogously, the spectrum of first order Fermi acceleration can be written as the ratio of the two time scales, the acceleration time scale and the escape time scale (e.g. Berezinsky et al. 1990) and again they have to be of similar order of magnitude to reproduce the observed spectra, and that of course is the case in standard first order Fermi acceleration (Axford et al. 1977; Krymskii 1977; Bell 1978a, b; Blandford \& Ostriker 1978).

\subsection{Step 1:  Pitch angle scattering}

The distribution in pitch angle starts with a narrow distribution, a torus, that is slowly broadened.   Considering the pitch angle scattering, the key dependence is on $Q_{0}$ and $A$, and the time step is given by some small multiple of the cycle time of the dominant ions.  The scattering in pitch angle in terms of the distribution function $f$ is given by the equation

\begin{equation}
\frac{\partial f}{\partial t} \; = \; \frac{\partial}{\partial \mu}\left( D_{\mu \mu} \frac{\partial f}{\partial \mu}\right) \, .
\label{Pitchanglescatt}
\end{equation}
\noindent We use again the following convention. $I(k)$ is the energy density per wave-number $k$ in magnetic irregularities, or $P(\omega)$ per frequency, for $\omega$ the resonant gyro frequency, $\omega \; = \, (Q \, e \, B)/(A \, m_p \, c)$, with $Q$ the charge, and $Q_{0}$ the initial charge state, $A$ the mass number, and $m_p$ the mass of the proton (obviously this is only correct to within the modifications due to nuclear binding energies).  Here we have $r_g \, = \; (p c)/(Z e B)$ and $k \, \sim r_g^{-1}$. $v$ is the velocity of the particle, with $p \, = \, v \, A \, m_p$ for sub-relativistic speeds, This is derived from scattering with $\kappa \; \sim \; r_g \, v $ for the Bohm case, and $\kappa \; \sim \; const$  in the Jokipii case and $v$ the velocity of the particle using 

\begin{equation}
\kappa \; = \; \frac{1}{3} \frac{p c}{Z e B} \, v \,  \frac{B^2/{[8 \pi]}}{I(k) k} \, .
\end{equation}

Following Jokipii (1966) we use here the energy in magnetic fluctuations as a function of resonant gyro frequency $\omega_{fluct} \, = \, (Q \, e \, \delta B)/(A \, m_p \, c)$, and with $\delta B$ the fluctuating part of the magnetic field in resonance.  Writing the ${<\delta B>}^2 \, = \, P(\omega)$ and $P$ the power in these fluctuations per frequency we have

\begin{equation}
D_{\mu \mu} \sim \frac{Q^2}{A^2} \, P(\omega) \, .
\end{equation}

For the Bohm case ($P \, \sim \, \omega^{-1}$) we obtain a constant by multiplying with $\omega/\omega$, and for the Jokipii case ($P \, \sim \, \omega^{-2}$) by multiplying with $(\omega/\omega)^2$, and therefore obtain for the sub-relativistic Bohm case, that  $D_{\mu \mu} \sim Q \, A^{-1} \, \sim r^{-1}$, and for the sub-relativistic as well as relativistic Jokipii case, that $D_{\mu \mu} \sim \;\; {\rm independent \; \; of} \; \; A, \; Q, \; r$.  

Here we adopt the sub-relativistic Bohm case. We see that the number of particles 
$\sim Q \, A^{-1} \;$.  This start of a flow in pitch angle space determines the character of the quasi steady flow (quasi steady over time scales much smaller than the over all flow turn-over time scale $r/\{4 U_{sh, 2}\} \, = \, r/U_{sh, 1}$).  

The rate $1/\tau_{\mu \, \mu}$ scales with $D_{\mu \mu}$, or as $U_{sh}/r_g \, \sim \, Q \, A^{-1}$, so:

\begin{equation}
1/\tau_{\mu \, \mu} \, \sim \, Q \, A^{-1} \, .
\end{equation}
\noindent The particle momentum distribution begins at momentum $p_a \, = \, A \, m_p \, U_{sh, 1} \, \sim \, A$.  We generally set $U_{sh, 1} \, = \, U_{sh}$ for simplicity in cases where no confusion can arise.

\subsection{Step 2: Injection to first order Fermi acceleration in the Bohm limit}

Next we consider in this model that at the shock particles scatter back and forth across the shock in a maximally turbulent regime; this means that the magnetic field irregularity spectrum can be described as $I(k) k \, \sim \, const$, the Bohm limit.  Then the acceleration time can be derived starting from the spatial scattering, required to scatter the particles back and forth across the shock:
\noindent The spatial scattering coefficient can be written as

\begin{equation}
D_{x x} = \frac{1}{3} \, r_g \, v \, \frac{B^2/(8 \pi)}{I(k) k} \, \sim \, \frac{p^2}{Q \, A} \, .
\end{equation}
\noindent We  note here again, that using the fast Jokipii limit of acceleration (Jokipii 1982, 1987) implies $D_{x x} \, = \, r_g \, U_{sh}$, and so does not modify the argument here.  The large scale Jokipii limit ($ D_{x x} \, = \, (1/4) \, r \, U_{sh}$ upstream: Biermann 1993) determines the spectrum, while the fast Jokipii limit gives the acceleration time.  We work in the sub-relativistic case: $D_{xx} \sim Q^{-1} \; A^{-1} \, p^2 \, r$.   The first order Fermi $\{1 \; F\}$-time-scale (acceleration at the shock), and assuming that $D_{x x, 1}/U_{sh, 1} \, = \; D_{x x, 2}/U_{sh, 2}$ is

\begin{equation}
\tau_{1 F} \; = \;\frac{1}{3} \, \frac{p \, c}{Q \, e \, B(r)} \, \frac{p}{A \, m_p} \, \frac{B^2/(8 \, \pi)}{I(k) \, k} \, \frac{8}{U_{sh, 1}^2} \, ,
\end{equation}
\noindent and we assume this to be the second stage, with a spectrum of particles in momentum of $p^{-2}$.  We also assume that at this stage the particles still have initially their original stage of ionization, $Q_{0}$.  The relevant term in the cosmic ray transport equation is then $- \frac{\partial}{\partial p} \left( \dot{p} f \right)$ and:

\begin{equation}
1/\tau_{1 \; F} \, \sim \, Q_{0} \, A \, .
\end{equation}
\noindent This gives the second rate in the full equations earlier, for $N_{1FJ}/N_{torus}$.  The flow in momentum space due to this process is proportional to $Q_{0} \, A$, so together with the flow in pitch angle space, we have here an enhancement by a factor of $Q_{0}^2$.  Allowing $Q_{0}$ to be either 1 or 2 immediately gives a factor of 4 between these two initial stages of ionization, consistent with our interpretation of the result of Murphy et al. (2016) as differentiating these two stages of ionization.  The spectrum is then, including this factor

\begin{equation}
Q_{0}^2 \, {\left(\frac{p}{p_a}\right)}^{-2} \, d p \, \sim \, Q_{0}^2 \, A^2 \, p^{-2} \, d p \, ,
\end{equation}
\noindent or, with the proper normalization to an initial momentum $p_{1FB}$, which might be quite close to the initial momentum $p_a$,

\begin{equation}
Q_{0}^2 \, A^2 \, p_{1FB}^{-2} \, {\left(\frac{p}{p_{1FB}}\right)}^{-2} \, d p \, .
\end{equation}

\subsection{Step 3:  Fermi acceleration across the entire affected region}

The first order Fermi $\{1 \; F\}$-time-scale (acceleration at the shock) is

\begin{equation}
\tau_{1\; F} \; = \;\frac{1}{3} \, \frac{p \, c}{Q \, e \, B(r)} \, \frac{p}{A \, m_p} \, \frac{B^2/(8 \, \pi)}{I(k) \, k} \, \frac{8}{U_{sh, 1}^2} \, ,
\end{equation}
\noindent as long as the particle is sub-relativistic, and 

\begin{equation}
\tau_{1\; F} \; = \;\frac{1}{3} \, \frac{p \, c}{Q \, e \, B(r)} \, c \, \frac{B^2/(8 \, \pi)}{I(k) \, k} \, \frac{8}{U_{sh, 1}^2} \, ,
\end{equation}
\noindent as soon as the particle is relativistic.  We assume this to be the final stage, with Jokipii limit irregularities, so $I(k) \, k \, \sim \, k^{-1}$ (see above, and Federrath 2013), and a spectrum of particles in momentum of $p^{-7/3}$ (Biermann 1993).  Both spectra, the polar cap component $p^{-2}$, and the $4 \, \pi$  component $p^{-7/3}$ occur concurrently (Biermann 1993), either as in a normal Parker (1958) configuration, separated spatially; an alternative is being separated episodically in regions where the local magnetic field is either temporarily radial or perpendicular to the shock normal, thus separated in momentum space. However, the maximum energy in the $p^{-2}$ component is the knee energy $E_{knee}$ discussed earlier.  The knee energy is related to the maximum energy overall by $E_{knee} \, \simeq \, (U_{sh, 1}/c)^2 \, E_{max}$. $E_{max} \; = \; E_{ankle}$ in turn is given by $E_{ankle} \, \simeq \, (1/8) \, Z \, e \, r \, B(r)$ to within factors of order unity.  This energy is  the maximum energy overall for GCRs, so corresponding to the $4 \, \pi$-component.  In all three equivalent pictures the fraction of the $p^{-2}$ component around GeV energies is at a level of a few percent relative to the $p^{-7/3}$ component, giving a switch over in the observed spectrum at several TeV (times charge $Z$) (Biermann et al. 2010).    The spectrum is then, without an intermediate stage with second order Fermi acceleration, 

\begin{equation}
Q_{0}^2 \, A^2 \, p_{1FB}^{-2} \, {\left(\frac{p_{1FJ}}{p_{1FB}}\right)}^{-2} 
\, {\left(\frac{p_A}{p_{1FJ}}\right)}^{-7/3} \, A \; \frac{d p}{A} \, ,
\end{equation}
\noindent where $p_{1FB}$ cancels out, and $p_{1FJ}$ is the transition momentum from a Bohm-like spectrum of magnetic irregularities to a Jokipii-like spectrum, and therefore independent of both $A$ and $Q$, since the rate of acceleration $1/\tau_{1FJ}$ is independent of both $A$ and $Q$.  Also, going from Bohm to Jokipii implies a switch of the spectrum from -2 to -7/3 (Biermann 1993, and above).

Here we scale the entire spectrum to energy (momentum) per nucleon, so use $p_A \, \sim \, A$, and also use a reference interval of energy per nucleon $\frac{d \, p}{A}$; this yields the factor $A^{-4/3}$.  All the rest of the factors here either cancel or are independent of both $Q$ and $A$;  as the transition into the $4 \, \pi$-component is independent of $A$ and $Q$ for both first and second oder Fermi acceleration (see, e.g., Seo \& Ptuskin 1994), $p_{1FJ}$ and also $p_{2FJ}$ are independent, while the other terms just cancel out, $p_{1FB}$.  Not allowing for any contribution from second order Fermi requires that there be a transition from Bohm to Jokipii mode, while the spectrum is still $p^{-2}$, here expressed in the condition that the spectrum from $p_{1FB} \, c$ to  $p_{1FJ} \, c$ is still -2. In this model an initial $p^{-2}$ contribution is necessary to get the $p_a^2 \, \sim \, A^2$ term as an initial factor at the lowest energies; it also shows the lower-energy intermingling of the polar-cap-component and $4 \, \pi$-component.  We have used here the simple approximations of using just power-law spectra of -2 (the polar cap component) and - 7/3 (the $4 \, \pi$ component), and will discuss below what influence it would have if we would allow for other power-law spectra.  As mentioned earlier, this initial $p^{-2}$ component in the Bohm case is required to explain the hardening of the spectra detected by CREAM and AMS (Ahn et al. 2010a, Yoon et al. 2017, Aguilar et al. 2017). There is a similar upturn in the total cosmic ray spectrum seen in the IceTop data around 20 PeV (Gaisser et al. 2016b), which is well reproduced (Todero Peixoto et al. 2015) by adding the polar cap and $4 \pi$ contributions below and above the their respective knee-energy from the different heavy elements (remember, that the knee energy is proportional to charge in this model).  In the model of Thoudam et al. (2016) they obtain a similar upturn and curvature of the total CR spectrum as in Todero Peixoto et al. (2015).

This then finally says for the ratio of CR source abundances over source material abundances:

\begin{equation}
\Phi_{source, i} \, Q_{i,0}^{+2} \, A_{i}^{+2/3} \,  {\left(\frac{p}{A_{i} p_{0}}\right)}^{-4/3} \, ,
\end{equation}
\noindent for element $i$, with $A_{i} \sim A$.  This matches the distributions of Binns et al. (e.g. Murphy et al. 2016).  Murphy et al. explicitly state that a factor of $\sim$ 4 is between the two lines in the graph, consistent with our claim that $Q_{i,0}$ either 1 or 2.

\subsection{The case of Carbon and Oxygen}

The numbers for Carbon and Oxygen recalculated for Fig.\ref{Binnsabundances2016} are $0.22 \pm 0.02$ and $0.123 \pm 0.02$, respectively (Binns 2017, priv. comm.; see also Rauch et al. 2009).  The model presented here says that an element with a high 2nd I.P. should be close to the lower line, labelled ``volatiles", and an element with a low 2nd I.P. should be close to the upper line, labelled ``refractories", assuming in the model proposed here that the switch between states of ionization is rapid.  Oxygen, with a 2nd I.P. of 35.1 eV, should therefore be close to the lower line. It is.  Carbon, with a 2nd I.P. of 24.4 eV, is a more ambivalent case, as its 2nd I.P. is close to the middle point between the two cases, as shown by the tendency of the break-point to shift down with mass $A$; it might be quite close to the case of Carbon at its relatively low mass.  Its number in this plot \ref{Binnsabundances2016} is just between the two cases.  An additional worry in using the plot of Murphy et al. (2016) is of course, that in the model of Binns and his group SN ejecta are mixed with ISM abundances, whereas in our model wind-abundances in BSG stars are mixed with wind-abundances of Red Super Giant (RSG) stars; these two mixes ought to be very similar. How much of the earlier ejecta are also included in the acceleration when the SN-shock slows down in the wind-shell is not clear; however, the relative sharpness of the {\it knee} feature in the spectrum suggests that the value of $r \, B(r)$ is the same for most observed accelerated particles; it ensues that the amount of freshly accelerated particles from the material beyond the wind-shock is relatively small. Considering the case of M82 this cannot be decided so easily.  We have argued above that all the compact radio sources in M82 correspond to the BSG star case: In contrast a shock will diminish in strength racing through RSG star winds: This slow-down was actually instrumental in explaining the AMS anti-proton abundance and its spectrum.  Furthermore, the abundances in RSG star winds are similar to ISM abundances, but not exactly the same (see below).  Above we have speculated that some spallation might affect Oxygen and Carbon nuclei in BSG stars, possibly explaining the low energy proton and Helium AMS spectral data; if that were really the case, Oxygen and Carbon would be even more problematic.

\subsection{The ratio of Solar System vs enriched injection}

Here we check another quantity determined from the data:  Murphy et al. (2016) give a ratio of 81 \% versus 19 \% for Solar System versus enriched ejecta.  The ratio is about 4.

Here we show that this ratio can naturally be understood as derived from the injection from RSG stars and BSG stars:

For instance, using Fig 16 in Heger et al. (2003) showing mass fractions of various chemical elements versus Zero Age Main Sequence (ZAMS) mass from simulations we can integrate the mass in the ejecta, that are (i) Solar System, (ii) enriched just in Helium, and (iii) enriched with Carbon, Oxygen and higher elements.  From the simulations we use the mass intervals 21 $M_{\odot}$, 30 $M_{\odot}$. 33 $M_{\odot}$, 52 $M_{\odot}$, and 80 $M_{\odot}$.  

We can make a similar argument using other stellar evolution calculations (Chieffi \& Limongi 2013), using the mass steps 25 $M_{\odot}$, 30 $M_{\odot}$, 35 $M_{\odot}$, 40, and 60 $M_{\odot}$.  Chieffi et al. give two extreme scenarios: 1) in which the final remnant is a neutron star over much of the relevant mass range, and 2) in which a relatively massive black hole develops, up to 7 $M_{\odot}$ in the mass range 25 - 60 $M_{\odot}$ ZAMS mass.  

Here we have used mass fraction in all integrals to derive these numbers , while in Murphy et al. (2016) they use number ratios, scaled to the Iron (Fe) abundance. This introduces a correction factor $> \, 1$, since a more enriched composition has fewer atoms for the same mass as a Solar System or only Helium enriched composition.  Using Carbon versus Helium gives a lower limit to this correction factor, a factor of 3.  Remembering that Oxygen is usually similar in abundance to Carbon, and that Helium is mixed in with a lot of Hydrogen increases this correction factor possibly to about 7.  This number is not precise since it depends critically on the uncertain aspects of stellar evolution.

Complications intrude here: First of all, in RSG stars there is a dredge-up process, modifying the abundances.  Second, rotation adds further uncertainty.  Third, even the BSG stars have an outer shell, which has Solar System abundance; this outer shell of the star may be part of the wind shell still touched by CR acceleration (we included this mass in the gamma-ray line slow-down argument earlier).  For the cosmic ray injection we mainly need the material in the wind at the time of the explosion, since the wind-shell is established over all the earlier time of the wind, rapidly decelerates the SN-shock.  As noted already above, activity episodes of the massive star close to its explosion evident in the data add even further uncertainty.  Finally, just distinguishing RSG and BSG (or Wolf Rayet, WR) stars and their winds is much too simple, as the WR stars themselves are formed showing a large variety of different mantle and wind compositions.

Counting then Helium with the enriched composition in Heger et al. (2003) gives an estimated corrected Solar System to enriched GCRS ratio of 14, and in Chieffi \& Limongi (2013) et al. about 4 in both scenarii.  Within the relatively large uncertainties these two numbers are consistent with each other.

The main conclusion is that both in the models by Chieffi \& Limongi (2013) and by Heger et al. (2003) the proposal is consistent with the data:  All the Solar System abundances above Helium contained in the abundances of cosmic rays and their sources allow the Red Super Giant (RSG) stars to provide the Solar System abundances used in the Binns-diagram of Murphy et al. (2016). The WR stars or BSG stars can provide the enriched abundances used in the Binns-diagram of Murphy et al. (2016).

This leads to a clear prediction: Helium ought to derive mostly from RSG or BSG stars, and so have the same spectrum essentially as Carbon, Oxygen etc. This is supported by the AMS data (Aguilar et al. 2017); these data even show an identical upturn interpreted here as the slow switch to the polar cap component. The CREAM and AMS data suggest that at very low energy Helium has a spectrum slightly different from Carbon and Oxygen, closer to hydrogen, and at higher energy the spectra converge to the same spectrum as Carbon and Oxygen, consistent with this argument  (Ahn et al. 2009, 2010a, b; Aguilar et al. 2015a, b).  The slight difference in spectrum at low energies for Helium may well correspond to the (Interstellar Medium, ISM) ISM-SN-CR component, which dominates the Hydrogen/proton component at lower energies, and may contribute weakly to Helium.  However, above we also speculate whether there could be a spallation component for both protons and Helium nuclei.

\subsubsection{Dependence on spectrum}

Next we need to check a few conditions of the idea proposed here:  First, the exponent of the mass number $A$, here $4/3$, is connected to the spectrum of cosmic rays, for which we took here $E^{-7/3}$ at source (Biermann 1993); this number was derived using a limiting argument on the scaling behavior of cosmic ray scattering; this spectral index may have to be modified to incorporate further non-linear effects.  Using the integrated spectrum gives $E^{-4/3}$, and the exponent $2/3$ is $2 \, - \, 4/3$.  So if the real spectrum were steeper, for example, then this exponent would be shallower.  Or if shallower, the dependence would be stronger, as already noted above for the polar cap component.  There is no influence implicit from the assumption that the polar-cap spectral component has a spectrum of $E^{-2}$. Obviously, this could also be slightly steeper due to non-linear effects (de Boer et al. 2017).

\subsubsection{Polar cap component?}

Second we need to check whether the injection conditions for the polar-cap component differ in an analogous way: If injection were just local, the isomorphic argument would suggest that the scaling of the flux is then with 
$Q_{i,0}^{+2} \, A_{i}^{+1}$ instead of with $Q_{i,0}^{+2} \, A_{i}^{+2/3}$.  A consequence would be that the transition energy $E_{\star}$ between the two components at higher energy would scale either as $E_{\star} \, \sim \, const(A)$ or as  $E_{\star} \, \sim \, A$.  The data (Ahn et al. 2010a, Fig.5) suggest the second case, $E_{\star}/A \, \sim \, const$, implying that for injection the polar-cap component fully mixes with the $4 \, \pi$-component, as already noted above.  Since this component is almost all of $4 \, \pi$ this is not surprising.

\subsubsection{Ionization rate}

Third, we also need to check that the ionization does not depend greatly on radius. The ionization time scale should be longer than the initial fast acceleration processes.  After all, the arguments on the dependence of injection on the initial degree of ionization need to work out over the entire range of radii which the SN-shock traverses:

The ionization rate $\zeta$ can be written as follows (see, e.g., Nath \& Biermann 1994),  using all the ions at post-shock velocity of typically $ \beta \, c = \, 0.1 \, c$ (summarized above)

\begin{equation}
\zeta \; \simeq \; 4 \, \pi \, c \, \beta \, n_{down} \, 10^{-17.3} \, {\rm sec^{-1}} \, ,
\end{equation}
\noindent where $n_{down}$ is the post-shock downstream density of the ionizing energetic ions.  This acts at most a flow-time of $10^{6.5} \, {\rm s} \, (r)/(10^{16} \, {\rm cm})$, using again the post-shock velocity.  The density is from above $10^{3.5} \, {\rm cm^{-3}} \, \, {\dot{M}}_{-5} \, V_{W, 8.3}^{-1}  \, (\{r\}/\{10^{16} \, {\rm cm}\})^2$, so that ionization can significantly proceed already in a time which is given by $10^{2.6} \, {\rm sec} \, (r)/(10^{16} \, {\rm cm})^2$, which is to be compared with the flow time; this shows that even at the largest radius ionization will be efficient within the time available.  However, this also shows that ionization is slow compared to the other time-scales near the shock, such as the second order Fermi or first order Fermi time-scales (see also, Morlino 2009).  This is required in this model, since we use the slow ionization to get the selection effects established.

We note that this is quite different from the $A$, $Q$ selection in Solar energetic ions (Reames et al. 2015).

\subsubsection{Alternative ?}

Fourth, we need to see whether this proposal is actually consistent with the idea that cosmic ray particles are injected and accelerated in OB-star-super-bubbles (e.g. Higdon \& Lingenfelter 2003, 2005, 2006, Higdon et al. 2004, Lingenfelter \& Higdon 2007). This proposal seems to be supported by the work of Binns et al. (Wiedenbeck et al. 1999, Rauch et al. 2009, Binns et al. 2011, 2013, Murphy et al. 2016).  We have used the notion at several steps in the argument that the concept derives from acceleration in winds.  First, we used it to derive the knee and the ankle energies, using observations of supernova shocks in winds.  It seems implausible that the same shocks would persist at the observed velocities far into any OB super-bubble, but it cannot be excluded at this time.  Second, we used the spectral index at injection of $-7/3$, derived quite explicitly from a concept of shocks in winds. Furthermore, in the concept proposed, this spectral index is directly connected to the power-law of the $A$-dependence of the abundance ratio in the Binns-diagram; this could be used as a test.  However, considering the uncertainties in spallation during propagation and source abundance corrections due to our incomplete understanding of stellar evolution, this is not a strong test. Again,  a similar argument could be constructed for the initial expansion of shocks into OB-star-super-bubbles.  And third, we have shown that the abundance ratios match the red super-giant star winds versus the blue super-giant star wind; that is just an alternative, however not an either/or argument. The one test which could be strong is calculating the abundances of radio-active isotopes that are injected and accelerated very early; $^{59}$Ni is not strong enough as a test (Binns et al. 2007), as it depends on the initial uncertain abundances.  The selection effects depending on the second Ionization Potential (I.P.) leading to a proposed explanation of the Binns-diagram are the strongest argument perhaps.  The positive detection of an isotope with a short lifetime could be a decisive test, but such an isotope may depend on spallation during early propagation.

\subsection{Injection conclusion}

This section is inspired by the data obtained by Binns et al. (e.g. Murphy et al. 2016).  The fit of this model to the data supports the view that the ionization structure from which these cosmic ray particles are drawn is sufficiently hot to allow the second ionization potential to become the arbiter for cosmic ray abundances.  The abundance mix supports a combination from RSG star winds, mostly un-enriched and so similar to Solar System abundances, and BSG winds, which are heavily enriched (ejecta).  This mix can readily reproduce the mix worked out by Binns et al. (e.g. Murphy et al. 2016).

\subsection{Ultra-heavies}

Next we discuss the consequences for the CR abundances of ultra heavy elements after detecting an abundant source in the neutron star merger GW170817 (Chornock et al. 2017, Murquia-Berthier et al. 2017, Pian et al. 2017, Rosswog et al. 2017, Tanvir et al. 2017). The observed features in the spectrum have tentatively been associated with heavy elements in the mass range of the third r-process peak, although alternatively lighter r-process elements could probably also explain the observed spectra and light curve. Clearly this would then also be the input for CR-injection.  But is that injection into high energies already at the neutron star merger event? Or do these heavy elements mix into the general medium in the next cycle of star formation and explosion?  Already the neutron star merger showed non-thermal emission, so injection and acceleration of energetic electrons. However, we do not know at this early stage how this plays out for energetic and ultra-heavy nuclei. Could the energetic electrons be secondary from interactions of nuclei?  The CR abundances, as, for example, summarized in Wiebel-Sooth \& Biermann (1999) in their Fig. 11b, of the ultra heavies as compared to Solar system abundances, show the following: (i) Spallation fills in abundance valleys, and (ii) these heavy element abundances scale with the Iron abundance (based on Byrnak et al. 1981, Binns et al. 1985, 1989, Fowler et al. 1987, and Westphal et al. 1998).  Most Iron derives from SN Ia, much of Carbon and most of Oxygen from WR (BSG) star explosions.  In addition, much if not all of the very heavy elements can be traced to neutron star mergers. This suggests that these elements are first incorporated into the ISM and then accelerated along with everything else in BSG star explosions, as explained above.  This can be tested with the Binns-diagram using the ultra heavy elements. One difficulty in the model proposed will be that most interaction happens in the wind-shells around the BSG star winds. So the spallation is dominated by interaction in the wind-shell after acceleration in the SN-shock of the ``old" ultra-heavy element nuclei.  Unstable isotopes may shine light on the various time scales involved for all the steps: These include production, nuclear reactions during the explosion, spallation during transport, immersion into a massive star, renewed explosion, nuclear reactions again, acceleration, and spallation again.  

\vskip2.0cm

\section{The starburst galaxy M82 with a recent binary black hole merger as a source of UHECRs?}


To understand the Telescope-Array (TA) data (Abbasi et al. 2014; and as cited in Biermann et al. 2016), suggesting the starburst galaxy M82 as a source of Ultra High Energy Cosmic Ray particles (UHECRs), we ask what particle energy can be reached in a relativistic Supernova (SN) explosion or Gamma Ray Burst (GRB).  We refer to the space available (see above: the Larmor motion has to fit well inside the post-shock shock shell). We also need to allow for the relativistic velocities suggested by the GRB data  of observed Lorentz $\Gamma$ factor about 20, using nuclei of charge $Z$ (de Hoffmann \& Teller 1950, Gallant \& Achterberg 1999, Achterberg et al. 2001, Meli \& Quenby 2003a, b, Soderberg et al. 2010a, Kamble et al. 2014, Ellison et al. 2016):

\begin{equation}
E_{max} \; = \; (Z/8) \, e \, B \, r \, \Gamma^2\; = 10^{20.1 \pm 0.2} \, Z \, {\rm eV} \, .
\end{equation}

\noindent This could explain the TA-Collaboration data already allowing protons near $10^{20}$ eV.  The only source for which such an idea might work is the strongest compact radio source, 41.9+58 (Kronberg et al. 1985). This source corresponds to a break-out south in the radio contours (Kronberg et al. 1985). In fact, it does seem unusual in its behavior, and Muxlow et al. (2005) have suggested that it is an off-center GRB (see also Gendre et al. 2013); this allows using other GRB data.  If it were an off-center GRB, one can explain the flux of UHECR protons seen by TA.  The flux roughly corresponds to a pure UHECR luminosity of $10^{40.5} \, {\rm erg/s}$ - not integrating down to GeV energies, which could add a substantial factor (Meli \& Biermann 2013). A GRB just might produce $10^{51.5}$ ergs of UHECRs, a very generous estimate, for a time scale of 3000 yrs.  The SN-rate in M82 has been estimated to be of order 1 SN every 5 years (Kronberg et al. 1985); the rate of stars above 20 - 25 $M_{\odot}$ to explode is about 1/5, so every 25 yrs (see, e.g., Muxlow et al. 1994).  The fraction of those exploding, as either a relativistic SN, or as a GRB has been estimated to be of order 1/100 (Soderberg et al., 2010b and references therein). In M82 this implies one every 2,500 years. The smearing during the transport is likely to be of a similar or even longer time scale, as TA can detect such protons all across the Northern sky, scattered in the magnetic galactic halo wind of our Galaxy (Biermann et al. 2014, 2015, 2016). We note above that we have estimated that an even larger SN rate for M82 is quite possible; so a restriction on the progenitor to the BSG stars is easily allowed.

This would support the now well-established idea that some UHECR particles derive from GRBs:  The analogy between active galactic nuclei and GRBs in producing UHECRs was noted in Biermann (1994a, 1994b), and worked out quantitatively in Milgrom \& Usov (1995, 1996), Vietri (1995, 1996), and Waxman (1995), Miralda-Escud{\'e} \& Waxman (1996), and Waxman \& Bahcall (1997), with a recent review on GRBs by Zhang \& M{\'e}sz{\'a}ros (2004). 

The difficulties in extracting $10^{52}$ ergs in total energy from a core collapse SNe mentioned in Kamble et al. (2014) might be alleviated by considering the magneto-rotational explosion mechanism for core-collapse SNe of Bisnovatyi-Kogan (1970), Akiyama et al. (2003), Ardeljan et al. (2005), Burrows et al. (2007), Bisnovatyi-Kogan et al. (2008, 2013), Moiseenko et al. (2006, 2010, 2012), Moiseenko \& Bisnovatyi-Kogan (2015), Bisnovatyi-Kogan et al. (2015).  This mechanism gives rise to jet formation.  As noted in Biermann (1993) and above, this mechanism allows understanding of the commonality of the magnetic fields observed in wind-SNe that is demonstrated above using common radio data: These numbers are all rather similar to each other regardless even of whether the progenitor star was a BSG or RSG star.

\begin{figure}[h!]
\centering
\includegraphics[bb=0cm 0cm 17.62cm 8.68cm,viewport=1.0cm 0.0cm 17.62cm 8.68cm,clip,scale=1.0]{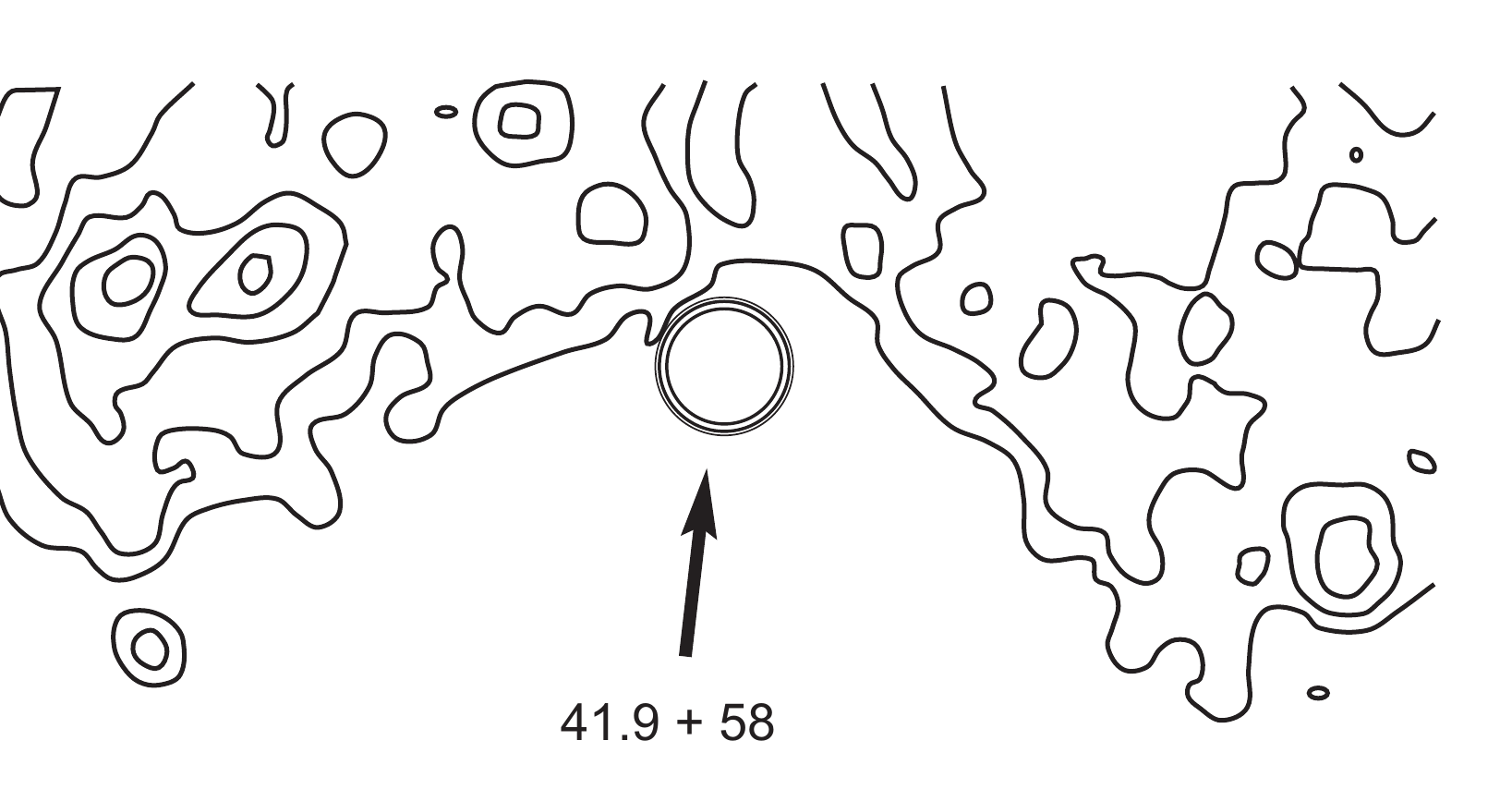}
\caption{{The compact radio source 41.9+58 in a map at 5 GHz with an angular resolution of about 0.35 arc second (VLA, Kronberg et al. 1985). 41.9+58 is emphasized here by a double circle (to suggest its observed VLBI radio structure, a broken shell, Bartel et al. 1987, not to scale),  has been proposed to be a GRB} (Muxlow et al. 2005). The width of the picture is about ten arc seconds, corresponding to about 150 pc.  Here we see that the compact radio source is very near the apex of an open cone of missing radio emission. Can this image be explained by conical sweeping and proper motion, as is possible in a merger of two stellar mass Black Holes (BHs) with uncorrelated spin (Gergely \& Biermann 2009)?  Source: excerpt of Fig. 1 in Kronberg et al. 1985, produced by P.P. Kronberg 2017.}
\label{1985KBSradioM82Fig1}
\end{figure}

\begin{figure}[h!]
\centering
\includegraphics[bb=0cm 0cm 21.0cm 29.71cm,viewport=0.5cm 9.5cm 21.0cm 20.5cm,clip,scale=0.8]{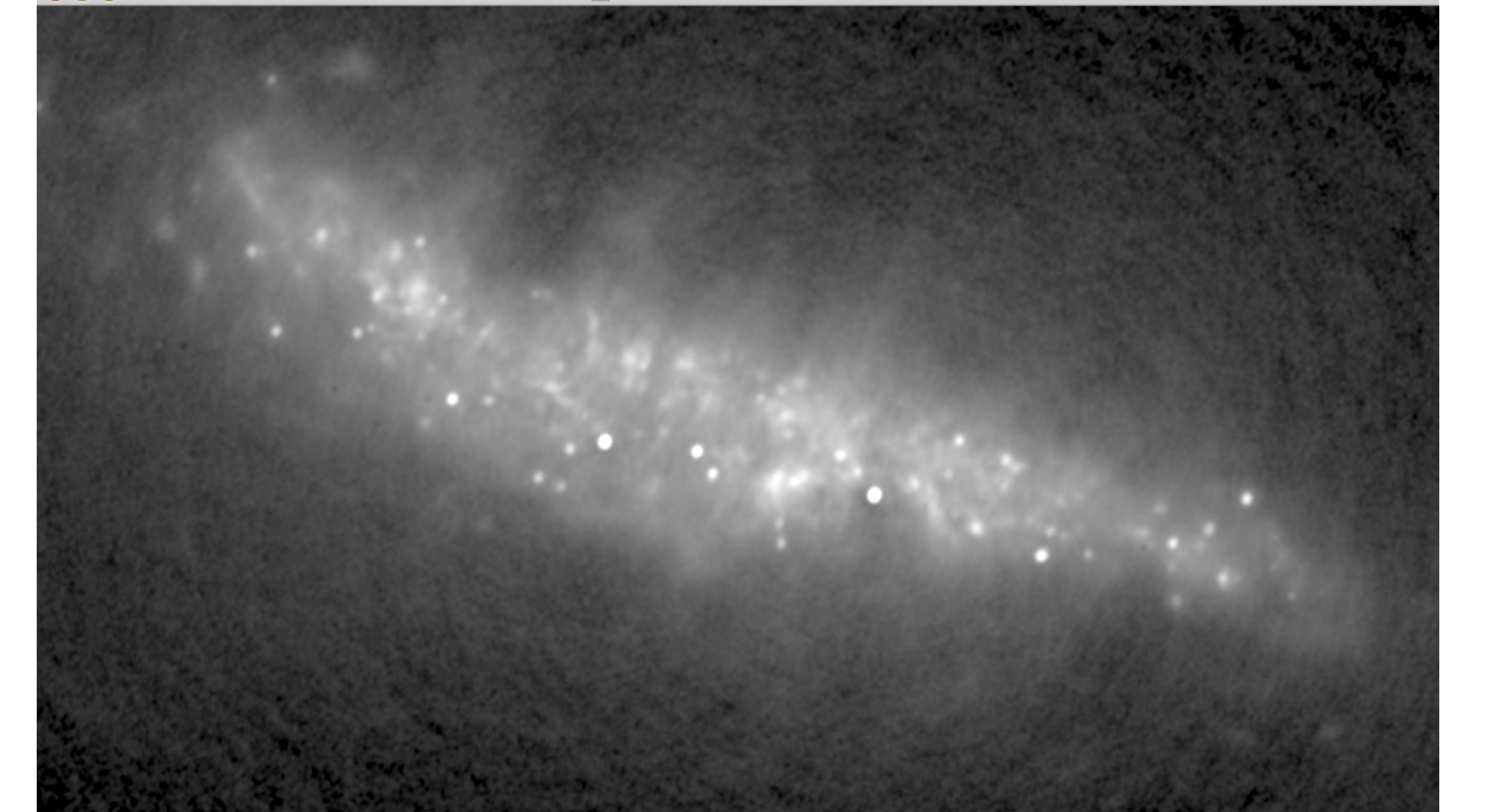}
\caption{A new radio-map of M82 at 8 GHz, with 0.25 arc second resolution (VLA), by M.L. Allen \& P.P. Kronberg, showing the stem-like outflows expected for normal explosions based on the work by, e.g., Kompaneets (1960), Laumbach \& Probstein (1969), Sakashita (1971), and Moellenhoff (1976). A stem-like outflow is a parallel outflow, going out perpendicular to the plane of the galaxy's disk, in contrast to a conical outflow.  The lateral scale of this image is about 60 arc seconds, corresponding to about 1 kpc; it shows many previously unknown features. The compact source distribution can be readily interpreted as a nearly edge-on ``ring", of about half the image scale (see, e.g., Weliachew, Fomalont  \& Greisen 1984, Kronberg \& Wilkinson 1975, and Kronberg et al. 1981).  The compact source 41.9+58 is the brightest compact source in this map, and its open cone downwards is also recognizable here, a geometry very different from stem-like outflows.  Source: M.L. Allen \& P.P. Kronberg not yet published 2017, and used with permission.}
\label{2016KronbergAllen8GHzM82}
\end{figure}

In fact, the unusual shape of the compact radio emission contours shown in Kronberg et al. (1985) Fig.\ref{1985KBSradioM82Fig1} allows us to speculate a bit more:  The morphology is that of an open cone, not the stem-like structure which a powerful explosion would give; see Fig.\ref{2016KronbergAllen8GHzM82} (e.g., Kompaneets 1960, Laumbach \& Probstein 1969, Sakashita 1971, Moellenhoff 1976). A stem-like outflow is a parallel outflow, going out perpendicular to the plane of the galaxy's disk, as opposed to a conical outflow widening from the disk.  In that figure, the compact source distribution, interpreted here as recent radio-SNe, can be readily interpreted as a nearly edge-on ``ring", with ring diameter $\sim$ 500 pc, thus about half the image scale, with a maximal rotational velocity $\sim$ 137 km s$^{-1}$, thickness $<$ 120 pc (see, e.g., Weliachew, Fomalont  \& Greisen 1984, Kronberg \& Wilkinson 1975, and Kronberg et al. 1981). Most of these compact radio sources seem to connect to stem-like features of outflow, as seen in the new 8 GHz radio data, Fig.\ref{2016KronbergAllen8GHzM82}. On the other hand, such an open cone (its width at the maximal opening visible about 10 arc seconds, or about 150 pc) as around the compact source 41.9+58 can be produced by the sweeping of powerful jets during the merger of two black holes (Gergely \& Biermann 2009). This sweeping action goes through a large spiral in its direction, finally settling down on the spin-axis of the orbit of the two black holes (see also Zier \& Biermann 2001, 2002), and so in projection runs through an open cone.  This speculative interpretation posits that two massive stars exploded and produced black holes, both in two different binary systems, so that the black holes can get fed and produce a powerful jet, such as a microquasar (Mirabel 2004, 2011).  Two different binary systems are required, so that the two jets are uncorrelated in direction. This  picture requires the encounter of the two binary systems to occur in a dense massive star cluster (for an example see Bally et al. 2017), with star exchange, so that the two black holes subsequently become bound to each other, and can have very different spin and jet directions.  Then they merge, and in doing so, the jets sweep around (Gergely \& Biermann 2009).  A similar picture has been used for super-massive black hole mergers in Kun et al. (2017) with their jet pointing at Earth and their radio observations, showing a flat radio spectrum extending to near THz.  So this speculation suggests three powerful events in sequence: two supernova explosions, both producing stellar mass black holes, and then a binary BH merger.  This stellar mass BH merger will have happened within the recent century, and would have been of order 100$\times$ closer than the gravitational wave events first detected (Abbott et al. 2016a, b, 2017a, b).  The third Gravitational Wave (GW) event seen (Abbott et al. 2017a) suggests that uncorrelated spins of the two merging BHs is a common occurrence.  Even the fifth observed GW event (numbering them with the dates of announcement), a neutron star merger, is much farther than M82 at 40 Mpc in NGC4993 (Abbott et al. 2017c, d).  Unfortunately, not in all cases do the data allow the determination of the spin directions of the original black holes with sufficient precision to make such a case (Abbott et al. 2017b).  This picture suggests that there could be two kinds of such mergers, one with parallel spins before the merger, and one with quite different spin directions before the merger. The fifth GW detection also suggests that there is a corresponding class of neutron star mergers, either with originally parallel or uncorrelated spins.  The effect on the environment is clearly quite different, in one case a straight shooting of the jet power as in the, for example the W50/SS433 system (Dubner et al. 1998), which shows some precession (Milgrom 1981), and in the other, a broad conical sweep and clean-out such as around the compact radio source 41.9+58.

Although statistical arguments based on a single well observed case are dubious, this does raise the question why we observe such a merger of two stellar mass black holes with uncorrelated spins once within a few months at a distance of order Gpc, and once at about 3 Mpc within order 2,500 years.  This may be explainable with the very high mass of the two stellar mass BHs observed at the large distance.  The mass function of massive stars is simply steep (Kroupa 2007). The two parent stars of the two black holes of the third GW event (Abbott et al. 2017a) must have included at least one very massive star.  This in turn suggests that the event in M82 relates to the lowest stellar mass black holes commonly produced in binary star evolution.  Speculating a bit further using the luminosity function of FIR-bright galaxies in Lagache et al. (2003) shows that galaxies like M82 exist at a space density of about $10^{-4} \; {\rm Mpc^{-3}}$, and so within 400 Mpc there are about $2.5 \cdot 10^{4}$ galaxies like M82. This distance is now comfortably covered by GW detection.  If the GW observatories detect an event such as we suggest 41.9+58 may have been about a hundred years ago, every $t_{3.4} \, \times \, 2,500$ years then the suggested rate at Earth would be  $\sim \, 10 \, t_{3.4}^{-1}$ per year.  This is about one every month, modulo the factor of the time scale $t_{3.4}$, with $t_{3.4}$ a highly uncertain parameter.  This corresponds to a present day rate of $40 \, t_{3.4}^{-1} \, {\rm Gpc^{-3} \, yr^{-1}}$, so very close to the estimated GRB rate of order $30 \, {\rm Gpc^{-3} \, yr^{-1}}$ (Piran 2004), or 1/6 times the rate of $250 \, {\rm Gpc^{-3} \, yr^{-1}}$ obtained by Zhang \& M{\'e}sz{\'a}ros (2004). This comparison allows that some, perhaps many, GRBs are in fact binary black hole mergers.  Finally one could ask whether detectable echoes could be caused through the conical sweeping (Gergely \& Biermann 2009) of a relativistic jet and its highly beamed radiation field during a binary black hole merger (e.g., Rampadarath et al. 2010).

We note that the gravitational wave event GW170817 has been identified as a neutron star merger, a slow GRB, and a compact radio source in the galaxy NGC4993 (Abbott et al. 2017c, d).  This is basically a similar interpretation as our speculation here, with the main difference that in M82 we suggest that it was a binary stellar mass black hole merger with uncorrelated spins, common for starburst galaxies, while the galaxy NGC4993 contains an older stellar population, where a neutron star merger may occur more often.

So we conclude here in summary, that the over 40 observed compact radio sources in M82 may all be young radio Radio Supernova Remnants (SNRs), most of them attributable to BSG star explosions, quite compatible with even younger radio SNe.  For one source, 41.9+58, it has been speculated (Kronberg et al. 1985, Muxlow et al. 2005) that it may not be a normal massive star explosion, but a misdirected GRB or relativistic SN, possibly even a recent stellar mass binary BH merger. This could explain the UHECR particle energies detected.  The expected frequency of such events, one in several hundred normal massive star SNe, would be sufficient to explain the TA data with the starburst galaxy M82, using the expected temporal smearing in the halo of our Galaxy, of the UHECR flux and the expected power.  There ought to be many other events like it in starburst galaxies, possibly already detected (Aab et al. 2018).  While many observed High Energy (HE) neutrinos may relate to super-massive black hole (SMBH) mergers (Kun et al. 2017) with their current relativistic jets pointed at Earth, stellar mass black hole mergers should also produce detectable HE neutrinos, if the associated jets point at Earth.

\vskip2.0cm

\section{Overall summary and conclusion}

Massive stars from slightly above about $10 \, {\rm M_{\odot}}$ (Heger et al. 2003) explode as supernovae that produce powerful shock-waves which in turn accelerate cosmic ray particles.  In the arguments presented here we have not addressed the cosmic ray population produced by explosions of star between $10 \, {\rm M_{\odot}}$ and $25 \, {\rm M_{\odot}}$, since we derived the explanation for the (i) CR {\it knee} and CR {\it ankle} energies, (ii) the CR spectra, and (iii) the CR abundances using only SN-shocks in winds. Powerful stellar winds are produced by stars above a ZAMS mass of $25 \, {\rm M_{\odot}}$.  However, these abundances in the Binns-diagram all refer to elements heavier than Helium, which implies that for those elements the wind-SN-CRs are the key for our understanding.  Explosions of the stars in this ZAMS mass range between $10 \, {\rm M_{\odot}}$ and $25 \, {\rm M_{\odot}}$ were dealt with in Biermann \& Strom (1993), and Biermann (1994a, b, 1997), giving a prediction of an observed spectrum of $E^{-2.75 \pm 0.04}$, and a cutoff well below the knee energy; if this ISM-SN-CR population also had a polar-cap component, it would extend over only a small energy range, and so be undetectable underneath the two wind-SN-CR components.  In Nath et al. (2012) we showed that the interaction of this CR population in the average ISM is small.  What is interesting is that the AMS data allow an understanding of the lower energy proton and Helium spectra as secondaries; if that were really true, then all the CRs from explosions of stars between $10 \, M_{\odot}$ and $25 \; M_{\odot}$ would contribute CRs, as visible in gamma-ray data, but not any significant contribution above about $20 \, Z$ GeV to the directly observed CRs.  In fact, most of the observed SN-explosions in our galaxy, with the exception of the PeV H.E.S.S. source in the Galactic Center (Abramowski et al. 2016) seem to refer to explosions of stars in the mass range  between $10 \, M_{\odot}$ and $25 \; M_{\odot}$; thus many attempts to reconcile the observed CRs spectra usually relate to the directly observed SN-explosions, and not to the rare explosions of stars above $25 \, M_{\odot}$.  So we conclude on this point that this needs a further extension of an analysis such as done in Dembinski et al. (2017) at somewhat lower energy.  We note that the PAMELA results indicate a flatter spectrum for both cosmic ray protons and Helium nuclei (Karelin et al. 2016), a possible discrepancy in data that needs resolving.  It could also be useful to compare the AMS data results with those obtained from the $\gamma$-ray data of SN-explosions in our Galactic neighborhood, for which we usually know whether the progenitor star was in the mass range of between $10 \, M_{\odot}$ and $25 \; M_{\odot}$. Mandelartz \& Becker Tjus (2015) and Becker Tjus et al. (2016) have done such analysis and find that the observed $\gamma$-ray spectra suggest a typical CR spectrum in the sources in the range $E^{-2.2}$ to $E^{-2.4}$, after including transport (Kolmogorov) $E^{-2.53}$ to $E^{-2.73}$, hence too flat as compared with an observed spectrum of $E^{-2.85}$, but quite compatible with the PAMELA spectra (Karelin et al. 2016).  Furthermore, in agreement with arguments by Zank \& V{\"o}lk (1988) they find that the spectra cannot be described by a single slope, but cover a fairly large range of slopes.  Therefore, at this time the uncertainties are too large to make any firm conclusion here.  Considering the systematics in the errors, all the data are consistent with the conclusion that the lower energy proton and Helium CR particles derived from this range of ZAMS masses of massive stars, between $10 \, M_{\odot}$ and $25 \; M_{\odot}$, and the predicted spectrum is consistent with the observed spectrum.

Regarding explosions of very massive stars, with a ZAMS mass above $25 \, M_{\odot}$, we consider the explosion (section 2), the ensuing shock racing through the stellar wind, and the CR acceleration in that shock.  The supernova shock speeds through the wind; the magnetic fields in the wind have been determined through radio interferometric observations, both for blue super-giant star (BSG) explosions as well as for red super-giant (RSG) star explosions. Numbers for both kinds of stars are essentially the same within the limited statistics available, with upstream supernova shock velocity $U_{sh, 1} \, \simeq \, 0.1 \, c$, at radius $r \, \simeq \, 10^{16} \, {\rm cm}$ the downstream magnetic field B$ \, \simeq \, {\rm 1 \, Gau{ss}}$, and a general run of the magnetic field with radius $r$ of about $r^{-1}$, supported by the data of the compact radio sources in the starburst galaxy M82 for BSG stars.  The implied characteristic particle energies are $E_{ankle}  \;  = \; (1/8) \, e \, B(r) \, r \; = \; 10^{17.5 \pm 0.2} \, Z$ eV; and $E_{knee} \; = \; E_{ankle} \, (U_{sh,1}/c)^2 \;  = \; 10^{15.9 \pm 0.2} \, Z$ eV, consistent with the energies for ankle and knee determined from CR data; these energies run with the charge $Z$ of a cosmic ray nucleus.  Here we generalize our approach of 1993 (Biermann 1993), to an alternative of a fully chaotic description, where we also have magnetic fields nearly perpendicular to the shock normal over most of $4 \pi$, and magnetic fields parallel to the shock normal for a small fraction of the surface, like a small number of magnetic islands; this implies, for instance, that drift acceleration contributes significantly to the energy gain of particles; curvature drifts and gradient drifts are both important (Jokipii 1982, 1987, Biermann 1993, Le Roux et al. 2015, Zank et al. 2015). Drift energy gains are reduced beyond the knee-energy $E_{knee}$, and so the power-law spectrum turns to a steeper power-law; it is just a fraction of the energy gain that is lost at higher energy at each shock crossing.  So we propose that there are always two components of wind-SN cosmic rays: the polar cap component with a $E^{-2}$ spectrum at source, and the $4 \, \pi$ component, which has $E^{-7/3}$ at source; the polar cap component cuts off at the knee energy, and the $4 \, \pi$ component turns down to a steeper power-law with $E^{-2.85}$ at the source; above we also discuss the errors inherent in such predictions.  The steep energetic electron spectrum observed for the early SN radio emission can be interpreted by noting that the electrons necessary to explain the observed radio emission are so low in energy, that they do not ``see" the shock, so that their energy gain is only from drifts.  

In section 3 we use the AMS data to differentiate the two cosmic ray spectral components of explosions into stellar winds and their environment: Here we use these two CR-components to derive the wave-spectra they excite: These wave-spectra in turn determine different secondary spectra of CRs. Thus there are four possible combinations of excited wave-spectra and CR-spectra that they act upon.  The anti-proton data can be explained by considering the slowing SN-shock going through the dense wind of a RSG star, and so interacting considerably more than in the tenuous wind of a BSG star wind.  The positrons can be explained by noting that the cosmic ray electrons of the polar cap component interact with the surrounding photon field with triplet-pair production. 

In section 4 we concentrate on testing this paradigm by proposing a theory of injection based on the combined effect of the first and second ionization potential, to reproduce the plots of the ratio of CR source abundances to the abundances in the putative source material obtained by Binns et al. (Binns et al. 2001, 2005, 2006, 2007, 2008, 2011, 2013; George et al. 2001; Wiedenbeck et al. 2003; Rauch et al. 2007; Murphy et al. 2016).  We can interpret the abundance ratio data by requiring the total number of ions to be enhanced by the simple factor of $(Q_{0} \, A)^2$ (where $Q_{0}$ is the initial degree of ionization, and $A$ is the mass number).  Normalizing to a spectrum in energy per nucleon we get an additional factor depending on $A^{-4/3}$, obtaining finally a factor between cosmic ray abundances and source abundances of $Q_{0}^2 \, A^{+2/3}$.  This interpretation implies the high temperature in the winds of blue super-giant stars,  and requires that cosmic ray injection happens in the shock travelling through such a wind. Using the radio observations of the radio supernovae in the starburst galaxy M82 we can check the magnetic field at the stage when the supernova shock stalls at the environmental wind shell (Kronberg et al. 1985, Bartel et al. 1987, Allen \& Kronberg 1998):  this confirms the numbers for radial scale and magnetic field, implying that the piston mass is in fact large enough and that the magnetic field in the SN-induced shock in the wind runs approximately as $r^{-1}$ all the way out for BSG star winds.   Further critical tests are given by other cosmic ray experiments such as e.g. CREAM, TIGER, Super-TIGER, PAMELA, ISS-CREAM, Kaskade-Grande, IceTop/IceCube, LOFAR, the Telescope Array (TA), and  Auger.  Thus, by studying the cosmic ray particles, their abundances at knee energies, and their spectra, we can learn about what drives these stars to produce the most powerful outbursts yet known in the universe.  A key expectation is that by determining cosmic ray isotopic abundances towards the knee, and possibly beyond, we learn more about what happens close to the core of the exploding star, in some cases close to a budding stellar black hole.

In a last major section (section 5) we explore the possible interpretation of the compact radio source 41.9+58 in the starburst galaxy M82 as a recent GW event, which occurred ca. 100 years ago.  Massive stars above about $25 \, {\rm M_{\odot}}$, depending on their initial heavy element abundance, commonly produce stellar black holes in their supernova explosions (Woosley \& Heger 2002, Heger et al. 2003, Chieffi \& Limongi 2013).  In dense stellar clusters of massive stars, encounters between binary systems can lead to an exchange of stars, and so may lead to a tight black hole binary (for a discussion of encounters in dense stellar systems, see Bally et al. 2017; and for a discussion of explosions in such an environment Bykov et al. 2017b). A black hole binary system resulting from such an exchange encounter between binary star systems is likely to produce uncorrelated spin directions for the two black holes: This is argued above (using Gergely \& Biermann 2009) to explain the compact radio source 41.9+58 in the starburst galaxy M82 (Kronberg et al. 1985, Bartel et al. 1987, Allen \& Kronberg 1998, Muxlow et al. 1994, 2005, 2010, Gendre et al. 2013); it is also required to explain the third GW event (Abbott et al. 2017a).  Two such black holes or neutron stars in a tight binary system finally merge in a gigantic emission of gravitational waves, now observed (Abbott et al. 2016a, b, 2017a, b, c, d).  If the spins of the two merging stellar mass black holes or neutron stars are uncorrelated, their relativistic jets carve out a cone, visible in the radio data (Kronberg et al. 1985) around the compact source 41.9+58 in the starburst galaxy M82. If accompanied by a GRB, following the merger a hyper-relativistic jet may have accelerated the UHECR protons detected by TA. There should be analogously high-energy neutrinos, possibly pointing elsewhere.   If this reasoning correctly interprets what happened in M82, the event ought to be detectable with interferometric radio data: it also follows that this type of event must be quite common in starburst galaxies.  This speculation leads us to expect that such an event will be detected via gravitational wave detectors, $\gamma$-ray detectors, in UHECR particles (Aab et al. 2018) and other telescopes within a few years.

\vskip2.0cm

\section{Acknowledgement}

Andreas Brunthaler supplied the first inspiring set of radio data of SNe.  W. Bob Binns gave a beautiful challenge in two lectures at Erice July 2014, and pursued an extensive and ongoing exchange over email; he also provided the final version of his graph and much further information.  Jon Paul Lundquist and Pierre Sokolsky gave a lot of background to the idea that TA may have detected UHECRs from the starburst galaxy M82.  During and after a lecture on the proposal to explain the TA results contained in the section on M82 above at the University of Utah on May 11, 2017, P. Sokolsky and G. Thomson gave many helpful comments.  Gabriela E. P{\v a}v{\v a}la{\c{s}} determined the energy budget for wind-SN-CRs versus ISM-SN-CRs in her M.Sc. thesis some years ago.  Eun-Suk Seo has accompanied the author PLB for many decades in discussions of CR physics.  The author PLB is grateful to all of them.

The author PLB wishes to thank N. Barghouty, N. Bartel, G. Bisnovatyi-Kogan, H. Bluemer,  R. Chini, R. Engel, H. Falcke, C. Galea, F. Giovanelli, N. Grevesse, B. Harms, A. Haungs,  M. Joyce, K.H. Kampert, N. Langer, A. Lazarian, I.F. Mirabel, J. Rachen, W. Rhode, P. Rochus, N. Sanchez, P. Sokolsky, V. de Souza, G. Thomson, and H. de Vega for intense discussions.

P.L. Biermann is member of the Auger-Coll., the JEM-EUSO-Coll., and the LOPES-Coll.; J. Becker Tjus is member of the IceCube-Coll. and H.E.S.S.-Coll.; W. de Boer and I. Gebauer are members of the AMS-Coll.; additional support from the DFG in Germany (Grant BO 1604/3-1) is warmly acknowledged; L.I. Caramete is member of the ANTARES-Coll., the KM3Net-Coll., and JEM-EUSO-Coll.; R. Diehl is Co-PI of the INTEGRAL gamma-ray spectrometer SPI, with support from the European Space Agency ESA, and DLR in Germany; L.{\'A.} Gergely is member of the LIGO-Coll.; A. Meli is member of the IceCube-Coll; T. Stanev is member of the IceCube-Coll., as well as the Sybill-Coll..

Comments by two referees greatly helped to improve the paper; we are grateful for their detailed work.

\end{document}